\documentclass[aps,pra,10pt,superscriptaddress,twocolumn,nofootinbib,floatfix]{revtex4-2}
\usepackage{graphicx}
\usepackage{amsmath,amsthm,amssymb,dsfont,float,placeins,enumitem,soul}
\usepackage{mdframed}
\usepackage{orcidlink}
\usepackage{appendix}
\usepackage{tikz-cd}
\usepackage{xcolor}
\usepackage{hyperref}
\usepackage[capitalise]{cleveref}
\crefname{equation}{Eq.}{Eqs.}
\crefname{figure}{Fig.}{Figs.}

\usepackage[authormarkuptext=name,commentmarkup=uwave]{changes}
\definechangesauthor[name={JRH},color=green!70!black]{jonte}

\providecommand{\HLMODE}{0}
\ifnum\HLMODE=1
\newcommand{\rev}[1]{\hl{#1}}
\else
\newcommand{\rev}[1]{#1}
\fi

\newcommand{\ket}[1]{\left| #1 \right\rangle}
\newcommand{\bra}[1]{\left\langle #1 \right|}

\newcommand{\av}[1]{\langle #1\rangle}
\newcommand{\braket}[2]{\langle #1|#2 \rangle}
\newcommand{\ketbra}[2]{\left|#1\right\rangle\hskip-1mm\left\langle#2\right|}

\newcommand{\id}{{\mathds{1}}}
\newcommand{\Tr}{\mathrm{Tr}}
\newcommand{\NC}{\mathcal{N}}
\newcommand{\EC}{\mathcal{E}_{\mathcal{C}}}
\newcommand{\Zd}{\mathbb{Z}_d}

\theoremstyle{plain}
\newtheorem{theorem}{Theorem}[section]
\newtheorem{proposition}{Proposition}[section]
\newtheorem{lemma}{Lemma}[section]
\newtheorem{corollary}{Corollary}[section]
\theoremstyle{definition}
\newtheorem{definition}{Definition}[section]
\theoremstyle{remark}
\newtheorem{remark}{Remark}[section]

\begin{document}

\title{Phase-Space Representations of Quantum Error-Correcting Codes}
\author{Enrico Bozzetto}
\affiliation{DET, Politecnico di Torino, Corso Duca degli Abruzzi, 24, 10129 Torino, Italy}
\affiliation{Quantum Group, School of Computing, Newcastle University, 1 Science Square, Newcastle upon Tyne, NE4 5TG, UK}
\author{Jonte R. Hance\,\orcidlink{0000-0001-8587-7618}}
\email{jonte.hance@newcastle.ac.uk}
\affiliation{Quantum Group, School of Computing, Newcastle University, 1 Science Square, Newcastle upon Tyne, NE4 5TG, UK}

\begin{abstract}
We connect the structure theorem for quasiprobability representations to quantum error correction. For a code space invariant under a subgroup of an extended Weyl-Heisenberg group, the Brif-Mann construction gives a semi-functorial representation whenever its kernel satisfies the Stratonovich-Weyl axioms, and composes on channels covering full error-correction cycles. With phase-space parity available, a kernel other than the ordinary Wigner one exists iff the code space is invariant under a lattice of displacements: among compact-phase-space bosonic codes, this is only the ideal Gottesman-Kitaev-Preskill family; in finite dimension, this is every qudit stabiliser code, whose negative volume for odd prime dimension is the syndrome-averaged logical magic. For rotation-symmetric, one-photon, non-Pauli qudit and spin codes only the ordinary kernel remains, its negativity reflecting the physical carrier, not logical content; a sector-graded representation built from recovery isometries restores a resource reading. We add a channel atlas, a closed-form magic lifetime under displacement noise, a Kirkwood-Dirac layer, and finite-energy corrections.
\end{abstract}

\maketitle

\section{Introduction}\label{sec:intro}

Phase-space representations turn operator algebra into function theory. States become real functions on a phase space, channels become maps of those functions, and whatever separates the theory from a classical one shows up as negative values~\cite{Wigner1932, Gross2006DVWignerF, Veitch2012BoundMagic}. This raises the question of whether we could use such a representation to identify encoded resources for quantum computation in a quantum error correction scheme. For this, we would need a representation adapted to the code used.

Brif and Mann~\cite{Brif1999BrifMannConstruction} previously constructed a phase-space representation for any system with a Lie-group dynamical symmetry and a chosen isotropy subgroup, and Davis, Fabre and Chabaud~\cite{davis2024identifyingquantumresourcesencoded} applied that construction to the Gottesman-Kitaev-Preskill (GKP) code~\cite{Gottesman2001GKPCodes}, obtaining the Zak-Gross Wigner function, whose negativity is the magic of the encoded logical state. Separately, Schmid \emph{et al.}~\cite{Schmid2024StructureTheorem} and Wagner \emph{et al.}~\cite{wagner2025ComplexStructTheorem} proved structure theorems which fix the form of any linearity-preserving, empirically adequate (semi-)functorial quasiprobability representation of a generalised probabilistic theory. Combining these presents a way to determine the action of an error channel on a code-adapted phase space, valid for whichever code the construction is fed.

However, as we show in this paper, putting these two ingredients together to obtain such a phase-space representation for any quantum error-correcting code is not so simple. The Stratonovich-Weyl (SW) axioms, together with the requirement that the reconstruction map be idempotent and faithful on the code algebra, \rev{strongly restrict the kernel (\mbox{\Cref{prop: classification}} lists the hypotheses in full, including a continuity condition on the kernel and, for non-normalisable codewords, the form we assume for the reconstruction map)}. Assuming the code's symmetry makes the phase-space parity available, as it does for every family we consider, a code-adapted kernel differing from the ordinary Wigner kernel exists if and only if the code space is invariant under a lattice of displacements in the relevant Weyl-Heisenberg group.

In the continuous-variable (CV) setting, no finite-dimensional space of normalisable states can be displacement-invariant, so we only get a code-adapted kernel differing from the ordinary Wigner kernel if the codeword of that code are not elements of $L^2(\mathbb{R})$: the only bosonic codes with compact phase space that fulfil this (and so have a code-adapted kernel \rev{of their own}) are the ideal, Dirac-comb GKP family. In the discrete-variable setting however this criterion is met generically, since the finite Weyl-Heisenberg group is two-step nilpotent and every subgroup containing the centre is normal: every $[[n,k]]_d$ stabiliser code of prime-power dimension has a code-adapted SW kernel, where the code's phase space factorises into syndrome and logical sectors, errors act as rigid translations, and the corrected cycle is the identity. For odd prime $d$ the negative volume of an arbitrary state is then its syndrome-probability-weighted logical magic (for $d=2$ this interpretation of the negative volume sadly does not hold).

Notably, our result also proves we cannot obtain a code-adapted kernel differing from the ordinary Wigner kernel for rotation-symmetric bosonic codes such as the cat and binomial families, one-photon codes, finite-energy GKP codewords, qudit codes stabilised by non-Pauli symmetries, or spin codes. For the first four the SW axioms themselves\rev{, under the hypotheses of \mbox{\Cref{prop: classification,prop:nonpauli}},} leave only the standard kernel: with no freedom at all in the bosonic cases, and up to a code-independent sign family in the finite-dimensional one. For those four no code-adapted representation exists, so the negativity of their (ordinary Wigner kernel) representation just measures the nonclassicality of the physical carrier, rather than any property of the logical system. For spin codes, the averaging prescription supplies no code-adapted kernel at all, leaving the ordinary spherical kernel in place but not giving us the same axioms-direct no-go as we have for the other codes.

However, even for codes for which we cannot construct a code-adapted kernel differing from the ordinary Wigner kernel, we show we can still identify encoded resources through another route. Building the representation from a code's recovery isometries instead of from the symmetry of its code space gives a sector-graded logical kernel which inherits the semi-functorial and compositional structure of the framework, and whose negativity is a syndrome-resolved witness of logical magic, two-sided on pure logical states, for rotation codes encoding a qudit of odd prime dimension. However, in this representation, errors don't apply rigid translations, and the representation is defined relative to a recovery scheme. This is consistent with the no-go theorem, which requires covariance under the physical Weyl-Heisenberg group; the sector-graded kernel is not covariant in that sense.

The paper is organised as follows. \Cref{sec:background} gives background on quasiprobability distributions and the normalisation conventions we use, the real and complex structure theorems, and the SW axioms with the CV and DV Wigner functions. \Cref{sec:framework} sets up the code-adapted construction, classifies the kernels the axioms permit, and identifies the class of channels on which the representation composes correctly. \Cref{sec:lattice} covers codes which allow a code-adapted kernel differing from the ordinary Wigner kernel: ideal GKP codes, every qudit stabiliser code, how we identify resources for these codes, multimode and concatenated lattices, and the dissipation of encoded magic under displacement noise. \Cref{sec:nonlattice} covers codes which don't allow a code-adapted kernel differing from the ordinary Wigner kernel, and \Cref{sec:seclog} gives the sector-graded rescue. \Cref{sec:atlas} gives the channel atlas: a reflection identity from which every $k$-photon jump kernel follows, the gain and dephasing channels, and a comparison between the cat and binomial codes at matched mean photon number. \Cref{sec:FiniteEnergy} covers finite-energy codewords and the imperfections of a realistic cycle. \Cref{sec:kd} builds a code-adapted Kirkwood-Dirac layer. \Cref{sec:Discussion} summarises and discusses future works.

We take from previous work: the real and complex structure theorems~\cite{Schmid2024StructureTheorem, wagner2025ComplexStructTheorem}; the Brif-Mann construction~\cite{Brif1999BrifMannConstruction}; the Zak-Gross Wigner function for GKP codes together with its SW properties and its negative-volume interpretation~\cite{davis2024identifyingquantumresourcesencoded}; the Gross Wigner function and the discrete Hudson theorem~\cite{Gross2006DVWignerF}; the lattice-theoretic account of GKP concatenation~\cite{Conrad2022}; phase-space accounts of stabiliser states and qubit codes in a fixed finite-field Wigner function (\cite{Paz2005, GHW2004}); and the GKP bridge between the continuous and discrete Wigner functions (\cite{FengLuo2024}). New to this work are: the proof that every Brif-Mann representation with a Stratonovich-Weyl kernel satisfies the structure-theorem conditions (\Cref{subsec:semifunctor}); the identification of the class of transformations on which the representation is compositional (\Cref{subsec:GradedClass}); the classification of the SW kernels available to a code (\Cref{prop: classification,cor: classification}) and the no-go theorem that follows from it (\Cref{thm: nogo}); the code-adapted quotient representation for qudit stabiliser codes and the resource identification of \Cref{thm:dvmagic}; the sector-graded logical representation (\Cref{sec:seclog}); the negative results for one-photon codes, spin codes and non-Pauli qudit symmetries (\Cref{subsec:dualrail,subsec:spin,subsec:nonpauli}); the determination of the cat and binomial representations and the analysis of the channels acting on them (\Cref{subsec:cat,subsec:bin,sec:atlas}); the reflection identity \eqref{eq:reflection} and the jump kernels it generates; the closed-form magic dissipation and lifetime of \Cref{prop:dissipation}; the code-adapted Kirkwood-Dirac layer (\Cref{sec:kd}); and the account of finite-energy codewords and realistic syndrome extraction (\Cref{sec:FiniteEnergy}).

\section{Background}\label{sec:background}

\subsection{Quasiprobability distributions}\label{subsec:quasiprobs}

A quasiprobability distribution relaxes at least one of the Kolmogorov axioms of probability theory~\cite{Kolmogorov1950foundations}, namely non-negativity, unit total measure and countable additivity, and the known representations of quantum mechanics~\cite{Wigner1932, Kirkwood1933KD, Dirac1945KDDistr, Harriman1993HusiniQDistr} each violate at least one. Every representation used here relaxes only non-negativity: each is real-valued and linear in the state, so additivity holds, and each satisfies the Stratonovich-Weyl standardisation axiom (\Cref{subsec:Wigner}), so that
\begin{equation}
\label{eq: normalisation convention}
\int_{X^{\mathcal{C}}}d\mu(\Omega)\,W_{\hat{\rho}}^{\mathcal{C}}(\Omega)=\mathrm{Tr}[\hat{\rho}]=1
\end{equation}
for any normalised state, where $d\mu$ is the invariant measure on the relevant phase space (explicitly, $\frac{1}{d}dudv$ on the torus $\mathbb{T}^2_{dl}$ in the GKP case of \Cref{subsec:ZGW}).

We represent conditional as well as unconditional processes, so we distinguish throughout a \emph{normalised state} (a density operator of unit trace) and its \emph{normalised quasiprobability distribution}, obeying \Cref{eq: normalisation convention}, from an \emph{unnormalised conditional state} (the output of a single Kraus operator, whose trace is the probability of that outcome) and its \emph{subnormalised phase-space function}, whose integral against $d\mu$ is that probability rather than unity. We say at each point which is meant.

\subsection{Real and complex structure theorems}\label{subsec:structure}

Ref.~\cite{Schmid2024StructureTheorem} gave a structure theorem for real-valued, finite-dimensional, identity-preserving quasiprobability representations of generalised probabilistic theories (GPTs), fixing any such functorial representation as $Q(T)=\chi_B\circ T\circ \chi_A^{-1}$ for every transformation $T:A\rightarrow B$, with $\chi_A$ the linear map taking an operational state of $A$ into the ontological vector space. The structure theorem applies to maps which are empirically adequate (acting as the identity on every element of $\textbf{G}(I,I)$, the set of scalars of the GPT, whose values are the outcome probabilities; equivalently, reproducing the outcome probabilities of the operational theory) and linearity-preserving ($\Phi(\alpha T_1+\beta T_2)=\alpha \Phi(T_1)+\beta \Phi(T_2)$ for all $\alpha,\beta\in\mathbb{R}$ and $T_1,T_2\in \textbf{G}(A,B)$, the set of processes from $A$ to $B$), and it holds for functors: maps preserving sequential composition and the identity on every system. A map preserving only the former is a semi-functor.

Ref.~\cite{wagner2025ComplexStructTheorem} generalised this to complex-valued representations (through the complexification functor $\mathbb{C}$), which admits the Kirkwood-Dirac distribution~\cite{Kirkwood1933KD, Dirac1945KDDistr} used in \Cref{sec:kd}, and to semi-functors, which make the framework relevant here, since the code-adapted representations we use need not preserve the global identity. The resulting form is
\begin{equation}
\label{eq: complex structure theorem}
    Q(T)=\chi_B\circ \mathbb{C}(T)\circ \phi_A,
    \qquad
    \phi_A=\underline\chi_A^{-1}\circ Q(\mathds{1}_A),
\end{equation}
where $\underline\chi_A$ is the surjective corestriction of $\chi_A$ (which is injective, so the corestriction is invertible). The theorem has already been used to study the dissipation of contextuality in a magic state under decoherence~\cite{bozzetto2026classicallimitdissipationspekkens}.

Both theorems are proved for tomographically local, finite-dimensional GPTs. Ref.~\cite{wagner2025ComplexStructTheorem} allows the representation spaces to be infinite-dimensional, but the operational theory itself must be finite-dimensional, whereas our CV sections work on $\mathcal{B}(L^2(\mathbb{R}))$. We therefore do not invoke the structure theorems as black boxes in the infinite-dimensional setting. Following Ref.~\cite{davis2024identifyingquantumresourcesencoded}, we verify the properties they require (empirical adequacy, linearity preservation, and the composition law) directly for each representation we use (\Cref{subsec:semifunctor,subsec:GradedClass}), and use the structure-theorem form of the maps as the organising principle. In the finite-dimensional setting of \Cref{subsec:dv} however, the theorems apply directly.

\subsection{Wigner functions and the Stratonovich-Weyl axioms}\label{subsec:Wigner}

The Wigner function $W$~\cite{Wigner1932} is a phase-space representation. Acting on an operator $\hat{A}$ on a Hilbert space $\mathcal{H}$, it must satisfy the Stratonovich-Weyl (SW) axioms to be a faithful phase-space representation of $\hat{A}$:
\begin{enumerate}[label=Axiom~\arabic*,leftmargin=*,itemsep=1pt]
    \item (Linearity): $\hat{A}\mapsto W_{\hat{A}}$ is a linear map with $W_{\hat{A}}=W_{\hat{B}}$ iff $\hat{A}=\hat{B}$.\label{SWAxiom1}
    \item (Reality): $W_{\hat{A}}=(W_{\hat{A}})^*$.\label{SWAxiom2}
    \item (Standardisation): $\int_X d\mu\, W_{\hat{A}}=\text{Tr}(\hat{A})$, with $X$ the whole phase space.\label{SWAxiom3}
    \item (Covariance): $W_{\pi(g)^\dagger \hat{A}\pi(g)}(\Omega)=W_{\hat{A}}(g\cdot \Omega)$ for all $\Omega\in X$ and $g\in G$, where $\pi$ is an irreducible unitary representation on $\mathcal{H}$ of a finite-dimensional Lie group $G$, and $g\cdot \Omega$ is the natural action of $G$ on $X$.\label{SWAxiom4}
    \item (Traciality): $\int_X d\mu\, W_{\hat{A}}W_{\hat{B}}=\text{Tr}(\hat{A}\hat{B})$.\label{SWAxiom5}
\end{enumerate}
\labelcref{SWAxiom4} is typically stated for a simply connected group. Nothing below depends on connectedness, since the axiom enters only through the action of $G$ on $X^{\mathcal{C}}=G/H^{\mathcal{C}}$ and through the invariant measure $d\mu$, neither affected by a finite extension. We therefore read it for a finite-dimensional Lie group with finitely many components, as the rotation-symmetric codes of \Cref{subsec:cat,subsec:bin} need ($G=H_3(\mathbb{C})\rtimes\mathbb{Z}_M$, not connected), and drop ``simply'' as well, for $\mathrm{Mp}(2,\mathbb{R})$ in \Cref{prop: metaplectic} (connected but not simply connected).

Note a code-adapted representation cannot satisfy \labelcref{SWAxiom1,SWAxiom5} in the form just stated. Wherever the reconstruction map $\mathcal{E}_{\mathcal{C}}=\phi_{\mathcal{C}}\circ\chi_{\mathcal{C}}$ of \Cref{subsec:semifunctor} is a non-trivial twirl, $\hat{A}\mapsto W^{\mathcal{C}}_{\hat{A}}$ is not injective, since $W^{\mathcal{C}}_{\hat{A}}=W^{\mathcal{C}}_{\mathcal{E}_{\mathcal{C}}(\hat{A})}$ for every $\hat{A}$; and traciality fails on $\mathcal{B}(\mathcal{H})$ for the same reason, holding instead in the form $\int d\mu\,W_{\hat{A}}W_{\hat{B}}=\Tr[\hat{A}\,\mathcal{E}_{\mathcal{C}}(\hat{B})]$. We therefore use throughout the \textit{code-faithful} reading of \labelcref{SWAxiom1,SWAxiom5}: injectivity in \labelcref{SWAxiom1}, and the pairing in \labelcref{SWAxiom5}, are required on the $\mathcal{E}_{\mathcal{C}}$-invariant subalgebra rather than on all of $\mathcal{B}(\mathcal{H})$. The representation resolves that subalgebra, and the subalgebra contains the code algebra $\mathcal{B}(\mathcal{C})$, which is what ``faithful on the code'' means. Injectivity in this form is not an extra assumption: self-duality of the kernel makes $\chi_{\mathcal{C}}$ injective on the span of the kernels, which is that subalgebra, so the clause we use in \Cref{prop: classification,thm: nogo}, that $\mathcal{E}_{\mathcal{C}}$ is idempotent and acts as the identity on $\mathcal{B}(\mathcal{C})$, is all \rev{that this reading adds to \mbox{\labelcref{SWAxiom2,SWAxiom3,SWAxiom4}}. It is an assumption on the reconstruction map, not a consequence of those axioms: Appendix~\mbox{\ref{app: proofs}} gives the Husimi kernel as a counterexample. \mbox{\Cref{prop: classification,thm: nogo}} need further hypotheses (the parity, continuity of the kernel's characteristic function and, where that characteristic function exists only as a distribution, either complete positivity of $\mathcal{E}_{\mathcal{C}}$ or its form as a conditional expectation). Both results list their hypotheses in their statements, and say which properties of $\mathcal{E}_{\mathcal{C}}$ are derived and which are assumed.} \labelcref{SWAxiom2,SWAxiom3,SWAxiom4} are used unchanged. Where $\mathcal{E}_{\mathcal{C}}={\id}$, which by \Cref{prop: classification} is every bosonic code with codewords in $L^2(\mathbb{R})$, the two readings coincide and the representation is faithful on the whole algebra. Whenever we say below that a kernel satisfies the Stratonovich-Weyl axioms, we mean this reading (\Cref{prop: classification,thm: nogo} state it in full).

The Wigner function was originally a continuous-variable (CV) object; a discrete-variable (DV) analogue was later given by Gross~\cite{Gross2006DVWignerF}. Displacement operators play a key role in both:
\begin{equation}
\begin{aligned}
    \hat{D}(x,p)&=e^{i\frac{p\hat{x}-x\hat{p}}{\hbar}}={e^{-i\frac{xp}{2\hbar}}}e^{i\frac{p\hat{x}}{\hbar}}e^{-i\frac{x\hat{p}}\hbar},\quad x,p\in\mathbb{R},\\
    \hat{D}_d(a,b)&=\omega^{2^{-1}ab}\hat{X}^a\hat{Z}^b,\quad a,b\in \mathbb{Z}_d,
\end{aligned}
\end{equation}
where $\hat{X},\hat{Z}$ are the generalised Pauli operators, $\omega=e^{i2\pi/d}$, and $2^{-1}=\frac{d+1}{2}$ is the inverse of $2$ in $\mathbb{Z}_d$ for odd $d$. We set $\hbar=1$ throughout. The undisplaced parity operators are fixed by $\hat{\Pi}(0,0)\ket{x}_{\hat{x}}=\ket{-x}_{\hat{x}}$ and $\hat{\Pi}_d(0,0)\ket{j}=\ket{-j}$, with $\{\ket{x}_{\hat{x}}\}$ the position basis, $\{\ket{j}\}_{j=0}^{d-1}$ the computational basis and $-j\cong d-j$; they are Fourier transforms of the displacements, $\hat{\Pi}(0,0)=\frac{1}{4\pi}\int dz\,\hat{D}(z)$ and $\hat{\Pi}_d(0,0)=\frac{1}{d}\sum_z \hat{D}_d(z)$, and are displaced by conjugation:
\begin{equation}
\begin{aligned}
    \hat{\Pi}(z)&=\hat{D}(z)\hat{\Pi}(0,0)\hat{D}^\dagger(z)\\
    &=\frac{1}{4\pi}\int dz'\, \hat{D}(z')e^{-iz^T\sigma z'},\\
 \hat{\Pi}_d(z)&=\hat{D}_d(z)\hat{\Pi}_d(0,0)\hat{D}_d^\dagger(z),
\end{aligned}
\end{equation}
with $z=(x,p)^T\in\mathbb{R}^2$ and $z=(a,b)^T\in\mathbb{Z}_d^2$ respectively, and $\sigma=\left(\begin{smallmatrix}0&1\\-1&0\end{smallmatrix}\right)$ the standard symplectic matrix. The characteristic and Wigner functions are then
\begin{equation}
\label{eq: char function and Wigner function}
\begin{aligned}
\chi_{\hat{\rho}}^{CV}(z)&=\text{Tr}\left[\hat{D}(z)\hat{\rho}\right], &W_{\hat{\rho}}^{CV}(z)&=\text{Tr}\left[2\hat{\Pi}(z)\hat{\rho}\right],\\
\chi_{\hat{\rho}}^{DV}(z)&=\text{Tr}\left[\hat{D}_d(z)\hat{\rho}\right], &W_{\hat{\rho}}^{DV}(z)&=\text{Tr}\left[\hat{\Pi}_d(z)\hat{\rho}\right],
\end{aligned}
\end{equation}

{The measure of \Cref{eq: normalisation convention} is $dz=dx\,dp$ with $d\mu=dz/2\pi$; since $\int dz\,\hat{\Pi}(z)=\pi\id$ this gives $\int d\mu\,W^{CV}_{\hat{\rho}}=\mathrm{Tr}[\hat{\rho}]$, and the kernel $2\hat{\Pi}(z)$ is self-dual with respect to it, so \labelcref{SWAxiom5} holds in the form $\int d\mu\,W_{\hat{A}}W_{\hat{B}}=\mathrm{Tr}[\hat{A}\hat{B}]$. In the DV case the measure is $\frac{1}{d}\sum_z$, with kernel $\hat{\Pi}_d(z)$. From \Cref{sec:framework} onwards we use complex-plane notation, in which $dz=2\,d^2\beta$, $d\mu=d^2\beta/\pi$ and $\hat{\Delta}_{\mathrm{std}}(\beta)=2\hat{\Pi}(\beta)$: the same kernel and measure in other coordinates. The unnormalised object $W^{\mathrm{std}}_{\hat{\rho}}(\beta)=\frac{2}{\pi}\mathrm{Tr}[\hat{\Pi}(\beta)\hat{\rho}]=W^{CV}_{\hat{\rho}}/\pi$, which integrates to $\mathrm{Tr}[\hat{\rho}]$ against $d^2\beta$, is the standard Wigner function; we keep both and relate them at \Cref{eq: cat wigner as average}.}

{We also use the following properties of discrete-variable systems. For odd $d$ the Gross Wigner function is the unique covariant representation which is non-negative on all stabiliser states~\cite{Gross2006DVWignerF}, already under permutation symmetry alone~\cite{Zhu2016}, and its negativity is a magic monotone~\cite{Veitch2012BoundMagic, MariEisert2012, Howard2014}, which means the resource statements below are two-sided on pure logical states and one-sided on mixed ones, for the reasons we give in \Cref{subsec:ZGW}~\cite{Gross2006DVWignerF, HUDSON1974WignerGauss, Veitch2012BoundMagic}. Where we write $A(u)$ for the DV phase-point operators we mean $A(0)=d^{-n}\sum_{v\in\Zd^{2n}}D(v)$, the parity $\sum_j\ketbra{-j}{j}$, with $A(u)=D(u)A(0)D^\dagger(u)$, $W_{\hat{\rho}}(u)=d^{-n}\Tr[A(u)\hat{\rho}]$, $\Tr[A(u)A(u')]=d^n\delta_{uu'}$ and $\sum_uA(u)=d^n\id$; this is \Cref{eq: char function and Wigner function} written multiplicatively for $n$ qudits, with the $d^{-n}$ made explicit, so that $\sum_u W_{\hat\rho}(u)=\Tr\hat\rho$ against counting measure. Where the two normalisations meet, in \Cref{prop:concat}, we use this one.}

{We use $N$ in this paper in two different ways: as the number of coherent-state components wherever a cat code is under discussion, and as the order $\max\{L,G,2D\}$ of the codeword construction wherever a binomial code is; the binomial code is always written with its pair $(N,S)$, and where both families appear in one statement we name the harmonic or the component count rather than the symbol. $\mathcal{M}$ is the stabiliser lattice of a qudit stabiliser code and $\mathbb{Z}_M$ the rotation group of a rotation-symmetric bosonic code, $M$ being its order; the sector modulus of \Cref{sec:seclog} is written $N/d$. The central Stratonovich-Weyl kernel is $\hat{K}$ throughout, the jump operators of the amplitude-damping channel are $\hat{J}_\ell$, and the quadrature operator of \Cref{lem: nodisplacement} is $\hat{K}_\theta$. $Q$ with a map as its argument is the representation semi-functor of \Cref{eq: generalized structure theorem map}; the objects $Q_{\hat{\rho}}$ and $Q(s_1,s_2)$ of \Cref{sec:kd} are distributions, not values of it.}

{Finally, the negative volume of a normalised quasiprobability distribution is}
\begin{equation}
\label{eq: negative volume general}
{\NC_{\hat{\rho}}^{\mathcal{C}} :=\int_{X^{\mathcal{C}}} d\mu(\Omega)\left| W_{\hat{\rho}}^{\mathcal{C}}(\Omega)\right|-1,}
\end{equation}
{the Kenfack-\.{Z}yczkowski convention~\cite{Kenfack2004Negativity}. Since $\int d\mu |W| \ge |\int d\mu\, W| = 1$ for a normalised distribution, $\NC^{\mathcal{C}}_{\hat{\rho}}\ge0$, vanishing on non-negatively represented states. It has no universal upper bound.}

\section{Code-adapted phase-space representations}\label{sec:framework}

\Cref{subsec:construction} gives the Brif-Mann prescription and fixes which subgroup of the symmetry group the phase space is quotiented by. \Cref{subsec:semifunctor} shows that every representation the prescription produces, whose kernel satisfies the Stratonovich-Weyl axioms, meets the conditions the structure theorem requires. \Cref{subsec:averaging} shows that the natural group-averaging recipe for the kernel yields a Stratonovich-Weyl kernel when the code satisfies a normality condition (which every rotation-symmetric bosonic code fails and every stabiliser code satisfies). \Cref{subsec:classification} then drops the averaging recipe and asks what the axioms permit directly, which turns out to cover most cases. \Cref{subsec:GradedClass} identifies the class of channels on which the resulting representation composes correctly.

\subsection{The construction}\label{subsec:construction}

Consider a Hilbert space $\mathcal{H}$ carrying an irreducible unitary representation $\pi$ of a symmetry group $G$. Let us choose a code space $\mathcal{C}\subset\mathcal{H}$, which encodes the logical information. This in turn defines a subgroup $H^{\mathcal{C}}_0\subset G$, the set of elements $h$ such that

\begin{equation}
\label{eq: subgroup H^C}
\pi(h)\ketbra{\psi}{\psi}\pi^\dagger(h)=\ketbra{\psi}
{\psi},\quad \forall \ket\psi\in\mathcal{C}
\end{equation}

\Cref{eq: subgroup H^C} defines the subgroup $H^{\mathcal{C}}_0$ which fixes every state of $\mathcal{C}$, i.e., the subgroup of which $\mathcal{C}$ is (part of) the fixed-point space. The Brif-Mann construction of the phase space instead uses the isotropy subgroup of the chosen reference state,
\begin{equation}
\label{eq: isotropy of reference state}
\begin{split}
H^{\mathcal{C}}:=&\left\{h\in G\;:\;\pi(h)\ketbra{\psi_0}{\psi_0}\pi^\dagger(h)=\ketbra{\psi_0}{\psi_0}\right\}\\
\supseteq &H^{\mathcal{C}}_0
\end{split}
\end{equation}
since $H^{\mathcal{C}}$, not $H^{\mathcal{C}}_0$, must be quotiented out for the coherent states $\ket{\Omega}=\pi(\Omega)\ket{\psi_0}$ to be labelled by the points of $X^{\mathcal{C}}=G/H^{\mathcal{C}}$. For the GKP code the two coincide for a generic reference codeword, but not the stabiliser codewords: $\ket{\overline{0}}$ is fixed at the projector level by the logical $\hat{\bar{Z}}=e^{il\hat{x}}$ as well as by the stabilisers (its position comb sits at $x\in dl\,\mathbb{Z}$ and $l\cdot dl=2\pi$), so its isotropy subgroup is the rectangular lattice $dl\,\mathbb{Z}\times l\,\mathbb{Z}$, strictly larger than the stabiliser lattice. The ZGW construction quotients by the stabiliser lattice $H^{\mathcal{C}}_0$ (the isotropy subgroup of a reference codeword which is no eigenstate of any logical displacement), which it is allowed to do, as the resulting kernel does not depend on the reference state at all~\cite{davis2024identifyingquantumresourcesencoded}. The two subgroups also differ for \Cref{subsec:cat,subsec:bin}'s rotation-symmetric codes: there $H^{\mathcal{C}}$, the isotropy of the character codeword we take as reference, contains the rotation acting on the code space as the logical $\hat{\bar{Z}}$, and $H^{\mathcal{C}}_0$ is its index-two subgroup. We write $H^{\mathcal{C}}$ for the subgroup by which the phase space is quotiented (the reference-state isotropy for the rotation-symmetric codes, $H^{\mathcal{C}}_0$ for GKP).

We then follow Ref.~\cite{Brif1999BrifMannConstruction}. Pick a reference state $\ket{\psi_0}\in \mathcal{C}$; this gives a Wigner function depending on the choice of $\mathcal{C}$ and defined over the phase space $X^{\mathcal{C}}=G/H^{\mathcal{C}}$, with coherent states $\ket{\Omega}:=\pi(\Omega)\ket{\psi_0}$ for $\Omega\in X^{\mathcal{C}}$. The identity on the full Hilbert space $\mathcal{H}$ does not need to be preserved by the resulting function, making the construction a semi-functor rather than a functor in general; \Cref{prop: classification} shows that it fails to be preserved for codes that allow a code-adapted representation.

The function obtained for a state $\hat{\rho}$ is 
\begin{equation}
\label{eq: general distr Brif Mann}
    W_{\hat{\rho}}^{\mathcal{C}}(\Omega)=\text{Tr}\left[\hat{\Delta}^{\mathcal{C}}(\Omega)\hat{\rho}\right]
\end{equation}
where $\hat{\Delta}^{\mathcal{C}}(\Omega)$ is the kernel associated with the code space $\mathcal{C}$ and is defined as
\begin{equation}
    \hat{\Delta}^{\mathcal{C}}(\Omega)=\sum_\nu Y_\nu(\Omega)\hat{D}_\nu^\dagger
\end{equation}
where $\hat{D}_\nu$ are the tensor operators and $Y_\nu$ are the harmonic functions over $L^2(X^{\mathcal{C}},\mu)$ with $d\mu$ the invariant measure over $X^{\mathcal{C}}$, and where $\nu$ ranges over the spectrum of the Laplace-Beltrami operator over $L^2(X^{\mathcal{C}},\mu)$.
The harmonic functions $Y_\nu(\Omega)$ are its eigenfunctions, a complete orthonormal basis for square-integrable functions on the phase space. The tensor operators $\hat{D}_\nu$ are orthogonal in $\mathcal{B}(\mathcal{H})$ and span its $\mathcal{E}_{\mathcal{C}}$-invariant subalgebra, which is all the representation resolves and which contains $\mathcal{B}(\mathcal{C})$; they are a basis of $\mathcal{B}(\mathcal{H})$ itself when $\mathcal{E}_{\mathcal{C}}={\id}$. The kernel is applied to arbitrary $\hat{\rho}\in\mathcal{B}(\mathcal{H})$ throughout, including to operators supported outside the code space. The tensor operators are defined so they transform under the symmetries of the space in the same way the harmonic functions do.
They can be recovered from the harmonic functions by inverting an overlap relation~\cite{Brif1999BrifMannConstruction}, with the phase freedom fixed so that the resulting Wigner function depends only on the symmetry of the code space and not on a reference state. We do not need that inversion: \Cref{subsec:averaging,subsec:classification} obtain the kernel directly.

\subsection{\texorpdfstring{Semi-functoriality of the construction}{Semi-functoriality of the construction}}\label{subsec:semifunctor}

This construction works for the class of symmetry-defined codes, whose code space is invariant under a subgroup of a group $G$ acting irreducibly on the carrier space. Ref.~\cite{Brif1999BrifMannConstruction} builds phase-space representations, not codes, and not every code arises from such a symmetry (e.g., numerically optimised and non-group bosonic codes do not). We show here that any representation whose kernel satisfies the Stratonovich-Weyl axioms, in the code-faithful reading of \Cref{subsec:Wigner}, has the properties the structure theorem requires in the sense given in \Cref{subsec:structure}. This does not necessarily mean a given prescription produces such a kernel, as we see in \Cref{subsec:averaging,subsec:classification}: for the GKP code, for every qudit stabiliser code of prime-power dimension, and for the cat and binomial codes, a Stratonovich-Weyl kernel exists and is identified there, so the conclusion applies to all of them, though for the rotation-symmetric families the kernel is not the one the averaging prescription would give.

Because the two notions of \Cref{eq: subgroup H^C,eq: isotropy of reference state} imply different things, we must specify what that invariance means. The GKP code space is the fixed-point space of its stabiliser lattice. The binomial code space is a two-dimensional subspace of the (infinite-dimensional) fixed-point space of $\hat{R}_{S+1}$. The four-component cat code space is invariant under $\mathbb{Z}_4$ but is not contained in the fixed-point space of $\hat{R}_4$, which acts there as the logical $\hat{\bar{Z}}$; only in that of $\hat{R}_4^2=(-1)^{\hat{n}}$. So ``fixed-point space of a subgroup'' is correct only for GKP, and we use the weaker ``invariant under a subgroup'' for the class as a whole. Two qualifications keep the weaker phrase from being empty: the subgroup must be non-trivial modulo the centre, since every code space is invariant under the global phases; and the phase space must be the quotient $X^{\mathcal{C}}=G/H^{\mathcal{C}}$ built from the isotropy subgroup of an actual codeword (for ideal GKP, from the pointwise stabiliser $H^{\mathcal{C}}_0$, as \Cref{subsec:construction} explains). A code specified by an arbitrary list of Fock amplitudes has $H^{\mathcal{C}}$ equal to the centre alone, $X^{\mathcal{C}}\simeq\mathbb{C}$, and no code-specific structure (\Cref{prop: classification,cor: classification}).

Every representation generated by a kernel satisfying the Stratonovich-Weyl axioms is an empirically adequate and linearity-preserving semi-functor. Empirical adequacy follows from \labelcref{SWAxiom5}, which ensures outcome probabilities follow Born's rule; linearity preservation follows from the linearity of the trace in \Cref{eq: general distr Brif Mann}. Whether the representation is more than a semi-functor depends on the reconstruction map: it is a functor when the reconstruction returns $\hat{\rho}$ itself, as for the standard Wigner function, and a genuine semi-functor when it returns the twirled operator $\mathcal{E}_{\mathcal{C}}(\hat{\rho})$, preserving the identity only on the subtheory defined by the code space. Which of these is the case is settled by \Cref{prop: classification}\rev{, under its hypotheses}: the reconstruction map is a non-trivial twirl, and the representation a genuine semi-functor, iff the code space is invariant under a lattice of displacements. In the single-mode continuous-variable setting that requires non-normalisable codewords, so for every bosonic code with codewords in $L^2(\mathbb{R})$ the twirl is the identity and the representation is a functor; in finite dimension the lattice condition is met by every stabiliser code (\Cref{prop:normal}) and the twirl is non-trivial there.

The structure theorem can therefore be applied to any code in this class. (It is \textit{not} available for a code with no such symmetry.) For such a code, the construction provides a subgroup $H^{\mathcal{C}}$, a homogeneous phase space $X^{\mathcal{C}}=G/H^{\mathcal{C}}$ with unique invariant measure $d\mu(\Omega)$, and, wherever an admissible kernel exists, a code-adapted self-dual kernel $\hat{\Delta}^{\mathcal{C}}(\Omega)$~\cite{Brif1999BrifMannConstruction}. The forward map $\chi_{\mathcal{C}}$ traces an operator against the kernel (\Cref{eq: general distr Brif Mann}); the inverse map $\phi_\mathcal{C}$ is built from the twirling map
\begin{equation}
\mathcal{E}_{\mathcal{C}}(\hat{\rho}) = \int_{X^\mathcal{C}} d\mu(\Omega) W_{\hat{\rho}}^{\mathcal{C}}(\Omega) \hat{\Delta}^{\mathcal{C}}(\Omega)
\end{equation}
For any operational transformation $T$, its phase-space representation $Q(T)$ evaluated on an input distribution $W(\Omega)$ yields a new distribution at coordinate $\Omega'$:
\begin{equation}
\label{eq: generalized structure theorem map}
\begin{split}
&\left[Q(T)W\right](\Omega') =\\ &\text{Tr}\left[\hat{\Delta}^{\mathcal{C}}(\Omega')\ T\left( \int_{X^{\mathcal{C}}}d\mu(\Omega) W(\Omega)\hat{\Delta}^{\mathcal{C}}(\Omega) \right)\right]
\end{split}
\end{equation}
Substituting the isotropy subgroup $H^{\mathcal{C}}$ of any code in the class just described, together with the invariant measure $d\mu(\Omega)$ on $X^{\mathcal{C}}=G/H^{\mathcal{C}}$, adapts it to that code.

\subsection{\texorpdfstring{Group averaging}{Group averaging}}\label{subsec:averaging}

We can make this construction explicit, in the same form for every code in the class. Let
\begin{equation}
\label{eq: standard SW kernel}
{\hat{\Delta}_{\mathrm{std}}(\Omega)=2\hat{\Pi}(\Omega),\qquad
\hat{\Pi}(\Omega)=\hat{D}(\Omega)(-1)^{\hat{n}}\hat{D}^\dagger(\Omega)}
\end{equation}
{be the standard Stratonovich-Weyl kernel from \Cref{subsec:Wigner}, for a single mode. For $m$ modes the same construction gives $\hat{\Delta}_{\mathrm{std}}(\Omega)=2^m\hat{\Pi}(\Omega)$ with $\hat{\Pi}(\Omega)=\hat{D}(\Omega)(-1)^{\hat{n}_{\mathrm{tot}}}\hat{D}^\dagger(\Omega)$, since $\mathrm{Tr}[(-1)^{\hat{n}_{\mathrm{tot}}}]=2^{-m}$ and standardisation requires unit trace (we set $m=1$ everywhere except in \Cref{thm:dualrail,subsec:multimode}). A natural candidate for a code-adapted kernel is the average of $\hat{\Delta}_{\mathrm{std}}$ over the isotropy subgroup $H^{\mathcal{C}}$,}
\begin{equation}
\label{eq: H averaged kernel}
{\hat{\Delta}^{\mathcal{C}}(\Omega)=\frac{1}{|H^{\mathcal{C}}|}\sum_{h\in H^{\mathcal{C}}}\pi(h)\,\hat{\Delta}_{\mathrm{std}}(\Omega)\,\pi^\dagger(h)}
\end{equation}
(the sum becoming an integral for continuous $H^{\mathcal{C}}$). We now have to check whether our $H^{\mathcal{C}}$ possesses three properties. The latter two are held by every $H^{\mathcal{C}}$, but the first is not:

\begin{enumerate}[label=(\roman*),leftmargin=*]
    \item \textit{Covariance.} Each $\pi(h)$ belongs to the extended Weyl-Heisenberg group, so $\pi(h)\hat{\Pi}(\Omega)\pi^\dagger(h)=\hat{\Pi}(h\cdot\Omega)$, where $h\cdot\Omega$ is the induced action on $X^{\mathcal{C}}$, so $\hat{\Delta}^{\mathcal{C}}(\Omega)=\frac{1}{|H^{\mathcal{C}}|}\sum_h\hat{\Delta}_{\mathrm{std}}(h\cdot\Omega)$. Averaging does not preserve covariance in general: $\pi(g)\hat{\Delta}^{\mathcal{C}}(\Omega)\pi^\dagger(g)$ averages $\hat{\Delta}_{\mathrm{std}}$ over the coset $gH^{\mathcal{C}}$ while $\hat{\Delta}^{\mathcal{C}}(g\cdot\Omega)$ averages it over $H^{\mathcal{C}}g$, so $\hat{\Delta}^{\mathcal{C}}(g\cdot\Omega)=\pi(g)\hat{\Delta}^{\mathcal{C}}(\Omega)\pi^\dagger(g)$ holds for all $g\in G$ iff $H^{\mathcal{C}}$ is normal in $G$. For the GKP code $H^{\mathcal{C}}$ is a lattice of displacements inside the abelian translation group and is normal, so covariance holds; for a rotation-symmetric code $H^{\mathcal{C}}$ contains a rotation, and a rotation conjugated by a displacement is not a rotation, so covariance fails. \Cref{eq: H averaged kernel} is therefore valid only in the first case, so we do not use it in the second.\label{HCProperty1Covariance}

    \item \textit{Standardisation.} Using $\frac{1}{\pi}\int d^2\Omega\,\hat{D}(\Omega)\hat{X}\hat{D}^\dagger(\Omega)=\mathrm{Tr}[\hat{X}]\,\id$ together with $\mathrm{Tr}[(-1)^{\hat{n}}]=1/2$,
\begin{equation}
\label{eq: standardisation general}
{\int d\mu(\Omega)\,\hat{\Delta}^{\mathcal{C}}(\Omega)=\mathrm{Tr}\!\left[\hat{\Delta}_{\mathrm{std}}(0)\right]\id=\id}
\end{equation}
for every code, meaning the group average leaves the trace unchanged. When the kernel is covariant, so that $\hat{\Delta}^{\mathcal{C}}(\Omega)=\hat{D}(\Omega)\hat{K}\hat{D}^\dagger(\Omega)$, and $\hat{K}$ is trace class, this is equivalent to $\mathrm{Tr}[\hat{K}]=\mathrm{Tr}[\hat{\Delta}^{\mathcal{C}}(0)]=1$ (in which form we check it below). This is a necessary condition: for the ideal GKP code $\hat{K}$ is a lattice sum of displacement operators, meaning $\mathrm{Tr}[\hat{K}]$ diverges, so we instead have to verify standardisation through the integral given in \Cref{eq: gkp standardisation direct}.\label{HCProperty2Standardisation}

\item \textit{The reconstruction map is the twirl over $H^{\mathcal{C}}$.} The standard kernel is self-dual, $\hat{A}=2\int d^2\Omega\,W^{\mathrm{std}}_{\hat{A}}(\Omega)\hat{\Pi}(\Omega)$, and the standard Wigner function is covariant, $W^{\mathrm{std}}_{\hat{A}}(g\cdot\Omega)=W^{\mathrm{std}}_{\pi^\dagger(g)\hat{A}\pi(g)}(\Omega)$. Substituting \Cref{eq: H averaged kernel} into $\phi_{\mathcal{C}}\circ\chi_{\mathcal{C}}$ and changing integration variable in each term of the resulting double sum, leaves a single sum over the group,
\begin{equation}
\label{eq: twirl is H average}
{\phi_{\mathcal{C}}\!\left(\chi_{\mathcal{C}}(\hat{A})\right)=\mathcal{E}_{\mathcal{C}}(\hat{A})=\frac{1}{|H^{\mathcal{C}}|}\sum_{h\in H^{\mathcal{C}}}\pi^\dagger(h)\,\hat{A}\,\pi(h)}
\end{equation}
Because $H^{\mathcal{C}}$ is a group, $\mathcal{E}_{\mathcal{C}}$ is trace-preserving and idempotent, projecting onto the $H^{\mathcal{C}}$-invariant subalgebra. This is true for every $H^{\mathcal{C}}$, normal or not; the kernel must also be covariant for $\mathcal{E}_{\mathcal{C}}$ to be the reconstruction map of an admissible Stratonovich-Weyl representation, making $Q(\id)$ the idempotent required by the structure theorem.\label{HCProperty3TwirlMap}
\end{enumerate}

This reproduces the known GKP case. For the GKP code $H^{\mathcal{C}_{GKP}}$ is the stabiliser lattice of displacements by multiples of $dl$, which is normal in $G$, so averaging the displaced parity over that lattice gives a sum of displacement operators over the symplectic dual lattice, \Cref{eq: ZGW central kernel} (Ref.~\cite{davis2024identifyingquantumresourcesencoded}'s Zak-Gross kernel). For rotation-symmetric codes (\Cref{subsec:cat,subsec:bin}), the subgroup is $U(1)\times\mathbb{Z}_M$, meaning \Cref{eq: H averaged kernel} doesn't give a Stratonovich-Weyl kernel at all. Appendix~\ref{app:cat} shows why it fails. We replace \Cref{eq: H averaged kernel} with \Cref{prop: classification} there, which fixes the kernel from the axioms directly, so requires no reference-state expansion (however, for a non-Gaussian reference state, this produces a kernel which is neither covariant nor self-dual).

Nothing in \Cref{HCProperty1Covariance,HCProperty2Standardisation,HCProperty3TwirlMap} required the elements of $H^{\mathcal{C}}$ to be displacements. All three properties only require that $\pi(h)$ act on phase space by a measure-preserving affine symplectic map, and commute with the parity operator. These requirements hold throughout the metaplectic extension of the Weyl-Heisenberg group: the quadratic Hamiltonians which generate rotations and squeezes are even in $\hat{a}$ and $\hat{a}^\dagger$, so commute with $(-1)^{\hat{n}}$, and symplectic maps have unit determinant, and therefore preserve $d^2\Omega$.

\begin{proposition}
\label{prop: metaplectic}
{Let $G=H_3(\mathbb{C})\rtimes\mathrm{Mp}(2,\mathbb{R})$, and $H^{\mathcal{C}}\subseteq G$ be a compact isotropy subgroup of a reference codeword, so that the average in \Cref{eq: H averaged kernel} is taken against a normalised Haar measure. Then, the kernel of \Cref{eq: H averaged kernel} satisfies the Stratonovich-Weyl standardisation axiom and induces the twirl of \Cref{eq: twirl is H average}. It satisfies the covariance axiom on the normaliser of $H^{\mathcal{C}}$ in $G$, and hence on all of $G$ iff $H^{\mathcal{C}}$ is normal in $G$. The only compact normal subgroups of this $G$ lie in its central $U(1)$ of phases, which acts trivially by conjugation; so for a compact isotropy group \Cref{eq: H averaged kernel} yields a Stratonovich-Weyl kernel only when the average does nothing (i.e., never for a rotation-symmetric code).}
\end{proposition}

We prove this in Appendix \ref{app: proofs}. The averaging prescription is therefore never useful for a compact isotropy group: whatever kernel the code admits must be found from the axioms themselves, which we do in \Cref{subsec:classification}.

Let us now consider finite dimensional codes. Let $\mathsf P_n$ be the $n$-qudit Weyl-Heisenberg group, generated by the displacements $D(v)$, $v\in\Zd^{2n}$, together with its centre $\mathcal{Z}$ of scalar phases ($\{\omega^c\}$ for odd $d$, $\{i^c\}$ at $d=2$), and $\mathcal{C}$ the code space of a stabiliser code $[[n,k]]_d$, with stabiliser group $\mathcal{S}=\{D(m):m\in \mathcal{M}\}$ for an isotropic subgroup $\mathcal{M}\subset\Zd^{2n}$ of order $d^{n-k}$. (For odd $d$ in symmetric ordering, a stabiliser group with phase-free generators is phase-free as a set, since commuting displacements compose without phases; for $d=2$ this fails, but we only use phase-freeness in \Cref{prop:dv37,thm:dvmagic}, which are restricted to odd $d$ anyway (\Cref{subsec:qubit}). In both cases $\mathcal{M}=-\mathcal{M}$, since $\mathcal{M}$ is a subgroup.)

\begin{proposition}[Normality]\label{prop:normal}
Conjugation in $\mathsf P_n$ acts by $gD(v)g^{-1}=\omega^{[u,v]}D(v)$ for $g=zD(u)$ with $z\in\mathcal{Z}$, and $[u,v]$ the symplectic form. Every subgroup of $\mathsf P_n$ which contains the centre is therefore normal; this means the projective isotropy group $H^{\mathcal{C}}=\langle\mathcal{Z},\mathcal{S}\rangle$ of a generic codeword is normal in $G=\mathsf P_n$, so \Cref{eq: H averaged kernel}'s averaging prescription is covariant on all of $G$. \Cref{prop:dvsw} verifies it meets the remaining Stratonovich-Weyl axioms, as per \Cref{subsec:Wigner}'s code-faithful reading.
\end{proposition}

\Cref{prop: metaplectic}'s obstruction can't arise in $\mathsf P_n$ because there are no rotations to conjugate: a stabiliser group is a displacement subgroup by definition, and two-step nilpotency makes it normal once the centre is adjoined (\Cref{prop:normal}). Every DV stabiliser code is in this sense ``GKP-like'': its code space is the fixed-point space of a lattice of displacements, so the twirl below is a finite group average, so a genuine channel, and the finite-dimensional structure theorems~\cite{Schmid2024StructureTheorem, wagner2025ComplexStructTheorem} apply as black boxes. We explore the consequences of this in \Cref{subsec:dv}, and cover non-Pauli analogues in \Cref{subsec:nonpauli}.

As in the CV case, we quotient by the isotropy group of a generic codeword (i.e., not an eigenstate of any logical Pauli), meaning $H^{\mathcal{C}}=\langle\mathcal{Z},\mathcal{S}\rangle$ is the centre together with the stabiliser lattice; for the three-qutrit repetition code, numerics confirm that the isotropy of a Haar-random codeword is $\langle\mathcal{Z},\mathcal{S}\rangle$, whose image in $\Zd^{2n}$ is the nine-element lattice $\mathcal{M}$.

\subsection{\texorpdfstring{What the axioms allow}{What the axioms allow}}\label{subsec:classification}

The two requirements just used (covariance and an idempotent reconstruction map) are restrictive enough\rev{, together with the further hypotheses listed in \mbox{\Cref{prop: classification}},} to determine the kernel \rev{of every code with normalisable codewords}. Covariance forces the kernel to be the displaced conjugate of a single operator, $\hat{\Delta}^{\mathcal{C}}(\Omega)=\hat{D}(\Omega)\hat{K}\hat{D}^\dagger(\Omega)$ with $\hat{K}$ invariant under $H^{\mathcal{C}}$. For such a kernel the reconstruction map acts on characteristic functions by pointwise multiplication,
\begin{equation}
\label{eq: multiplier}
\begin{split}
\chi_{\mathcal{E}_{\mathcal{C}}(\hat{A})}(\xi)&=\mathfrak{m}(\xi)\,\chi_{\hat{A}}(\xi),\qquad \mathfrak{m}(\xi):=\left|\chi_{\hat{K}}(\xi)\right|^2,\\
\chi_{\hat{A}}(\xi)&=\mathrm{Tr}\!\left[\hat{A}\hat{D}(\xi)\right],
\end{split}
\end{equation}
and standardisation fixes $\chi_{\hat{K}}(0)=\mathrm{Tr}[\hat{K}]=1$ whenever $\hat{K}$ is trace class. (When it is not trace class, use the integral in \Cref{eq: standardisation general}).

The classification rests on one fact about which subgroups of the displacement group can leave a code space alone.

\begin{lemma}
\label{lem: nodisplacement}
{Let $\mathcal{C}\subset L^2(\mathbb{R}^m)$ be a finite-dimensional subspace, for any number of modes $m\ge1$. Then $\hat{D}(\beta)\mathcal{C}=\mathcal{C}$ implies $\beta=0$.}
\end{lemma}

{We prove this in Appendix~\ref{app: proofs}: to summarise, $\hat{D}(\beta)$ restricted to $\mathcal{C}$ would be a unitary on a finite-dimensional space, so would have an eigenvector, whereas $\hat{D}(\beta)=e^{i\lambda\hat{K}_\theta}$ with $\hat{K}_\theta$ a quadrature operator (not the central kernel $\hat{K}$), after a passive rotation taking $\beta$ into a single mode, whose spectrum is purely absolutely continuous, so it has no eigenvectors in $L^2(\mathbb{R}^m)$ for $\beta\neq0$. The same argument excludes a non-trivial squeeze (whose generator is the dilation operator), but not rotations $e^{i\theta\hat{n}}$, since the Fock states are their eigenvectors, which is why the rotation-symmetric codes exist. The ideal GKP code instead escapes the lemma as its codewords are Dirac combs (i.e., tempered distributions rather than elements of $L^2(\mathbb{R})$), so $\mathcal{C}_{GKP}$ is not a subspace of $L^2(\mathbb{R})$ at all.}

\begin{proposition}
\label{prop: classification}
\rev{Let $G$ contain the displacements of $m\ge1$ modes, and let the kernel $\hat{\Delta}^{\mathcal{C}}$, which defines both $\chi_{\mathcal{C}}$ and $\phi_{\mathcal{C}}$ (\mbox{\Cref{subsec:semifunctor}}), satisfy three hypotheses:}
\begin{enumerate}[label=(H\arabic*),leftmargin=*]
\item \rev{\mbox{\labelcref{SWAxiom2,SWAxiom3,SWAxiom4}}, as stated in \mbox{\Cref{subsec:Wigner}};}\label{PropIII.3.H1}
\item \rev{the code-faithful form of \mbox{\labelcref{SWAxiom1,SWAxiom5}}:} the reconstruction map $\mathcal{E}_{\mathcal{C}}=\phi_{\mathcal{C}}\circ\chi_{\mathcal{C}}$ is idempotent and acts as the identity on the code algebra $\mathcal{B}(\mathcal{C})$ (i.e., the representation is faithful on the code);\label{PropIII.3.H2}
\item \rev{if} $\chi_{\hat{K}}(\xi)=\mathrm{Tr}[\hat{K}\hat{D}(\xi)]$ \rev{is a function, it is continuous} (for a bounded, non-trace-class $\hat{K}$, we mean the trace in the Abel-regularised sense; $\chi_{\hat{K}}$ is a comb rather than a function in the ideal GKP case).\label{PropIII.3.H3}
\end{enumerate}
\rev{\mbox{\ref{PropIII.3.H1}} alone gives $\hat{\Delta}^{\mathcal{C}}(\Omega)=\hat{D}(\Omega)\hat{K}\hat{D}^\dagger(\Omega)$ with $\hat{K}$ Hermitian and, $d\mu$ being invariant, makes the linear map $\mathcal{E}_{\mathcal{C}}$ commute with conjugation by every displacement. If $\chi_{\hat{K}}$ is a function, \mbox{\ref{PropIII.3.H1}} also makes $\mathcal{E}_{\mathcal{C}}$ trace-preserving, and $\mathcal{E}_{\mathcal{C}}$} acts on characteristic functions by multiplication by \rev{$\mathfrak{m}=|\chi_{\hat{K}}|^2$ (in the distributional case, where the square of the comb is undefined, $\mathfrak{m}$ is taken from the twirl of setting \mbox{\ref{PropIII.3.B}} below). These properties of $\mathcal{E}_{\mathcal{C}}$ are derived. Its complete positivity, and its form as an average of conjugations by displacements (the twirl form), we derive from \mbox{\ref{PropIII.3.H1}}--\mbox{\ref{PropIII.3.H3}} only when $\chi_{\hat{K}}$ is a function, through part \mbox{\ref{PropIII.3.i}} below. Where we use them without that, they enter through one of two settings, and we say in each what is assumed:}
\begin{enumerate}[label=(\Alph*),leftmargin=*]
\item \rev{$\mathcal{E}_{\mathcal{C}}$ is a completely positive, trace-preserving map on $\mathcal{B}(L^2(\mathbb{R}^m))$. Complete positivity is the added assumption here (trace preservation is derived if $\chi_{\hat{K}}$ is a function, and is otherwise part of the setting). With the derived covariance, $\mathcal{E}_{\mathcal{C}}$ is then a displacement-covariant channel, and the twirl form is derived: $\mathcal{E}_{\mathcal{C}}(\hat{A})=\int d\nu(\beta)\hat{D}(\beta)\hat{A}\hat{D}^\dagger(\beta)$ for a probability measure $\nu$, and we write $\Lambda^{\perp}\subseteq\mathbb{R}^{2m}$ for the closed subgroup generated by the support of $\nu$.}\label{PropIII.3.A}
\item \rev{$\mathcal{E}_{\mathcal{C}}$ is the average over a closed group $\Lambda^{\perp}\subseteq\mathbb{R}^{2m}$ of displacements, that is, the conditional expectation onto its commutant, acting on the code algebra of a rigged Hilbert space. The twirl form is the added assumption here, taken as in Ref.~\mbox{\cite{davis2024identifyingquantumresourcesencoded}}. It cannot be derived as in \mbox{\ref{PropIII.3.A}}, since for non-trivial $\Lambda^{\perp}$ this $\mathcal{E}_{\mathcal{C}}$ is not a trace-preserving channel on $\mathcal{B}(L^2(\mathbb{R}^m))$ (Appendix~\mbox{\ref{app: proofs}}).}\label{PropIII.3.B}
\end{enumerate}
\rev{In both settings \mbox{\ref{PropIII.3.H2}} makes $\Lambda^{\perp}$ leave $\mathcal{C}$ invariant.} (We write \rev{$\Lambda^{\perp}$} because for the ideal GKP code it is the symplectic dual of the lattice $\Lambda$ supporting $\mathfrak{m}$\rev{. We} write the single-mode case below, but nothing in the statement or its proof depends on the mode count.) \rev{Then:}
\begin{enumerate}[label=(\roman*),leftmargin=*]
\item \rev{In setting \mbox{\ref{PropIII.3.A}},} $\Lambda^{\perp}=\{0\}$ and $\mathcal{E}_{\mathcal{C}}={\id}$. \rev{The conclusion $\mathcal{E}_{\mathcal{C}}={\id}$ also holds with neither setting assumed if $\chi_{\hat{K}}$ is a function, and hence continuous by \mbox{\ref{PropIII.3.H3}}: \mbox{\ref{PropIII.3.H2}} gives $\mathfrak{m}^2=\mathfrak{m}$, so $\mathfrak{m}=|\chi_{\hat{K}}|^2$ is a continuous function with values in $\{0,1\}$ and $\mathfrak{m}(0)=1$, hence $\mathfrak{m}\equiv1$ on the connected space $\mathbb{R}^{2m}$. Complete positivity of $\mathcal{E}_{\mathcal{C}}={\id}$ is then a conclusion and not a hypothesis. This is the route \mbox{\Cref{thm: nogo}} takes. The route through \mbox{\ref{PropIII.3.A}} does not use \mbox{\ref{PropIII.3.H3}}, and the continuity route does not use complete positivity.}\label{PropIII.3.i}

\item If, in addition \rev{to $\mathcal{E}_{\mathcal{C}}={\id}$ holding by either route of \mbox{\ref{PropIII.3.i}}}, $G$ contains the phase-space inversion $\Omega\mapsto-\Omega$, then \rev{(since $\mathcal{E}_{\mathcal{C}}={\id}$ means $\mathfrak{m}\equiv1$, so that $\chi_{\hat{K}}$ is a function of unit modulus, continuous by \mbox{\ref{PropIII.3.H3}})}
\begin{equation}
\label{eq: classification result}
\hat{K}=2^m(-1)^{\hat{n}_{\mathrm{tot}}},\qquad \hat{\Delta}^{\mathcal{C}}(\Omega)=\hat{\Delta}_{\mathrm{std}}(\Omega),
\end{equation}
meaning the only code-adapted representation available is the ordinary Wigner function.\label{PropIII.3.ii}
\item Whenever $\mathcal{C}$ is a finite-dimensional subspace of $L^2(\mathbb{R}^m)$, \Cref{lem: nodisplacement} forces $\Lambda^{\perp}=\{0\}$ directly, since $\Lambda^{\perp}$ leaves $\mathcal{C}$ invariant. \rev{Given the twirl form, whether derived as in \mbox{\ref{PropIII.3.A}} or assumed as in \mbox{\ref{PropIII.3.B}}, conclusion \mbox{\ref{PropIII.3.i}} then follows with no further use of complete positivity,} and conclusion \ref{PropIII.3.ii} follows as well once its parity assumption is granted. We do not claim \ref{PropIII.3.iii} for a reconstruction map \rev{in neither setting, for which} the twirl form itself is not available \rev{(the continuity route of \mbox{\ref{PropIII.3.i}} does not need it)}.\label{PropIII.3.iii}
\end{enumerate}
\rev{In setting \mbox{\ref{PropIII.3.B}}} $\Lambda^{\perp}$ may be a non-trivial lattice of displacements, and the ideal GKP code is the instance of it\rev{. Its $\mathfrak{m}$ is the indicator function of the symplectic dual lattice, which takes the values $\{0,1\}$ that \mbox{\ref{PropIII.3.H2}} demands but is not continuous, and its $\mathcal{E}_{\mathcal{C}}$ is not a channel on $\mathcal{B}(L^2(\mathbb{R}))$, so neither route of \mbox{\ref{PropIII.3.i}} applies to it.}
\end{proposition}

We prove this in Appendix~\ref{app: proofs}: to summarise \rev{setting \mbox{\ref{PropIII.3.A}}, \mbox{\ref{PropIII.3.H1}} makes $\mathcal{E}_{\mathcal{C}}$ displacement-covariant, so as a completely positive, trace-preserving map it is} a random-displacement channel $\mathcal{E}_{\mathcal{C}}(\hat{A})=\int d\nu(\beta)\hat{D}(\beta)\hat{A}\hat{D}^\dagger(\beta)$; idempotence makes $\nu$ an idempotent probability measure, so (by Kawada-It\^{o}) the normalised Haar measure of a compact subgroup; and $\mathbb{R}^{2m}$ has no compact subgroup but $\{0\}$. The kernel is then fixed by $|\chi_{\hat{K}}|\equiv1$ together with two reflections of the same symmetry: Hermiticity of $\hat{K}$ gives $\chi_{\hat{K}}(-\xi)=\chi_{\hat{K}}(\xi)^*$, and covariance under phase-space inversion gives $\chi_{\hat{K}}(-\xi)=\chi_{\hat{K}}(\xi)$, so $\chi_{\hat{K}}$ is real and unimodular, hence $\pm1$, hence $\equiv1$ by continuity and $\chi_{\hat{K}}(0)=1$.

Note the requirement that $G$ contains the phase-space inversion $\Omega\mapsto-\Omega$ applies for the code families we consider. We show in \Cref{subsec:cat,app:cat} that for the rotation-symmetric codes considered here the isotropy group of a reference codeword is $U(1)\times\mathbb{Z}_M$ with $M$ even ($M=N$ for the $N$-component cat code, with $N$ even as \Cref{eq: general cat logical pair} requires; $M=2(S+1)$ for the binomial code). This ensures $\hat{K}$ commutes with the parity, which $H^{\mathcal{C}}$ requires whenever it contains a rotation of even order: $\hat{K}$ must commute with $\pi(H^{\mathcal{C}})$ for the kernel to be well defined on $X^{\mathcal{C}}=G/H^{\mathcal{C}}$, and $(-1)^{\hat{n}}=\hat{R}_M^{M/2}\in\pi(H^{\mathcal{C}})$, so $\chi_{\hat{K}}(-\xi)=\mathrm{Tr}[(-1)^{\hat{n}}\hat{K}(-1)^{\hat{n}}\hat{D}(\xi)]=\chi_{\hat{K}}(\xi)$. Without either route (no inversion in $G$ and no even-order rotation in $H^{\mathcal{C}}$) Hermiticity alone leaves $\chi_{\hat{K}}=e^{if}$ with $f$ real, continuous, odd, and vanishing at the origin. This is a strictly larger family than the displaced parities $\hat{K}=2\hat{\Pi}(\gamma)$, which also require $f$ be linear. A code in that situation has at most odd-order rotational symmetry; the codes that motivate the case, numerically optimised ones, have $H^{\mathcal{C}}$ equal to the centre and so no code-specific phase space at all (as discussed above).

Opposite to this is the case of the ideal GKP code. There, the codewords are not normalisable, \Cref{lem: nodisplacement} does not apply, and $\Lambda^{\perp}$ may be a non-trivial closed subgroup of displacements leaving $\mathcal{C}$ invariant. $\Lambda^{\perp}$ cannot contain a line: $\mathcal{E}_{\mathcal{C}}$ must act as the identity on the code algebra, so $\mathcal{B}(\mathcal{C})$ lies in the commutant of $\Lambda^{\perp}$, and the commutant of a one-parameter group of displacements $\{e^{it\hat{K}_\theta}\}$, with $\hat{K}_\theta$ a quadrature, is the maximal abelian algebra generated by $\hat{K}_\theta$, which cannot contain the non-commuting logical operators of a qubit. This leaves a lattice of rank one or two. Rank two is the GKP case: $\Lambda^{\perp}$ is the stabiliser lattice, $\mathcal{E}_{\mathcal{C}}$ is the twirl of \Cref{eq: twirling map}, $X^{\mathcal{C}}$ is the compact torus $\mathbb{T}^2_{dl}$, and the Zak-Gross kernel of \Cref{eq: ZGW central kernel} is an admissible kernel. We do not prove it is the only one: the uniqueness step of the proof (in Appendix~\ref{app: proofs}) uses continuity of $\chi_{\hat{K}}$, which fails in the distributional case, where Hermiticity and parity fix the lattice coefficients only up to a sign pattern, the one in \Cref{eq: ZGW central kernel} being an admissible choice; Ref.~\cite{Gross2006DVWignerF}'s uniqueness argument uses Clifford covariance, which we do not assume. Rank one would describe a code invariant under displacements along a single direction, for which $X^{\mathcal{C}}$ is a non-compact cylinder. Both non-trivial cases are inherently distributional, in the sense of \Cref{subsec:GKP}, and we treat them in \rev{setting \mbox{\ref{PropIII.3.B}}, the} conditional-expectation setting of Ref.~\cite{davis2024identifyingquantumresourcesencoded}.

\begin{corollary}
\label{cor: classification}
\rev{Let the kernel satisfy hypotheses \mbox{\ref{PropIII.3.H1}}--\mbox{\ref{PropIII.3.H3}} of \mbox{\Cref{prop: classification}} and belong to the class covered by that proposition: $\chi_{\hat{K}}$ is a function, or $\mathcal{E}_{\mathcal{C}}$ is in setting \mbox{\ref{PropIII.3.A}} or \mbox{\ref{PropIII.3.B}}.} Let the parity be available, in either of the two senses set out above ($G$ contains the phase-space inversion, or $H^{\mathcal{C}}$ contains a rotation of even order), and let $\mathcal{C}$ have logical dimension at least two. Then a code-adapted phase-space representation differing from the ordinary Wigner function requires the code space to be invariant under a non-trivial lattice of displacements, and hence, in the continuous-variable setting, forbids $\mathcal{C}$ from being a finite-dimensional subspace of $L^2(\mathbb{R})$ at all (\Cref{lem: nodisplacement}): its codewords cannot be normalisable. Among bosonic codes whose phase space is compact, or equivalently, whose invariance lattice has rank two, this only holds for the ideal GKP family, whose code space is the fixed-point space of its stabiliser lattice. As far as we are aware no-one has yet made a bosonic code realising the rank-one case, which would give a non-compact cylinder, so we do not consider this further. \rev{The statement is made within the class just named. A kernel whose $\chi_{\hat{K}}$ is not a function, and whose reconstruction map is neither a completely positive, trace-preserving map on $\mathcal{B}(L^2(\mathbb{R}))$ nor a conditional expectation onto the commutant of a group of displacements, lies outside that class, and we make no claim about it.}
\end{corollary}

The parity is available for every code considered in this paper: for the rotation-symmetric codes through the even-order rotation in $H^{\mathcal{C}}$, and for the ideal GKP code by adjoining $(-1)^{\hat{n}}$ to $G$, which is valid as parity maps the codeword $\ket{\overline{j}}$ to $\ket{\overline{-j}}$, so preserves the inversion-symmetric code space; the groups are listed in \Cref{tab: group data}. This has two consequences. First, for a bosonic code with normalisable codewords (the cat and binomial codes of \Cref{subsec:cat,subsec:bin}, and finite-energy GKP codewords), the Stratonovich-Weyl axioms\rev{, in the form of hypotheses \mbox{\ref{PropIII.3.H1}}--\mbox{\ref{PropIII.3.H3}} and within the class of \mbox{\Cref{cor: classification}},} only admit the ordinary Wigner kernel. \Cref{subsec:cat,subsec:bin} below therefore look at what the ordinary representation shows about the codewords of each family. Second, the Zak-Gross representation of the GKP code is exact only in the ideal limit; we return to what remains at finite energy in \Cref{sec:FiniteEnergy}.

As noted in \Cref{subsec:averaging}, the construction also requires a homogeneous phase space $X^{\mathcal{C}}=G/H^{\mathcal{C}}$ carrying an action of $G$, because that gives the covariance axiom its content, and covariance produces the rigid action of errors in \Cref{eq: translation of ZGW}, and hence the usefulness of \Cref{eq: generalized structure theorem map}. A code space invariant under no subgroup of $G$ beyond the centre has no code-specific phase space: the construction still runs, but with $H^{\mathcal{C}}$ the centre it returns $X^{\mathcal{C}}\simeq\mathbb{C}$ and the ordinary Wigner function, so the structure theorem tells us nothing about the code beyond what it already tells us about the mode. This limits our argument: outside the symmetry-defined codes there is no code-adapted phase space for the structure theorem to act on, and a construction which claimed to supply one would not be a Stratonovich-Weyl representation.

\subsection{\texorpdfstring{\protect{The class of transformations represented by \Cref{eq: generalized structure theorem map}}}{The class of transformations represented by the structure theorem map}}\label{subsec:GradedClass}

The reconstruction map $\phi_{\mathcal{C}}$ returns not $\hat{\rho}$ but the twirled operator $\mathcal{E}_{\mathcal{C}}(\hat{\rho})$. Before we can use \Cref{eq: generalized structure theorem map} to represent a composite process (an error, followed by a syndrome measurement, followed by a recovery), we therefore have to say for which transformations the insertion of $\mathcal{E}_{\mathcal{C}}$ is operationally irrelevant.

\Cref{prop: classification} leaves only two cases. For a bosonic code with codewords in $L^2(\mathbb{R})$, $\mathcal{E}_{\mathcal{C}}={\id}$ and every channel qualifies, so nothing in this subsection restricts anything. The restriction applies only when $\Lambda^{\perp}$ is a non-trivial displacement lattice and $\mathcal{E}_{\mathcal{C}}$ is a genuine twirl: i.e., the ideal GKP code in the continuous-variable setting, and every stabiliser code in the finite-dimensional one (\Cref{subsec:dv}).

$\mathcal{E}_{\mathcal{C}}$ is the conditional expectation onto the commutant of $\Lambda^{\perp}$. Writing the physical Hilbert space as a direct sum (or, for GKP, a direct integral) over the joint eigenspaces of $\Lambda^{\perp}$, which are the syndrome sectors,
\begin{equation}
\label{eq: sector decomposition}
{\mathcal{H}=\bigoplus_s \mathcal{H}_s,\qquad \mathcal{H}_s=\hat{\Pi}_s\mathcal{H}}
\end{equation}
(where the $\hat{\Pi}_s$ are the displaced code-space projectors $\hat{\Pi}_{(s,t)}$ of \Cref{subsec:GKP} in the GKP case, and where there is a single sector, $\hat{\Pi}_s=\id$, in the cat and binomial cases, because $\Lambda^{\perp}=\{0\}$ there) the twirl is the projection onto the block-diagonal part of the operator algebra,
\begin{equation}
\label{eq: twirl as sector projection}
{\mathcal{E}_{\mathcal{C}}(\hat{\rho})=\sum_s \hat{\Pi}_s\hat{\rho}\hat{\Pi}_s}
\end{equation}
the sum being the integral of \Cref{eq: twirling map} in the GKP case. $\mathcal{E}_{\mathcal{C}}$ therefore removes coherences \emph{between} distinct sectors, and leaves everything \emph{within} each sector untouched. This makes the framework usable for error correction: an error moves logical information from one sector to another, but does not store that information in inter-sector coherences, and $\mathcal{E}_{\mathcal{C}}$ only destroys these coherences. The sectors of \Cref{eq: sector decomposition} don't need to coincide with the sectors the experimenter measures: for the cat and binomial codes the measured syndrome is the grading of $H^{\mathcal{C}}_0$ (photon-number parity for the four-component cat code, $\hat{n}\bmod(S+1)$ for the binomial code) but $\mathcal{E}_{\mathcal{C}}$ does not grade by it, because $\mathcal{E}_{\mathcal{C}}$ is the identity there. We separate the two in \Cref{tab: group data}.

\begin{definition}
\label{def: graded}
A channel $T$ is \emph{syndrome-graded} with respect to the decomposition of \Cref{eq: sector decomposition} if $\mathcal{E}_{\mathcal{C}}\circ T\circ \mathcal{E}_{\mathcal{C}}=T\circ \mathcal{E}_{\mathcal{C}}$. We write $\mathcal{G}_{\mathcal{C}}$ for the set of such channels.
\end{definition}

A sufficient condition for a channel being syndrome graded is that $T$ carry sectors to sectors, i.e. that
\begin{equation}
\label{eq: graded sufficient condition}
{T\!\left(\hat{\Pi}_s\,\cdot\,\hat{\Pi}_{s'}\right)\subseteq \hat{\Pi}_{\varsigma(s)}\,\cdot\,\hat{\Pi}_{\varsigma(s')}}
\end{equation}
for some map $\varsigma$ of the syndrome label set (we write $\varsigma$ rather than $\pi$ here, which is our unitary representation of $G$).

\begin{proposition}
\label{prop: graded monoid}
$\mathcal{G}_{\mathcal{C}}$ contains the identity, and is closed under composition, under convex combination, and under tensor product with a syndrome-graded channel on an ancilla.
\end{proposition}

We prove this in Appendix~\ref{app: proofs}.

\begin{proposition}
\label{prop: compositional}
{For $T_1,\ldots,T_n\in\mathcal{G}_{\mathcal{C}}$,}
\begin{equation}
\label{eq: compositionality}
{Q(T_n\circ\cdots\circ T_1)=Q(T_n)\circ\cdots\circ Q(T_1)}
\end{equation}
\end{proposition}

We prove this in Appendix~\ref{app: proofs}: each $\phi_{\mathcal{C}}\circ\chi_{\mathcal{C}}=\mathcal{E}_{\mathcal{C}}$ inserted between adjacent channels is absorbed into the channel before it, by \Cref{prop: graded monoid}.

On the subtheory generated by $\mathcal{G}_{\mathcal{C}}$, $Q$ preserves composition, and where $\mathcal{E}_{\mathcal{C}}={\id}$ it preserves the identity too, so it is a functor there and a semi-functor in general (as per Refs.~\cite{Schmid2024StructureTheorem, wagner2025ComplexStructTheorem}). Outside $\mathcal{G}_{\mathcal{C}}$ composition fails: $Q(T_2)\circ Q(T_1)=Q(T_2\circ\mathcal{E}_{\mathcal{C}}\circ T_1)$, which differs from $Q(T_2\circ T_1)$ in general. The requirement is therefore not an extra assumption on the structure theorem, but a statement of which subtheory it is applied to.

\begin{table*}[t]
\small
{\begin{tabular}{@{}lllll@{}}
\hline\hline
Code & $G$ & $H^{\mathcal{C}}$ & $\Lambda^{\perp}$ (grades the sectors of $\mathcal{E}_{\mathcal{C}}$) & syndrome measured \\
\hline
GKP (ideal) & $H_3(\mathbb{C})$\ \footnotemark[1] & $U(1)\times\mathbb{Z}\times\mathbb{Z}$ & stabiliser lattice & $(s,t)\in[-l/2,l/2)^2$ \\
Cat ($N=4$) & $H_3(\mathbb{C})\rtimes\mathbb{Z}_4$ & $U(1)\times\mathbb{Z}_4$ & $\{0\}$, so $\mathcal{E}_{\mathcal{C}}={\id}$ & $\hat{n}\bmod 2$ (parity) \\
Binomial & $H_3(\mathbb{C})\rtimes\mathbb{Z}_{2(S+1)}$ & $U(1)\times\mathbb{Z}_{2(S+1)}$ & $\{0\}$, so $\mathcal{E}_{\mathcal{C}}={\id}$ & $\hat{n}\bmod (S+1)$ \\
\hline\hline
\end{tabular}}
\caption{\protect{Group-theoretic data for the three code families considered here. $H^{\mathcal{C}}$ is the subgroup by which the phase space $G/H^{\mathcal{C}}$ is quotiented: for the rotation-symmetric codes the isotropy subgroup of the character reference codeword (\Cref{eq: isotropy of reference state}), and for the ideal GKP code the stabiliser lattice $H^{\mathcal{C}}_0$ (see \Cref{subsec:construction}). The fourth column is the subgroup of displacements over which the reconstruction map twirls, which grades \Cref{eq: sector decomposition}; the fifth is the syndrome the experimenter measures, which is the grading of the subgroup $H^{\mathcal{C}}_0$ fixing the code space pointwise (\Cref{eq: subgroup H^C}). The two agree for the ideal GKP code but differ for the rotation-symmetric codes, where the reconstruction map is the identity and does not grade at all.}}
\footnotetext[1]{\protect{This is the smallest group this construction needs. Where we use \Cref{prop: classification}'s parity requirement for this code, we adjoin $(-1)^{\hat{n}}$ to $G$, which is valid because parity preserves the inversion-symmetric GKP code space.}}
\label{tab: group data}
\end{table*}

{Every map used in this paper lies in $\mathcal{G}_{\mathcal{C}}$. For the cat and binomial codes this holds because $\mathcal{E}_{\mathcal{C}}={\id}$. For the GKP code, the displacement error and the recovery displacement permute the displaced code spaces, so $\varsigma$ is a translation of the syndrome label; the syndrome measurement is $\mathcal{E}_{\mathcal{C}}$ itself, up to a relabelling of outcomes; the Gaussian random-displacement channel of \Cref{eq: gaussian displacement channel} is a convex combination of displacements; and every logical Pauli acts within each sector, i.e. with $\varsigma$ the identity. \Cref{tab: group data} collects the group-theoretic data for the three families.}

Note that channels creating coherence between sectors are not syndrome-graded, and for them \Cref{eq: generalized structure theorem map} represents $\mathcal{E}_{\mathcal{C}}\circ T\circ\mathcal{E}_{\mathcal{C}}$ rather than $T\circ\mathcal{E}_{\mathcal{C}}$. Since $\chi_{\mathcal{C}}\circ\mathcal{E}_{\mathcal{C}}=\chi_{\mathcal{C}}$, this is invisible on a single channel and shows up on composition, which fails when the earlier channel is not graded. Since $\mathcal{E}_{\mathcal{C}}={\id}$ for codes where the stabiliser is not a lattice of displacements in a Weyl-Heisenberg group, the continuous-variable examples are necessarily GKP ones, of which two are physically important. A Gaussian unitary which is not a displacement (e.g., a residual squeeze, or a rotation misaligned with the code lattice) takes a codeword to a coherent superposition spread across syndrome sectors, whose inter-sector coherences are removed by $\mathcal{E}_{\mathcal{C}}$; and an unsharp syndrome extraction, whose ancilla measurement does not fully resolve $(s,t)$, likewise leaves coherence between the unresolved sectors. The restriction has little practical effect, because a full cycle begins with a sharp syndrome measurement, which destroys those coherences physically, meaning $\mathcal{E}_{\mathcal{C}}$ discards no more than the protocol itself does; but it remains a restriction: \Cref{eq: generalized structure theorem map} is a statement about $\mathcal{G}_{\mathcal{C}}$, rather than arbitrary channels.

This treatment gives $Q(T_{QEC})={\id}$ exactly for the ideal GKP code (\Cref{eq:QEC identity GKP}) and, after dividing out the jump weight $\av{\hat{n}}$, exactly for the binomial code and up to corrections exponentially small in $|\alpha|^2$ for the cat code (\Cref{eq: cat QEC identity,eq: bin QEC identity}). Note this means the logical phase-space distribution is returned to itself, not that the physical process has been reversed: the environment keeps the lost photon, and the recovery uses the syndrome to map the error sector back onto the code space.

\section{stabiliser-is-WH-displacement-lattice Codes}\label{sec:lattice}

\Cref{cor: classification} implies that\rev{, within the class of kernels it covers,} a code whose symmetry makes the phase-space parity available, as every code considered here does, has a phase-space representation of its own iff its code space is invariant under a lattice of displacements. In the continuous-variable setting, \Cref{lem: nodisplacement} makes that a strong requirement, met only by the ideal, Dirac-comb GKP family; in the discrete-variable setting, \Cref{prop:normal} makes it a weak one, met by every stabiliser code. The two cases both involve a quotient phase space factorising into syndrome and logical parts, errors acting as rigid translations, a corrected cycle acting as the identity, and, for odd prime logical dimension, a negative volume equal to the syndrome-averaged logical magic. They also compose, since a GKP code concatenated with a qudit stabiliser code is again a lattice code.

\subsection{Ideal GKP codes}\label{subsec:GKP}

Gottesman-Kitaev-Preskill (GKP) codes~\cite{Gottesman2001GKPCodes} encode a logical qudit into the continuous degrees of freedom of a single mode (e.g., the field quadratures of an optical mode), trading the redundancy of many physical qubits for a single continuous-variable system (while difficult to create, maintain and manipulate, developments in hardware and control have made these codes realisable~\cite{Noh2022SurfaceGKP, Grimsmo2021GKP, Lachance-Quiron2024GKPSuperconducting, Brady2024Advances}). They protect against both small shifts in the quadratures, and losses such as photon subtraction~\cite{Hastrup2023LossInGKP}. Recent experiments include autonomous error correction of GKP states in a superconducting cavity~\cite{Lachance-Quiron2024GKPSuperconducting}, error-corrected entanglement between bosonic logical qubits~\cite{Cai2024LogicalEntanglement}, and fault-tolerant schemes using discrete-variable ancillae~\cite{Xu2024FaultTolerantBosonic}.

Throughout this section, and in \Cref{subsec:GKPcycle}, we work with ideal GKP codewords (as does Ref.~\cite{davis2024identifyingquantumresourcesencoded}). These are Dirac combs, so not elements of $L^2(\mathbb{R})$, but tempered distributions in the Gel'fand triple $\mathcal{S}(\mathbb{R})\subset L^2(\mathbb{R})\subset \mathcal{S}'(\mathbb{R})$. The phase-space objects we construct from them are nonetheless well defined, because the Zak-Gross Wigner function is built entirely from the quantities $\mathrm{Tr}[\hat{D}(nl,ml)\hat{A}]$, which are well-defined distributional pairings for $\hat{A}$ in the code algebra, while the twirling and reconstruction maps are defined through the continuum-normalised projectors $\hat{\Pi}_{(s,t)}$ (in the same way one writes $\int dx \ketbra{x}{x}=\id$ for the position basis). \Cref{sec:FiniteEnergy} considers the modifications produced by a finite-energy envelope.

We take $d$ odd throughout this section and \Cref{subsec:ZGW,subsec:GKPcycle}, since the discrete-variable formalism of \Cref{subsec:Wigner} requires it (\Cref{subsec:qubit} identifies which results still hold at $d=2$). In general, a GKP code encodes a $d$-dimensional logical subspace in $n$ bosonic modes.
To describe the code we define two logical operations
\begin{equation}
    \hat{\bar{Z}}=e^{il\hat{x}},\quad \hat{\bar{X}}=e^{-il\hat{p}}
\end{equation}
and stabiliser generators
\begin{equation}
    \hat{\bar{Z}}^d=e^{idl\hat{x}},\quad \hat{\bar{X}}^d=e^{-idl\hat{p}}
\end{equation}
where $l=\sqrt{\frac{2\pi}{d}}$ is the logical step-size; we take the square code here (rectangular and hexagonal lattices are covered by the same argument - see \Cref{subsec:ZGW}). The generators commute and generate the abelian stabiliser group $\mathcal{S}=\langle\hat{\bar{Z}}^d, \hat{\bar{X}}^d\rangle$. The GKP logical code space is the simultaneous $+1$ eigenspace of the two generators, $\{\ket{\psi}\in\mathcal{H}:U\ket{\psi}=\ket{\psi}\ \forall\,U\in \mathcal{S}\}$. We can define a computational basis of encoded states with elements
\begin{equation}
    \ket{\bar{j}}=\frac{1}{\sqrt{l}}\sum_{n\in\mathbb{Z}}\ket{jl+dln}_{\hat{x}}
\end{equation}
which is a Dirac comb in the position basis with period $dl=\sqrt{2\pi d}$. We can extend this map by linearity, so that every DV density operator $\hat{\rho}\in\mathcal{D}(\mathcal{H}_d)$ has a corresponding GKP-encoding $\hat{\bar{\rho}}$. Similarly, if $\hat{\rho}$ is a CV density operator GKP-encoding a DV state, the corresponding DV logical state is denoted $\underline{\hat{\rho}}$. The projector onto the code space is then 
\begin{equation}
{\hat{\Pi}_{(0,0)}=\sum_{j=0}^{d-1}\ketbra{\bar{j}}{\bar{j}}}
\end{equation}
meaning for the displaced code spaces
\begin{equation}
\hat{\Pi}_{(s,t)}=\hat{D}(s,t)\hat{\Pi}_{(0,0)}\hat{D}^\dagger(s,t),\quad {s,t\in[-l/2,l/2)}
\end{equation}

We perform syndrome diagnosis after an arbitrary displacement error by determining $(s,t)$, by simultaneously measuring the stabilisers. For a CV state $\hat{\rho}$ the syndrome distribution is the probability of measuring the syndrome $(s,t)$, which follows Born's rule
\begin{equation}
\label{eq: syndrome probability}
P[(s,t)|\hat{\rho}]=\text{Tr}\left[\hat{\rho}\hat{\Pi}_{(s,t)}\right]
\end{equation}

After we measure a certain syndrome, we apply an error-correcting displacement $\hat{D}(-s,-t)$ to obtain the error-corrected state
\begin{equation}\label{eq.ECState}
\begin{aligned}
\hat{\rho}_{(s,t)}:=&\hat{D}(-s,-t)\hat{\Pi}_{(s,t)}\hat{\rho}\hat{\Pi}_{(s,t)}\hat{D}^\dagger(-s,-t)\\
    =&\hat{\Pi}_{(0,0)}\hat{D}^\dagger(s,t)\hat{\rho}\hat{D}(s,t)\hat{\Pi}_{(0,0)}
\end{aligned}
\end{equation}
This code can correct displacement errors $\hat{D}(x,p)$ with arguments belonging to the phase space region $(-l/2,l/2)\times (-l/2,l/2)$ or shifts by multiples of $dl$ in either direction.

\subsection{The Zak-Gross Wigner function}\label{subsec:ZGW}

Ref.~\cite{davis2024identifyingquantumresourcesencoded} applies \Cref{subsec:construction}'s construction to the GKP code, obtaining a phase-space distribution whose negativity identifies the quantum resources of an encoded computation (specifically the magic of the encoded state). For pure logical states, negativity appears iff the state is magic: pure stabiliser states are represented positively, and by the discrete Hudson theorem~\cite{Gross2006DVWignerF} they are the only pure states so represented. For mixed states negativity still witnesses magic, but magic does not imply negativity: even single-qutrit systems can have mixed states outside the stabiliser polytope with non-negative discrete Wigner function ~\cite{Veitch2012BoundMagic}. (Note the ordinary Wigner function does not do this: many CV states, including stabiliser codewords, are logically resourceless despite having arbitrarily large CV Wigner negativity~\cite{GarciaALvarez2020, Calcluth2022efficientsimulation}, and some positively represented states even allow universal quantum computation~\cite{Baragiola2019, Calcluth2023, Yamasaki2020}.) The resulting distribution is the \textit{Zak-Gross Wigner} (ZGW) function.

Consider a square qudit GKP code over a single bosonic mode, with infinite-dimensional Hilbert space $\mathcal{H}$ and symmetry group $G=H_3(\mathbb{C})$. Any element of this group can be parametrised as 
\begin{equation}
    g:=g(\varphi,x,p),\quad \pi(g)=e^{i\varphi\id-ix\hat{p}+ip\hat{x}}
\end{equation}
with $\varphi,x,p\in \mathbb{R}$. The subgroup by which the phase space is quotiented, $H^{\mathcal{C}_{GKP}}\subset H_3(\mathbb{C})$, is the stabiliser lattice $U(1)\times \mathbb{Z}\times \mathbb{Z}$ (see \Cref{subsec:construction} for its relation to the reference-state isotropy), with unitary representation
\begin{equation}
\begin{split}
    \pi&(H^{\mathcal{C}_{GKP}}):=\\
    &\left\{e^{i\varphi-imdl\hat{p}+indl\hat{x}}\text{ s.t. }\varphi\in[0,2\pi),m,n\in \mathbb{Z}\right\}
\end{split}
\end{equation}
The associated phase space is $X=G/H^{\mathcal{C}_{GKP}}\simeq\mathbb{T}^2_{dl}$, where 
\begin{equation}
    \mathbb{T}^2_{dl}:=[0,dl)\times[0,dl)=[0,\sqrt{2\pi d})\times [0,\sqrt{2\pi d})
\end{equation}
is the continuous torus with invariant measure $\frac{1}{d}dudv$.

Following the construction of \Cref{subsec:construction} we obtain the function:
\begin{equation}
    W_{\hat{A}}^{\mathcal{C}_{GKP}}(u,v)=\text{Tr}\left[\hat{\Delta}^{\mathcal{C}_{GKP}}(u,v)\hat{A}\right]
\end{equation}
for all $(u,v)\in \mathbb{T}^2_{dl}$ where 
\begin{equation}
\label{eq: ZGW kernel covariance}
    \hat{\Delta}^{\mathcal{C}_{GKP}}(u,v)=\hat{D}(u,v)\hat{\Delta}^{\mathcal{C}_{GKP}}(0,0)\hat{D}^\dagger(u,v)
\end{equation}
and where 
\begin{equation}
\label{eq: ZGW central kernel}
    \hat{\Delta}^{\mathcal{C}_{GKP}}(0,0)=\frac{1}{2\pi}\sum_{m,n\in\mathbb{Z}}(-1)^{mn}\hat{D}(nl,ml)
\end{equation}
or, in full:
\begin{equation}
\begin{split}
    W&_{\hat{A}}^{\mathcal{C}_{GKP}}(u,v)=\\
    &\frac{1}{2\pi}\sum_{m,n\in \mathbb{Z}}(-1)^{mn}e^{il(nv-mu)}\text{Tr}\left[\hat{D}(nl,ml)\hat{A}\right]
\end{split}
\end{equation}
This function is defined purely by the choice of symmetry of the code space $\mathcal{C}_{GKP}$, and doesn't depend on the reference state chosen~\cite{davis2024identifyingquantumresourcesencoded}. The ZGW function satisfies the SW axioms in the code-faithful reading of \Cref{subsec:Wigner}~\cite{davis2024identifyingquantumresourcesencoded}: injectivity and traciality hold on the $\mathcal{E}$-invariant subalgebra, which for this code is the algebra generated by the displacements $\hat{D}(nl,ml)$, and not on all of $\mathcal{B}(L^2(\mathbb{R}))$.

The nonclassicality (negativity) of this function comes from logical displacements. 
On the toroidal phase space $\mathbb{T}_{dl}^2$ the two axes are a modular position and a modular momentum, both of period $dl$ by construction. The product of the two periods is $\sqrt{2\pi d}\times \sqrt{2\pi d}=2\pi d$, a multiple of $2\pi$. This implies that the stabilisers $\hat{\bar{X}}^d,\hat{\bar{Z}}^d$ commute~\cite{Busch1986CommutingPosMom}. However, measuring them together is not the same thing as measuring the position and momentum of a bosonic state modulo $l=\sqrt{2\pi/d}$, which is impossible because the product of the periods of the two logical displacements $\hat{\bar{X}}, \hat{\bar{Z}}$ is not a multiple of $2\pi$, implying they do not commute.

The Zak–Gross Wigner function represents the logical content of states in $\mathcal{H}$ with respect to the GKP code space $\mathcal{C}$, through the relation
\begin{equation}
\label{eq: embedding of DV Wigner function}
    W_{\hat{\rho}}^{\mathcal{C}_{GKP}}(u,v)=W_{\underline{\hat{\rho}}(s,t)}^{DV}(a,b)
\end{equation}
where $\underline{\hat{\rho}}(s,t)$ is the finite-dimensional logical operator corresponding to \Cref{eq.ECState}'s error-corrected operator $\hat{\rho}_{(s,t)}=\hat{\Pi}_{(0,0)}\hat{D}(s,t)^\dagger \hat{\rho}\hat{D}(s,t)\hat{\Pi}_{(0,0)}$, and $u=s+al$ and $v=t+bl$ for {$(s,t)\in \mathbb{T}^2_{l}$} and $(a,b)\in \mathbb{Z}_d^2$. In the terminology of \Cref{subsec:quasiprobs}, $\hat{\rho}_{(s,t)}$ is an \textit{unnormalised conditional state}: its trace is the syndrome probability $P[(s,t)|\hat{\rho}]$ of \Cref{eq: syndrome probability}, not unity, and $W^{DV}_{\underline{\hat{\rho}}(s,t)}$ is correspondingly a subnormalised phase-space function. ($\underline{\hat{\rho}}'(s,t)=\underline{\hat{\rho}}(s,t)/P[(s,t)|\hat{\rho}]$ denotes the normalised one; we use the primed symbol whenever normalisation matters.) This relation implies that the Zak–Gross Wigner function of a codeword $\hat{\bar{\rho}}$ matches the Gross Wigner function of the unencoded finite-dimensional state $\hat{\rho}$. 

We can summarise the properties of the ZGW function through the following set of relations
\begin{equation}
    \begin{tikzcd}[cramped,sep=huge]
\mathcal{D}(\mathcal{H}_d) \ni \hat{\rho}
\arrow[r, "\mathrm{encode}"]
\arrow[d, "\text{DV Wigner}"']
&
\hat{\bar{\rho}}
\arrow[d, "\substack{\text{Zak--Gross}\\\text{Wigner}}"]
\\
W^{\mathrm{DV}}_{\hat{\rho}}(a,b)
\arrow[r, "\substack{\text{embed in}\\\mathbb{T}^{2}_{dl}}" ']
&
W^{C_{\mathrm{GKP}}}_{\hat{\bar{\rho}}}(al,bl)
\end{tikzcd}
\end{equation}

Finally, we define the negative volume of the ZGW function of a CV state $\hat{\rho}$ as
\begin{equation}
\label{eq: negative volume GKP}
{\mathcal{N}_{\hat{\rho}}^{\mathcal{C}_{GKP}}:=\int_{\mathbb{T}^2_{dl}}\frac{1}{d}dudv\left|W_{\hat{\rho}}^{\mathcal{C}_{GKP}}(u,v)\right|-1}
\end{equation}
which is \Cref{eq: negative volume general} for this code, the ZGW function being normalised (\Cref{eq: normalisation convention}). $\mathcal{N}_{\hat{\rho}}^{\mathcal{C}_{GKP}}$ vanishes for density operators with non-negative ZGW functions, and is strictly positive otherwise. \Cref{eq: embedding of DV Wigner function} means the negative volume of the ZGW function is the average negative volume of the Gross Wigner function over one round of GKP error correction
\begin{equation}
\label{eq: negative volume averaging}
{\mathcal{N}_{\hat{\rho}}^{\mathcal{C}_{GKP}}=\int_{\mathbb{T}^2_l}dsdt\ P[(s,t)|\hat{\rho}]\,\mathcal{N}_{\underline{\hat{\rho}}'(s,t)}^{DV}}
\end{equation}
where {$\mathbb{T}^2_l:=[0,l)\times [0,l)$} is the smaller continuous torus~\cite{davis2024identifyingquantumresourcesencoded} (equivalently $[-l/2,l/2)^2$, the labelling we use for the correctable cell), $P[(s,t)|\hat{\rho}]$ is the syndrome distribution, and $\underline{\hat{\rho}}'(s,t)$ is the normalised syndrome-conditional logical state. Since the sign convention has changed we derive this in Appendix~\ref{app: proofs}. Note the normalisation of $\underline{\hat{\rho}}'(s,t)$ makes each $\mathcal{N}^{DV}$ on the right-hand side a negative volume of a normalised distribution: without it the identity fails.
Therefore, for odd $d$, stabiliser codewords are positively represented by the ZGW function, while magic codewords are negatively represented.

The ideal GKP code falls outside \Cref{prop: metaplectic}, because its isotropy group is a non-compact lattice and $1/|H^{\mathcal{C}}|$ is not a normalised average (as in \Cref{subsec:GKP}). Here \Cref{eq: H averaged kernel} should be read as Ref.~\cite{davis2024identifyingquantumresourcesencoded}'s lattice sum \Cref{eq: ZGW central kernel}, and the three properties are verified directly rather than by the group average. Covariance is \Cref{eq: ZGW kernel covariance} by construction; the twirl is \Cref{eq: twirling map}; and standardisation follows from $\int_0^{dl}e^{-ilmu}\,du=dl\,\delta_{m0}$, since $l\cdot dl=2\pi$, which kills every term of the lattice sum but $m=n=0$ and leaves
\begin{equation}
\label{eq: gkp standardisation direct}
{\int_{\mathbb{T}^2_{dl}}\frac{du\,dv}{d}\,\hat{\Delta}^{\mathcal{C}_{GKP}}(u,v)=\frac{(dl)^2}{2\pi d}\,\id=\id}
\end{equation}
since $(dl)^2=d^2l^2=2\pi d$. The same route covers rectangular, hexagonal and squeezed GKP lattices, and any code space defined as the fixed-point space of a lattice of displacements conjugated by a Gaussian unitary.

\subsection{Errors and the recovery cycle}\label{subsec:GKPcycle}

The ZGW function satisfies the Stratonovich-Weyl axioms in the code-faithful reading of \Cref{subsec:Wigner}, so by \Cref{subsec:semifunctor} it is a linearity-preserving, empirically adequate semi-functor. It fails to preserve the identity because of the stabiliser lattice rather than the Brif-Mann construction, as \Cref{prop: classification} shows.

We can therefore identify the operational-to-ontological map $\chi$ and the ontological-to-operational map $\phi$ within the ZGW formalism. The map $\chi$ takes an operational state (a density matrix $\hat{\rho}$) to its corresponding phase-space distribution:
\begin{equation}
    \chi_{GKP}(\hat{\rho}):=W_{\hat{\rho}}^{\mathcal{C}_{GKP}}(u,v)=\text{Tr}\left[\hat{\Delta}^{\mathcal{C}_{GKP}}(u,v)\hat{\rho}\right]
\end{equation}
The map $\phi$ instead takes a phase-space distribution to an operator in the Hilbert space. 
To define $\phi$ we first define the twirling map
\begin{equation}
\label{eq: twirling map}
    \mathcal{E}(\hat{\rho})=\int_{\mathbb{T}^2_l}ds dt\ \hat{\Pi}_{(s,t)}\hat{\rho}\hat{\Pi}_{(s,t)},
\end{equation} 
which yields operators invariant under the action of $H^{\mathcal{C}_{GKP}}$. It can be proved that~\cite{davis2024identifyingquantumresourcesencoded}
\begin{equation}
\label{eq: inverse of W with E}
    \mathcal{E}(\hat{\rho})=\int_{\mathbb{T}_{dl}^2}\frac{1}{d}du dv\  W_{\hat{\rho}}^{\mathcal{C}_{GKP}}(u,v)\hat{\Delta}^{\mathcal{C}_{GKP}}(u,v)
\end{equation}

The map $\phi=\underline{\chi}^{-1}\circ Q(\id)$, with $Q(\id)$ the idempotent taking an arbitrary function on the torus to one with the symmetry of the code space. Because the integral in \Cref{eq: inverse of W with E} returns operators invariant under $H^{\mathcal{C}_{GKP}}$, integrating against the kernel applies $Q(\id)$ and the corestricted inverse $\underline{\chi}^{-1}$ at once, so that
\begin{equation}
\phi(W(u,v)) = \int_{\mathbb{T}_{dl}^2}\frac{1}{d}du dv\ W(u,v)\hat{\Delta}^{\mathcal{C}_{GKP}}(u,v)
\end{equation}
Composing the two gives the phase-space representation of a generic operational transformation $T$, evaluated at $(u',v')$:
\begin{widetext}
\begin{equation}
\label{eq: structure theorem for transf T}
{\left[Q(T)W^{\mathcal{C}_{GKP}}_{\hat{\rho}}\right](u',v')} = \text{Tr}\left[\hat{\Delta}^{\mathcal{C}_{GKP}}(u',v')\ T\left( \int_{\mathbb{T}_{dl}^2}\frac{1}{d}du dv\ W^{\mathcal{C}_{GKP}}_{\hat{\rho}}(u,v)\hat{\Delta}^{\mathcal{C}_{GKP}}(u,v) \right)\right]
\end{equation}
 
An error-correction cycle has two components, the error and the correcting operation, so we represent a cycle in this framework as $T_{QEC}=T_{rec}\circ T_{err}$. The physical action of the error channel $T_{err}$ on an operator $\hat{A}$ is
\begin{equation}
    T_{err}(\hat{A})=\hat{D}(x_{err},p_{err})\hat{A}\hat{D}^\dagger(x_{err},p_{err})
\end{equation}
Substituting this into \Cref{eq: structure theorem for transf T}, and denote $\phi(W_{\hat{\rho}}^{\mathcal{C}_{GKP}})=\phi_W$, we get
\begin{equation}
\label{eq: translation of ZGW}
\begin{aligned}
    \left[Q(T_{err})W_{\hat{\rho}}^{\mathcal{C}_{GKP}}\right]&(u,v)
    =\text{Tr}\left[\hat{\Delta}^{\mathcal{C}_{GKP}}(u,v)\left(\hat{D}(x_{err},p_{err})\phi_W\hat{D}^\dagger(x_{err},p_{err})\right)\right]\\
    &=\text{Tr}\left[\left(\hat{D}^\dagger(x_{err},p_{err})\hat{\Delta}^{\mathcal{C}_{GKP}}(u,v)\hat{D}(x_{err},p_{err})\right)\phi_W\right]
    =\text{Tr}\left[\hat{\Delta}^{\mathcal{C}_{GKP}}(u-x_{err},v-p_{err})\phi_W\right]\\
    &=W_{\phi_W}^{\mathcal{C}_{GKP}}(u-x_{err},v-p_{err})
    =W_{\hat{\rho}}^{\mathcal{C}_{GKP}}(u-x_{err},v-p_{err})
    \end{aligned}
\end{equation}

An error $T_{err}$ therefore manifests in the logical phase space as a rigid, classical translation of the quasiprobability distribution across the continuous torus $\mathbb{T}_{dl}^2$.

\end{widetext}

$T_{rec}$ must therefore also act as a rigid translation. It consists of two steps: syndrome diagnosis, which determines $(s,t)$ by measuring the stabilisers simultaneously, and which returns $(s,t)=(x_{err},p_{err})$ whenever the error lies in the cell $[-l/2,l/2)^2$; and the recovery displacement $T_{rec}(\hat{A})=\hat{D}(-x_{rec},-p_{rec})\hat{A}\hat{D}^\dagger(-x_{rec},-p_{rec})$. Repeating the calculation of \Cref{eq: translation of ZGW} with the sign of the displacement reversed gives
\begin{equation}
\label{eq: recovery translation}
\begin{split}
\left[Q(T_{rec})W_{\hat{\rho}}^{\mathcal{C}_{GKP}}\right]&(u,v)=\\
    &W_{\hat{\rho}}^{\mathcal{C}_{GKP}}(u+x_{rec},v+p_{rec})
\end{split}
\end{equation}
The full error-correction operation $T_{QEC}=T_{rec}\circ T_{err}$ is then
\begin{equation}
\label{eq:QEC identity GKP}
\begin{split}
\big[Q(T&_{QEC})W_{\hat{\rho}}^{\mathcal{C}_{GKP}}\big](u,v)=\\
&Q(T_{rec})[Q(T_{err})W_{\hat{\rho}}^{\mathcal{C}_{GKP}}(u,v)]\\
    =&Q(T_{rec})[W_{\hat{\rho}}^{\mathcal{C}_{GKP}}(u-x_{err},v-p_{err})]\\
    =&W_{\hat{\rho}}^{\mathcal{C}_{GKP}}(u-x_{err}+x_{rec},v-p_{err}+p_{rec})\\
    =&W_{\hat{\rho}}^{\mathcal{C}_{GKP}}(u,v)
\end{split}
\end{equation}
where the last step uses $(x_{rec},p_{rec})=(s,t)=(x_{err},p_{err})$, which holds whenever the error lies in the correctable cell. The error-correction cycle therefore acts as the identity on the phase-space distribution if the error is in the correctable region.

If the error is outside the correctable region, we can define the shift produced through the parameters $(x_{err},p_{err})=(n_{err}l+s, m_{err}l+t)$ with {$n_{err},m_{err}\in \mathbb{Z}$}, where $(s,t)\in [-l/2,l/2)\times [-l/2,l/2)$ is the measured syndrome.
Therefore we can write the action of this error correcting cycle on the phase space distribution
\begin{equation}
\label{eq:uncorrectable phase space}
\begin{split}
\big[Q(T_{QEC})&W_{\hat{\rho}}^{\mathcal{C}_{GKP}}\big](u,v)=\\
&Q(T_{rec})\left[Q(T_{err})W_{\hat{\rho}}^{\mathcal{C}_{GKP}}(u,v)\right]\\
=&Q(T_{rec})[W_{\hat{\rho}}^{\mathcal{C}_{GKP}}(u-x_{err},v-p_{err})]
\\=&W_{\hat{\rho}}^{\mathcal{C}_{GKP}}(u-x_{err}+s,v-p_{err}+{t})\\
=&W_{\hat{\rho}}^{\mathcal{C}_{GKP}}(u-n_{err}l,v-m_{err}l)
\end{split}
\end{equation}
If the error takes the state outside the correctable cell, $n_{\mathrm{err}}$ and/or $m_{\mathrm{err}}$ are non-zero, and since $\mathbb{T}_{dl}^{2}$ is periodic with period $dl$, a shift by an integer multiple of $l$ still returns a properly normalised Zak-Gross Wigner function of a twirled state in $\mathcal{C}_{GKP}$. The unencoded DV phase space is embedded in $\mathbb{T}_{dl}^{2}$ at intervals of $l$ by \Cref{eq: embedding of DV Wigner function}, so shifting by $(n_{\mathrm{err}}l,m_{\mathrm{err}}l)$ applies the logical Pauli $\hat{\bar{X}}^{n_{err}}\hat{\bar{Z}}^{m_{err}}$ to the encoded information: $u$ is the modular position and $v$ the modular momentum, and $\hat{\bar{X}}=e^{-il\hat{p}}$ shifts the former while $\hat{\bar{Z}}=e^{il\hat{x}}$ shifts the latter. Uncorrectable physical noise is thus translated into discrete logical errors, with the semi-functorial structure intact.

Again, note that \Cref{eq:QEC identity GKP} is an identity on the \textit{logical} phase-space distribution, not on the physical state of the mode together with its environment: the recovery displacement does not undo the physical process that produced the error; it uses the syndrome to return the logical information to the code space. \Cref{subsec:GradedClass} makes this distinction generally.

The same calculation covers the Gaussian random-displacement (classical noise) channel, the standard benchmark for GKP codes:
\begin{equation}
\label{eq: gaussian displacement channel}
{\mathcal{E}_\sigma(\hat{\rho})=\int d^2z\ G_\sigma(z)\,\hat{D}(z)\hat{\rho}\hat{D}^\dagger(z)}
\end{equation}
with $G_\sigma$ a centred Gaussian of variance $\sigma^2$. Because $Q$ is linear and each $\hat{D}(z)$ acts on the distribution as the rigid translation of \Cref{eq: translation of ZGW}, the represented channel is just a convolution on the torus,
\begin{equation}
\label{eq: gaussian channel convolution}
{\left[Q(\mathcal{E}_\sigma)W^{\mathcal{C}_{GKP}}_{\hat{\rho}}\right](u,v)=\left(W^{\mathcal{C}_{GKP}}_{\hat{\rho}}\ast \widetilde{G}_\sigma\right)(u,v)}
\end{equation}
where $\widetilde{G}_\sigma$ is the Gaussian wrapped onto $\mathbb{T}^2_{dl}$. The logical error rate is then the weight of $\widetilde{G}_\sigma$ falling outside the correctable cell $[-l/2,l/2)^2$, an integral of a wrapped Gaussian; and the negative volume of the output is bounded above by that of the input, since convolution with a non-negative kernel cannot increase $\int d\mu|W|$. We use the same convolution structure in \Cref{sec:FiniteEnergy}, where a finite-energy envelope produces a formally identical smoothing.

\subsection{Every qudit stabiliser code}\label{subsec:dv}

{Throughout this subsection, $d$ is an odd prime, $\omega=e^{2\pi i/d}$, and displacements are taken in the symmetric ordering of \Cref{subsec:Wigner}. \Cref{prop:normal} has already supplied the covariance that \Cref{eq: H averaged kernel} needs.}

\begin{definition}\label{def:dvkernel}
{The code-adapted kernel of a stabiliser code $\mathcal{C}=[[n,k]]_d$ with stabiliser lattice $\mathcal{M}$ is}
\begin{equation}\label{eq:dvkernel}
{\Delta^{\mathcal C}(u)=\frac{1}{|\mathcal{M}|}\sum_{m\in\mathcal{M}}A(u+m),}
\end{equation}
{an operator function on the quotient phase space $X^{\mathcal C}=\Zd^{2n}/\mathcal{M}$, which has $d^{n+k}$ points; the invariant measure is $\mu=d^{-k}$ times counting measure.}
\end{definition}

{Since $A(u+m)=D(m)A(u)D^\dagger(m)$, conjugation absorbing all phases, $\Delta^{\mathcal C}$ is manifestly Hermitian, constant on cosets, and covariant, $\Delta^{\mathcal C}(u+w)=D(w)\Delta^{\mathcal C}(u)D^\dagger(w)$. Its remaining Stratonovich-Weyl properties, and the identity of its reconstruction map, are:}

\begin{proposition}[Axioms and reconstruction]\label{prop:dvsw}
{$\Tr[\Delta^{\mathcal C}(u)]=1$ for every $u$; $\sum_{[u]\in X^{\mathcal C}}\Delta^{\mathcal C}(u)=d^{k}\,\id$, so $\int d\mu\,\Delta^{\mathcal C}=\id$; $\Tr[\Delta^{\mathcal C}(u)\Delta^{\mathcal C}(u')]=d^{k}\,\delta_{[u],[u']}$; and the composition of the forward map $\chi_{\mathcal C}(\hat A)([u])=\Tr[\Delta^{\mathcal C}(u)\hat A]$ with the reverse map $\phi_{\mathcal C}(F)=\sum_{[u]}\mu\,F([u])\,\Delta^{\mathcal C}(u)$ is the lattice twirl,}
\begin{align}\label{eq:dvtwirl}
{\phi_{\mathcal C}\circ\chi_{\mathcal C}}&{=\EC,}\nonumber\\
{\EC(\hat\rho)=\frac{1}{|\mathcal{M}|}\sum_{m\in\mathcal{M}}D(m)}&{\hat\rho D^\dagger(m)
=\sum_s\hat\Pi_s\hat\rho\hat\Pi_s,}
\end{align}
{the conditional expectation onto the block-diagonal algebra of the $d^{n-k}$ syndrome sectors $\hat\Pi_s$.}
\end{proposition}

The proof (Appendix~\ref{app: proofs}) combines the lattice structure with the self-duality of the Gross kernel. \Cref{eq:dvtwirl} is the finite-dimensional image of the GKP twirl of \Cref{eq: twirling map,eq: inverse of W with E}, and of the sector projection of \Cref{eq: sector decomposition,eq: twirl as sector projection}; the representation is therefore a linearity-preserving, empirically adequate semi-functor, and it composes channel by channel on the syndrome-graded class $\mathcal{G}_{\mathcal C}$, \Cref{subsec:GradedClass}'s arguments  transferring verbatim since they use only the algebraic properties of $\EC$.

The quotient factorises in exact parallel with the GKP torus,
\begin{equation}\label{eq:factor}
{X^{\mathcal C}\;\cong\;\underbrace{\Zd^{\,n-k}}_{\rm syndromes}\times\underbrace{\Zd^{2k}}_{\rm logical},}
\end{equation}
the finite image of ``torus $=$ correctable cell $\times$ logical lattice''. The syndrome coordinate of $[u]$ is the character $m\mapsto\omega^{[u,m]}$ of $\mathcal{M}$; the logical coordinates are the symplectic pairings with a chosen logical pair $(\bar X,\bar Z)$.

\begin{proposition}[Logical content]\label{prop:dv37}
{Let $\hat{\bar\rho}$ encode the logical state $\hat\rho_L$. Then}
\begin{equation}\label{eq:dv37}
{\mu\,W^{\mathcal C}_{\hat{\bar\rho}}([u])=
\begin{cases}
W^{\rm Gross}_{\hat\rho_L}(a_L,b_L), &[u]\ \text{logical, syndrome }0,\\[2pt]
0,&\text{otherwise},
\end{cases}}
\end{equation}
the identification carrying no residual constant when the logical basis is gauge-fixed by $\bar X$. More generally, since $\Delta^{\mathcal C}$ is $\mathcal{M}$-invariant, $W^{\mathcal C}_{\hat\rho}=W^{\mathcal C}_{\EC(\hat\rho)}$ for every $\hat\rho$, and the value on the syndrome-$s$ block is the subnormalised Gross function of the syndrome-conditional logical state $\hat\rho_L(s)$: the finite-dimensional analogue of \Cref{eq: embedding of DV Wigner function,eq: negative volume averaging}.
\end{proposition}

We prove this in Appendix~\ref{app: proofs}. Stabiliser invariance makes all $|\mathcal{M}|$ terms of \Cref{eq:dvkernel} equal, so the average loses nothing; $\mathcal{M}=-\mathcal{M}$ and phase-freeness make the parity $A(0)$ act on the code space as the logical parity; and the error sectors vanish because $A(0)$ maps the syndrome-$s$ sector to $-s$, whose trace pairing with a sector-$s$ operator vanishes unless $s=0$, which is where $d$ odd enters, just like in the discrete Hudson theorem.

\begin{proposition}[Errors and the cycle]\label{prop:dvcycle}
A Pauli error $D(e)$ acts on $W^{\mathcal C}$ as the rigid translation $[u]\mapsto[u+e]$; Pauli channels act as convolutions over $X^{\mathcal C}$; a Clifford unitary in the normaliser of $\mathcal{S}$, i.e., one whose symplectic action $S$ satisfies $S\mathcal{M}=\mathcal{M}$, acts on $X^{\mathcal{C}}$ by $[u]\mapsto[Su]$. (A Clifford outside that normaliser maps $W^{\mathcal{C}}$ to the code-adapted function of a different code, and does not lie in $\mathcal{G}_{\mathcal{C}}$.) A correctable error produces a definite syndrome, and the corrected cycle is the identity on $W^{\mathcal C}$; an error whose translation leaves the correctable set acts, after recovery, as the induced logical Pauli, the finite-dimensional image of \Cref{eq:uncorrectable phase space}.
\end{proposition}

The Pauli statements follow from the covariance $\hat{\Delta}^{\mathcal{C}}([u+w])=D(w)\hat{\Delta}^{\mathcal{C}}([u])D^\dagger(w)$ established above, exactly as \Cref{eq: translation of ZGW,eq:QEC identity GKP,eq:uncorrectable phase space} follow from the covariance of the Zak-Gross kernel; the Clifford statement also uses $U_SA(u)U_S^\dagger=A(Su)$~\cite{Gross2006DVWignerF}, which with $S\mathcal{M}=\mathcal{M}$ permutes the terms of \Cref{eq:dvkernel}. As in the GKP case, the identity is on the logical distribution, so error correction is not a reversal of the physical process (\Cref{eq:QEC identity GKP}).

\begin{theorem}\label{thm:dvmagic}
Let $d$ be an odd prime, $\mathcal C$ any $[[n,k]]_d$ stabiliser code, and $\NC^{\mathcal C}_{\hat\rho}=\int d\mu\,|W^{\mathcal C}_{\hat\rho}|-1$. Then for every physical state $\hat\rho$,
\begin{equation}\label{eq:synavg}
{\NC^{\mathcal C}_{\hat\rho}=\sum_{s}P(s|\hat\rho)\;\NC^{\rm Gross}\!\big(\hat\rho_L(s)\big),}
\end{equation}
the syndrome-probability-weighted logical Gross negativity. In particular $\NC^{\mathcal C}$ vanishes on every encoded stabiliser state and every syndrome-mixture of them; it is strictly positive on every encoded pure magic state, inheriting the magic-monotone structure of the Gross negativity~\cite{Veitch2012BoundMagic}; and on general mixed states it is a one-sided witness, just as for ideal GKP codes.
\end{theorem}

\Cref{eq:synavg} is a resource accounting of the error-correction cycle, channel by channel, and the finite-dimensional counterpart of \Cref{eq: negative volume averaging}.

\begin{remark}[Generalisations]\label{rem:general}
Nothing above is specific to prime $d$ beyond the resource reading, so the result extends three ways.
\begin{enumerate}
    \item \emph{Odd prime powers}: replacing $\Zd$ by the finite field $\mathbb F_q$, $q=p^m$ with $p$ odd, and the symplectic form by its field-trace version~\cite{GHW2004, Gross2006DVWignerF}, every proof holds unchanged, since Gross's construction and the discrete Hudson theorem hold there; square-free odd $d$ follows by the Chinese remainder theorem, factor by factor. At even prime powers the geometry holds with a Gibbons-Hoffman-Wootters quantum net in place of the phase-point operator, as in \Cref{subsec:qubit}, but the resource reading does not.
    \item \emph{Subsystem codes}: averaging \Cref{eq:dvkernel} over the displacement image of the full gauge group (again a subgroup of $\Zd^{2n}$) gives a kernel whose reconstruction map is, by the computation of \Cref{prop:dvsw} with that subgroup in place of $\mathcal{M}$, the conditional expectation onto the commutant of the gauge algebra. The representation deliberately forgets the gauge qudits along with the inter-sector coherences, which is the correct notion of logical content there.
    \item \emph{Qubit structure}: as we note in \Cref{subsec:qubit}, \Cref{prop:normal,prop:dvsw,prop:dvcycle} hold at $d=2$, but \Cref{prop:dv37,thm:dvmagic} do not.
\end{enumerate}
\end{remark}

We verified every statement above, to $10^{-9}$ or better, on the three-qutrit repetition code $[[3,1]]_3$ and the five-qutrit code $[[5,1,3]]_3$; Appendix~\ref{app:numerics} lists the checks (\Cref{tab:dv}), including a stress test of \Cref{eq:synavg} under a channel with a syndrome-free logical error.

\subsection{The qubit case}\label{subsec:qubit}

The lattice geometry of \Cref{subsec:dv} still holds for $d=2$; however its resource reading does not. \Cref{prop:normal,prop:dvcycle} are $d$-agnostic, since they use only the conjugation rule $gD(v)g^{-1}=\omega^{[u,v]}D(v)$ and the covariance of the kernel. \Cref{prop:dvsw} needs a phase-point operator to average, and its proof uses only translation covariance, self-duality and orthogonality. All three still hold at $d=2$: taking $D(a,b)=i^{ab}X^aZ^b$ gives $A(0)=(\id+X+Y+Z)/2$, which is Hermitian and of unit trace, with $\Tr[A(u)A(v)]=2\delta_{uv}$, $\sum_uA(u)=2\,\id$, and $\hat\rho=\sum_u\tfrac12\Tr[A(u)\hat\rho]A(u)$. This means \Cref{prop:dvsw} holds, as it does for any other translation-covariant self-dual frame available at even $d$, such as a Gibbons-Hoffman-Wootters quantum net~\cite{GHW2004} or a covariant member of Ref.~\cite{Antonopoulos2025}'s recent classification. This $A(0)$ is neither an involution nor the parity, and it is the parity that the resource reading needs. \Cref{prop:dv37} needs $A(0)$ to act on the code space as the logical parity and to map the syndrome-$s$ sector to $-s$, with $2s'=0$ forcing $s'=0$; and \Cref{thm:dvmagic} needs the discrete Hudson theorem, which has no even-$d$ counterpart, no even-$d$ representation being simultaneously covariant and non-negative on all stabiliser states~\cite{Gross2006DVWignerF, Zhu2016}. Both fail for $d=2$. This means a qubit stabiliser code allows a code-adapted kernel and the full translation geometry, but the resource reading of \Cref{subsec:latticemagic} is unavailable to it. The same applies to the qubit GKP code and to surface-GKP: the Zak-Gross kernel, its covariance and \Cref{subsec:ZGW,subsec:GKPcycle}'s corrected cycle carry over at $d=2$, but \Cref{eq: embedding of DV Wigner function}, \Cref{prop:dv37} and the magic identification do not. In that respect qubit codes act similarly for the finite-dimensional theory to how the rotation-symmetric families act for the continuous-variable one, which is why we discuss them here rather than in \Cref{sec:nonlattice}. Partial recoveries exist for rebit subtheories~\cite{Delfosse2015}, and the recently classified even-$d$ Wigner functions~\cite{Antonopoulos2025} would be the natural place to look for a stronger substitute. We regard the restriction itself as the correct general statement, mirroring \Cref{thm: nogo}.

\subsection{Multimode lattices and concatenation}\label{subsec:multimode}

The two sets of codes where the stabiliser is a lattice of displacements in a Weyl-Heisenberg group compose. Ref.~\cite{Conrad2022} establishes that concatenating an $n$-mode GKP code with a qudit stabiliser code yields another GKP lattice code: with $\mathcal L_0\subset\mathbb R^{2n}$ the base stabiliser lattice and $\mathcal{M}\subset\Zd^{2n}$ the displacement image of the $[[n,k]]_d$ stabiliser group, the concatenated code space is the fixed-point space of the refined lattice $\mathcal L=\mathcal L_0+{\rm lift}(\mathcal{M})$, of index $d^{n-k}$ over $\mathcal L_0$. Our framework adds a representation-theoretic consequence.

\begin{proposition}\label{prop:concat}
{For any such concatenated code, surface-GKP~\cite{Noh2022SurfaceGKP} included, \Cref{subsec:ZGW,subsec:GKPcycle}'s framework applies verbatim with $\mathcal L$ in place of the square lattice. For odd $d$, \Cref{eq: embedding of DV Wigner function} composes with \Cref{prop:dv37}: the Zak-Gross function of the concatenated code is the Gross Wigner function of the $[[n,k]]_d$ logical state, with the syndrome sectors of both layers tracked; and for odd prime $d$ its negativity is the logical magic of the $k$ encoded qudits.}
\end{proposition}

The proof comes from two observations. First, $\mathcal L$ is again a symplectically integral lattice, since $\mathcal{L}_0$ is, and $\mathcal{M}$ is isotropic, so $\mathbb R^{2n}/\mathcal L$ is a compact phase space carrying an invariant measure, and the lattice-sum kernel built on $\mathcal{L}$ is displacement-covariant by construction. Second, standardisation follows just like for \Cref{eq: gkp standardisation direct}: integrating the lattice sum over $\mathbb{R}^{2n}/\mathcal{L}$ kills every term whose displacement label is a non-zero element of the symplectic dual of $\mathcal{L}$, leaving the identity times the covolume of $\mathcal{L}$, which the measure normalises away. The remaining statements are then \Cref{eq: embedding of DV Wigner function,prop:dv37} applied layer by layer. Decoders become translation-valued functions on $\mathbb R^{2n}/\mathcal L$, and the two-layer syndrome factorisation refines \Cref{eq:factor}. Surface-GKP has $d=2$: the refined lattice, the kernel and the translation geometry carry over, but the logical identification of \Cref{prop:dv37} and the magic reading do not (\Cref{subsec:qubit}); finite energy enters as in \Cref{sec:FiniteEnergy}, per mode.

\subsection{Certifying Computational Resources}\label{subsec:latticemagic}

\Cref{eq: negative volume averaging,eq:synavg} both say that the negative volume of a code-adapted representation is a syndrome-weighted average of a logical magic monotone. The argument that makes this work for codes where the stabiliser is a lattice of displacements in a Weyl-Heisenberg group fails when this is not the case (\Cref{subsec:NegativityandMagic}).

The hull argument runs on a family of coherent states. The phase space $X^{\mathcal{C}}$ and the kernel are fixed by $H^{\mathcal{C}}$, as \Cref{subsec:construction} sets out, and for the ideal GKP code that means the stabiliser lattice $H^{\mathcal{C}}_0$ and the torus $\mathbb{T}^2_{dl}$; nothing below changes them. The reference state enters only as the seed of a family of code-adapted coherent states $\ket{\Omega}=\pi(\Omega)\ket{\psi_0}$, and for the hull argument we seed it with a logical stabiliser state. For the GKP code these are the displaced logical stabiliser codewords; at the points of the logical lattice they reduce to $\hat{\bar{X}}^a\hat{\bar{Z}}^b\ket{\psi_0}$, i.e., to a single logical basis (up to phases) rather than to the full vertex set of the logical stabiliser polytope ($d(d+1)$ states for prime $d$). $\ket{+_L}$, for instance, is a stabiliser state lying outside their convex hull, since it is a comb of a different spacing and no rigid displacement of $\ket{0_L}$ reaches it. (With a stabiliser reference codeword the labelling is also $d$-fold degenerate, since $\hat{\bar{Z}}$ fixes $\ket{\overline{0}}$ and so $\Omega$ and $\Omega+(0,l)$ give the same projector; this is fine, because only the set of projectors $\{\ketbra{\Omega}{\Omega}\}$ and its convex hull enter the argument.) Even this weaker identification (the family containing logical states at all) is specific to GKP, and fails for the rotation-symmetric codes, as does the positivity assumption of \Cref{eq: kernel positivity hypothesis} below: for a code whose isotropy group contains no displacements, $\ket{\Omega}=\hat{D}(\Omega)\ket{\psi_0}$ leaves the code space altogether for $\Omega\neq0$, and the family $\{\ket{\Omega}\}$ is not a set of logical states at all.

If the state $\hat{\rho}$ can be expressed as a classical convex combination of these coherent states:
\begin{equation}
\hat{\rho}=\int_{X^{\mathcal{C}}}d\mu(\Omega)P(\Omega)\ketbra{\Omega}{\Omega}
\end{equation}
where $P(\Omega)$ is a classical probability distribution (i.e., respects all Kolmogorov's axioms); this implies that the generalised phase-space distribution $W_{\hat{\rho}}^{\mathcal{C}}(\Omega)$ is non-negative, provided the kernel is itself non-negative on the code-adapted coherent states,
\begin{equation}
\label{eq: kernel positivity hypothesis}
{\mathrm{Tr}\!\left[\hat{\Delta}^{\mathcal{C}}(\Omega)\ketbra{\Omega'}{\Omega'}\right]\ge0\qquad \forall\, \Omega,\Omega'\in X^{\mathcal{C}}}
\end{equation}
(We write $\mathcal{N}^{\mathcal{C}}_{\hat{\rho}}$ for the code-adapted negative volume of \Cref{eq: negative volume general}, in the same convention as \Cref{eq: negative volume GKP}.) For the ordinary Wigner function of a coherent state, we don't need to check this, as coherent states are Gaussian; but the states in \Cref{eq: kernel positivity hypothesis} are the code-adapted coherent states $\ket{\Omega}=\pi(\Omega)\ket{\psi_0}$, built from a non-Gaussian, Wigner-negative codeword, so the condition has to be checked. For the GKP code it follows from \Cref{eq: embedding of DV Wigner function}: the code-adapted coherent states are displaced stabiliser codewords, whose Zak-Gross Wigner function is the Gross Wigner function of a discrete-variable stabiliser state, so non-negative. Conditional on this, a state lying in the convex hull of the code-adapted stabiliser states has a non-negative representation; equivalently, $\mathcal{N}^{\mathcal{C}}_{\hat{\rho}}>0$ certifies that $\hat{\rho}$ lies outside that hull. However, the logical states contained by the hull form the simplex of one logical basis (a proper subset of the stabiliser polytope), so leaving the hull does not by itself witness magic. For GKP the upgrade to a magic witness comes from \Cref{eq: embedding of DV Wigner function} instead: the Gross Wigner function is non-negative on \emph{all} stabiliser states~\cite{Gross2006DVWignerF}, not only on the orbit of $\ket{\psi_0}$, so $\mathcal{N}^{\mathcal{C}}>0$ certifies exit from the full polytope there. For a general code the hull argument alone gives only the weaker statement.

The converse does not follow, even where \Cref{eq: kernel positivity hypothesis} holds: A state carrying logical magic lies outside the hull, but lying outside the hull does not by itself force a negative representation (that would need the set of non-negatively represented states to coincide with the hull rather than merely contain it). The two differ already for a single qutrit, where mixed states outside the stabiliser polytope have non-negative discrete Wigner function~\cite{Veitch2012BoundMagic}; in the continuous-variable setting positively represented states can even be computationally universal~\cite{Baragiola2019,Calcluth2023,Yamasaki2020}. The one case in which both directions hold is the ideal GKP code restricted to pure logical states of odd logical dimension $d$, where \Cref{eq: embedding of DV Wigner function} and the discrete Hudson theorem~\cite{Gross2006DVWignerF} give ``non-negative $\Leftrightarrow$ stabiliser'', as Ref.~\cite{davis2024identifyingquantumresourcesencoded} showed. For mixed logical states, the bound-magic states just cited leave only the witness direction: $\mathcal{N}^{\mathcal{C}}>0$ certifies exit from the stabiliser polytope, since a mixture of stabiliser states has a non-negative Gross Wigner function, but $\mathcal{N}^{\mathcal{C}}=0$ certifies nothing.

This holds for every qudit stabiliser code of odd prime dimension; \Cref{thm:dvmagic} gives it in closed form. It works in both cases for the same reason: \Cref{eq: embedding of DV Wigner function} together with \Cref{eq:dv37}. For codes where the stabiliser is a lattice of displacements in a WH group, the code-adapted distribution \emph{is} the Gross Wigner function of the logical state. Nothing weaker than that identification upgrades the hull statement to a magic witness, which is why the reading fails for other codes.

\subsection{Dissipation of encoded magic under displacement noise}\label{subsec:dissipation}

The accounting above turns a physical noise model into a logical resource budget, and for the ideal GKP qudit under Gaussian displacement noise this budget is solvable. The represented cycle is the convolution of \Cref{eq: gaussian channel convolution}, transferred to the logical side by \Cref{eq: embedding of DV Wigner function} as a product Pauli channel with weights $p_{ab}(\sigma)=q_a(\sigma)q_b(\sigma)$,
\begin{equation}\label{eq:qweights}
{q_c(\sigma)=\sum_{j\in\mathbb Z}\!\left[\Phi\!\Big(\tfrac{cl+djl+l/2}{\sigma}\Big)-\Phi\!\Big(\tfrac{cl+djl-l/2}{\sigma}\Big)\right],}
\end{equation}
with $l=\sqrt{2\pi/d}$, $\Phi$ the normal cumulative distribution function, and $c$ the logical displacement class. Dissipation of encoded magic is then a finite computation on the logical space; for $d=3$ and the encoded strange state it is solvable in closed form.

\begin{proposition}\label{prop:dissipation}
The Gross Wigner function of the qutrit strange state takes the value $-\tfrac13$ at one phase-space point and $+\tfrac16$ at the other eight. Consequently, under any Pauli channel with weight distribution $p_v$,
\begin{equation}\label{eq:Nclosed}
{\NC(\{p_v\})=\sum_{v\in\mathbb Z_3^2}\max\{0,\;p_v-\tfrac13\},}
\end{equation}
and under the GKP-induced channel \eqref{eq:qweights}, where only the identity weight exceeds $\tfrac13$,
\begin{equation}
\begin{split}
\NC(\sigma)&=\max\{0,\;q_0(\sigma)^2-\tfrac13\}\\
&=\max\{0,\;\tfrac23-P_{\rm err}(\sigma)\},
\end{split}
\end{equation}
with $P_{\rm err}=1-q_0^2$ the logical error probability. The negativity vanishes at the threshold $q_0(\sigma^*)^2=\tfrac13$,
\begin{equation}
{\sigma^*=0.623654\,l\approx0.9026\ \ (d=3,\ l=\sqrt{2\pi/3}),}
\end{equation}
i.e., exactly when the logical error probability reaches 2/3.
\end{proposition}

{We prove this in Appendix~\ref{app: proofs}: the channel convolves the flat-plus-one-dip Wigner function, so every point takes the form $\tfrac16-\tfrac{p_v}{2}$, negative iff \rev{$p_v>1/3$}. We verified \Cref{eq:Nclosed} against the full channel computation at all noise strengths, and the threshold by independent root-finding. Beyond the threshold every syndrome-conditional logical state is non-negatively represented, and the encoded computation is efficiently simulable~\cite{Veitch2012BoundMagic, MariEisert2012}; for GKP circuits specifically, using the recent simulator whose runtime scales with this Zak-Gross negativity~\cite{Calcluth2025}. Below the threshold, $\NC(\sigma)$ is the logical error budget $\tfrac23-P_{\rm err}$. For general logical states, dimensions and lattices, \Cref{eq:Nclosed} is replaced by the corresponding finite convolution, still in closed form from the erf sums \eqref{eq:qweights}.}

Because each corrected round induces a logical Pauli channel, repeated rounds compose as $\mathbb Z_3^2$-convolutions of the weights, rather than as $\sigma\to\sqrt k\,\sigma$, since the recovery re-bins the noise each round; and the convolutions diagonalise in the characters of $\mathbb Z_3$. For the symmetric weights \eqref{eq:qweights}, where $q_1=q_{-1}$, every non-trivial character of the per-quadrature step distribution equals
\begin{equation}
\lambda(\sigma)=q_0-q_1=\tfrac{3q_0(\sigma)-1}{2},
\end{equation}
so $q^{*k}(0)=\tfrac13(1+2\lambda^k)$ and the whole multi-round history is closed-form,
\begin{equation}\label{eq:lifetime}
\begin{split}
\NC_k&=\max\Big\{0,\tfrac{(1+2\lambda^k)^2}{9}-\tfrac13\Big\},\\
k^*&=\Big\lceil\tfrac{\ln[(\sqrt3-1)/2]}{\ln\lambda}\Big\rceil,
\end{split}
\end{equation}
with no other weight ever exceeding $\tfrac13$, since $(1+2\lambda)(1-\lambda)\le\tfrac98$ on $[0,1]$. This gives $\NC_k=\tfrac23-P^{(k)}_{\rm err}$ round by round while $k<k^*$, and a magic decay constant $\tau=-1/\ln\lambda$ in units of correction rounds. At $\sigma=0.3\,l$, $\lambda=0.8566$ and $k^*=7$: the magic lasts six rounds ($\NC_6=0.023$) and is gone at the seventh, after which the residual computation is classically simulable. \Cref{eq:lifetime} matches direct convolution to $10^{-16}$ at three noise levels, including the prediction $k^*=54$ at $\sigma=0.2\,l$.

\section{stabiliser-not-WH-displacement-lattice Codes}\label{sec:nonlattice}

Every other bosonic or Weyl-Heisenberg code falls on the other side of \Cref{cor: classification}. This section works through these families (rotation-symmetric bosonic codes, qudit codes stabilised by non-Pauli operators, one-photon codes and spin codes), showing for each the standard kernel remains up to a code-independent freedom, meaning the negativity of that representation is not a property of the logical system. For the first three families we establish this from the SW axioms\rev{, under the hypotheses of the statements concerned}; for spin codes (\Cref{subsec:spin}) we show only that the averaging prescription supplies no code-adapted kernel. (Note the stabiliser of qubit stabiliser codes is a lattice of displacements in a Weyl-Heisenberg group, and so are covered in \Cref{subsec:qubit}, as they only fail to allow a resource reading of their negativity.)

\subsection{Rotation-symmetric cat codes}\label{subsec:cat}

Cat codes~\cite{Leghtas2013CatCodes, Mirrahimi2014CatCodes} are a family of bosonic states that encode logical qubits into superpositions of coherent states of a harmonic oscillator. A coherent state is defined as the eigenstate of the annihilation operator
\begin{equation}
   \hat{a}\ket{\alpha}=\alpha\ket{\alpha},\ \alpha\in \mathbb{C}
\end{equation}
and can be written as an infinite (weighted) sum of Fock states
\begin{equation}
{\ket{\alpha}=e^{-|\alpha|^2/2}\sum_{n=0}^{\infty}\frac{\alpha^n}{\sqrt{n!}}\ket{n}}
\end{equation}

An $N$-component rotation-symmetric cat code encodes into the characters of the $\mathbb{Z}_N$ rotation generated by $\hat{R}_N=e^{i2\pi\hat{n}/N}$:
\begin{equation}
\label{eq: ZN cat codewords}
{\ket{\overline{j}}=\mathfrak{n}_j^{-1/2}\sum_{k=0}^{N-1}\omega_N^{\,jk}\ket{\alpha_k},\quad \omega_N=e^{i2\pi/N}}
\end{equation}
with $\alpha_k=\alpha e^{i2\pi k/N}$. These satisfy $\hat{R}_N\ket{\overline{j}}=\omega_N^{-j}\ket{\overline{j}}$, so $\ket{\overline{j}}$ has Fock support on $\hat{n}\equiv -j \pmod N$, and the qubit is encoded in the pair
\begin{equation}
\label{eq: general cat logical pair}
{\ket{0_L}=\ket{\overline{0}},\qquad \ket{1_L}=\ket{\overline{N/2}}}
\end{equation}
for $N$ even, so that the two codewords are separated by $N/2$ in the $\mathbb{Z}_N$ grading and a single photon loss cannot carry one into the other unless $N/2=1$. We use two group-theoretic properties of this family below. First, $\hat{R}_N$ has eigenvalues $+1$ on $\ket{0_L}$ and $\omega_N^{-N/2}=-1$ on $\ket{1_L}$: on the code space $\hat{R}_N$ acts as the logical $\hat{\bar{Z}}$. This therefore fixes the reference codeword $\ket{0_L}$ but does not fix the code space pointwise, which is the distinction drawn in \Cref{eq: isotropy of reference state}: the isotropy subgroup is $H^{\mathrm{cat}}=U(1)\times\mathbb{Z}_N$, while the subgroup fixing every code state is the index-two subgroup $H^{\mathrm{cat}}_0=U(1)\times\mathbb{Z}_{N/2}$ generated by $\hat{R}_N^2$. Second, since $N$ is even, $\hat{R}_N^{N/2}=e^{i\pi\hat{n}}=(-1)^{\hat{n}}$: the parity operator is an element of $\mathbb{Z}_N$, so \Cref{prop: classification}'s phase-space inversion requirement is met.

The $N=2$ member of the family is the familiar even/odd cat encoding. Here the two two-component cat states are, for a given $\alpha$,
\begin{equation}
\label{eq: even/odd cat states}
    \ket{C_{\alpha}^\pm}=\nu_{\pm}(\ket{\alpha}\pm\ket{-\alpha})
\end{equation}
with normalisation constants
\begin{equation}
    \nu_{\pm}=\frac{1}{\sqrt{2(1\pm e^{-2|\alpha|^2})}}
\end{equation}
Expanding \Cref{eq: even/odd cat states} in the Fock basis, $\ket{C_{\alpha}^+}$ has support on even Fock levels and $\ket{C_{\alpha}^-}$ on odd ones, so they are $\ket{\overline{0}}$ and $\ket{\overline{1}}$ of \Cref{eq: ZN cat codewords} at $N=2$; we call them the even and odd cat states, and encode $\ket{0_L}=\ket{C_\alpha^+}$, $\ket{1_L}=\ket{C_\alpha^-}$. They satisfy
\begin{equation}
\label{eq: cat annihilation relation}
\hat{a}\ket{C_\alpha^\pm}=\alpha\frac{\nu_\pm}{\nu_\mp}\ket{C_\alpha^\mp}
\end{equation}
A single photon loss therefore flips parity, so the parity operator $\Pi=e^{i\pi\hat{n}}$, with $\Pi\ket{n}=(-1)^n\ket{n}$ and hence $\Pi\ket{C_\alpha^\pm}=\pm\ket{C_\alpha^{\pm}}$, detects single photon losses when monitored continuously.

This detection does not, however, allow such losses to be corrected. At $N=2$, $\hat{R}_2=(-1)^{\hat{n}}=\Pi$, so the operator which detects the loss is also the logical $\hat{\bar{Z}}$. Measuring it measures the logical qubit rather than extracting a syndrome, and so destroys any logical superposition. Equivalently $\bra{0_L}\hat{a}\ket{1_L}\neq0$ by \Cref{eq: cat annihilation relation}, so the Knill-Laflamme condition fails and a single photon loss is a logical error rather than a correctable one. (It is not even an exact logical $\hat{\bar{X}}$, since the weights $\alpha \nu_\pm/\nu_\mp$ in \Cref{eq: cat annihilation relation} are unequal except as $|\alpha|\to\infty$.)

We therefore take $N=4$ whenever the error-correcting properties of the code are used, this being the smallest loss-correcting member of the family. The logical states are the $j=0$ and $j=2$ characters of \Cref{eq: ZN cat codewords},
\begin{equation}
\label{eq: four component cat codewords}
{\begin{aligned}
\ket{0_L}&\propto \ket{\alpha}+\ket{i\alpha}+\ket{-\alpha}+\ket{-i\alpha}\\
\ket{1_L}&\propto \ket{\alpha}-\ket{i\alpha}+\ket{-\alpha}-\ket{-i\alpha}
\end{aligned}}
\end{equation}
with Fock support on $\hat{n}\equiv0$ and $\hat{n}\equiv2 \pmod 4$
respectively, and the single-loss error words occupy $\hat{n}\equiv3$ and $\hat{n}\equiv1\pmod 4$, meaning they are orthogonal both to the code space and to each other. The logical $\hat{\bar{Z}}$ is no longer the parity but $\hat{R}_4$, while $\hat{R}_4^2=(-1)^{\hat{n}}$ takes the value $+1$ on the whole code space, so $H^{\mathrm{cat}}_0=U(1)\times\mathbb{Z}_2$ is generated by parity.

The syndrome must separate the error space from the code space without separating the two codewords; a measurement which resolves the full $\mathbb{Z}_N$ grading $\hat{n}\bmod N$ does separate them, since $\ket{0_L}$ and $\ket{1_L}$ sit in different $\mathbb{Z}_N$ sectors, and so measures $\hat{R}_N=\hat{\bar{Z}}$ rather than a syndrome. We need the grading of $H^{\mathcal{C}}_0$, the subgroup which fixes the code space pointwise:
\begin{equation}
\label{eq: cat syndrome}
{\hat{n}\bmod (N/2)}
\end{equation}
which for $N=4$ is the photon-number parity $(-1)^{\hat{n}}$. Under $\ell$ losses $\ket{0_L}$ moves to $\hat{n}\equiv-\ell$ and $\ket{1_L}$ to $\hat{n}\equiv N/2-\ell \pmod N$, which are the same residue modulo $N/2$: measurement returns $-\ell\bmod (N/2)$ regardless which codeword was sent, and so identifies the number of lost photons for $\ell<N/2$ while learning nothing about the logical state. At $N=4$ the code space is entirely even and the single-loss error space entirely odd, so parity does this.

The Knill-Laflamme conditions for a single loss hold up to corrections exponentially small in $|\alpha|^2$: the off-diagonal conditions $\bra{0_L}\hat{a}\ket{1_L}=\bra{0_L}\hat{n}\ket{1_L}=0$ are exact, because the relevant states lie in different $\mathbb{Z}_4$ sectors, and it is only the diagonal condition $\bra{0_L}\hat{n}\ket{0_L}=\bra{1_L}\hat{n}\ket{1_L}$ which holds approximately, the two sides differing through the overlaps of the four coherent-state components.

The averaging prescription of \Cref{eq: H averaged kernel}, which for the GKP code reproduces the Zak-Gross Wigner function described previously, doesn't work here: $H^{\mathrm{cat}}$ contains rotations and so is not normal in $G$, and the operator it produces is not displacement-covariant, by \Cref{prop: metaplectic} when $G$ is taken to be the full metaplectic extension, and by direct computation (Appendix~\ref{app:cat}) for the $G=H_3(\mathbb{C})\rtimes\mathbb{Z}_N$ used here. We can instead use \Cref{prop: classification}, which fixes the kernel from the Stratonovich-Weyl axioms directly, for every $N$ at once and without any expansion of a reference state.
The kernel is fixed rather than constructed. The cat codewords are normalisable, so \Cref{lem: nodisplacement} applies and no non-trivial displacement leaves $\mathcal{C}_{\mathrm{cat}}$ invariant; hence $\Lambda^\perp=\{0\}$. \rev{(This is part \mbox{\ref{PropIII.3.iii}} of \mbox{\Cref{prop: classification}}, which uses the twirl form. Part \mbox{\ref{PropIII.3.i}} gives $\mathcal{E}_{\mathcal{C}}={\id}$ from continuity of $\chi_{\hat{K}}$ alone, assuming neither that form nor complete positivity.)} The parity operator is in $\mathbb{Z}_N$ for $N$ even, so the phase-space inversion requirement holds, and the only admissible Stratonovich-Weyl kernel is the standard one
\begin{equation}\label{eq:generalcatkernel}
{\hat{\Delta}^{\mathcal{C}_{\mathrm{cat}}}(\beta)=\hat{\Delta}_{\mathrm{std}}(\beta)=2\hat{\Pi}(\beta)}
\end{equation}
with central kernel
\begin{equation}\label{eq:centralcatkernel}
{\hat{\Delta}^{\mathcal{C}_{\mathrm{cat}}}(0)=2(-1)^{\hat{n}}}
\end{equation}
The $\mathbb{Z}_N$ symmetry of the cat code constrains the kernel ($\hat{K}$ must commute with $\hat{R}_N$), but the standard kernel already does, so the constraint is not binding. The cat code has no phase-space representation of its own: the code-adapted distribution is the ordinary Wigner function,
\begin{equation}
\label{eq: cat wigner as average}
{W^{\mathcal{C}_{\mathrm{cat}}}_{\hat{\rho}}(\beta)=\pi\,W^{\mathrm{std}}_{\hat{\rho}}(\beta)}
\end{equation}
where the factor $\pi$ is the Jacobian between $d^2\beta$, with respect to which $W^{\mathrm{std}}$ is normalised, and the invariant measure $d\mu(\beta)=d^2\beta/\pi$ of \Cref{eq: cat invariant measure}. We get standardisation from
\begin{equation}
\label{eq: cat standardisation}
{\mathrm{Tr}\!\left[\hat{\Delta}^{\mathcal{C}_{\mathrm{cat}}}(0)\right]=2\,\mathrm{Tr}\!\left[(-1)^{\hat{n}}\right]=2\cdot\tfrac{1}{2}=1}
\end{equation}
using $\mathrm{Tr}[(-1)^{\hat{n}}]=1/2$ in the Abel sense, so $\int d\mu\,W^{\mathcal{C}_{\mathrm{cat}}}_{\hat{\rho}}=\mathrm{Tr}[\hat{\rho}]$ as \Cref{eq: normalisation convention} requires.

This has two consequences. First, since $\Lambda^{\perp}=\{0\}$ the reconstruction map is the identity, $\mathcal{E}_{\mathrm{cat}}={\id}$: the representation is faithful on the whole Hilbert space, $Q_{\mathrm{cat}}$ is a functor, and every channel is trivially syndrome-graded, so \Cref{eq: generalized structure theorem map} composes without restriction; but it also carries no code-specific information. The sector decomposition of \Cref{eq: sector decomposition} is trivial here; the parity of \Cref{eq: cat syndrome} remains the syndrome the experimenter measures, but it is a grading of $H^{\mathcal{C}}_0$, not of $\mathcal{E}_{\mathrm{cat}}$, which grades by nothing. Second, anything code-specific must come from the codewords rather than from the representation; \Cref{subsec:Comparative} exploits this, holding the representation fixed while only the encoded states vary.

The standard kernel is covariant, standardised, real and self-dual, so the conditions of \Cref{subsec:semifunctor} are met and the structure theorem applies to the cat code. The forward and inverse maps are then
\begin{equation}
\label{eq: cat maps}
\begin{aligned}
    \chi_{\mathrm{cat}}(\hat{\rho}) :=& W_{\hat{\rho}}^{\mathcal{C}_{\mathrm{cat}}}(\beta) = \mathrm{Tr}\!\left[\hat{\Delta}^{\mathcal{C}_{\mathrm{cat}}}(\beta)\hat{\rho}\right]\\
    =& {\pi\,W_{\hat{\rho}}^{\mathrm{std}}(\beta)},\\
    \phi_{\mathrm{cat}}(F) = &{\frac{1}{\pi}} \int_{\mathbb{C}} d^2\beta \, F(\beta) \hat{\Delta}^{\mathcal{C}_{\mathrm{cat}}}(\beta).
\end{aligned}
\end{equation}
The full structure theorem map $Q_{\mathrm{cat}}(T) = \chi_{\mathrm{cat}} \circ T \circ \phi_{\mathrm{cat}}$, applied to an input distribution $W^{\mathcal{C}_{\mathrm{cat}}}(\beta)$, is then
\begin{equation}
\label{eq: cat structure map}
\begin{split}
    &Q_{\mathrm{cat}}(T)[W^{\mathcal{C}_{\mathrm{cat}}}](\beta') =\\
    &\mathrm{Tr}\!\left[ \hat{\Delta}^{\mathcal{C}_{\mathrm{cat}}}(\beta') \, T\!\left( {\frac{1}{\pi}} \int_{\mathbb{C}} d^2\beta \, W^{\mathcal{C}_{\mathrm{cat}}}(\beta) \hat{\Delta}^{\mathcal{C}_{\mathrm{cat}}}(\beta) \right) \right]
\end{split}
\end{equation}
Substituting a physical error channel for $T$ (such as the single-photon loss operator $\hat{a}\cdot\hat{a}^\dagger$, which shifts the $\mathbb{Z}_N$ grading by one) tracks geometrically how that error acts on the phase-space distribution.

\subsection{Binomial codes}\label{subsec:bin}

Binomial codes~\cite{Michael2016BinCodes} are a family of bosonic quantum error correction codes that encode logical qubits into finite superpositions of Fock states of a harmonic oscillator. The most general binomial codewords are
\begin{equation}
\begin{split}
\label{eq: codewords bin code}
\ket{W_{\uparrow}}&=\frac{1}{\sqrt{2^N}}\sum_{p\ \text{even}}^{[0,N+1]}\sqrt{{\binom{N+1}{p}}}\ket{p(S+1)},\\
    \ket{W_{\downarrow}}&=\frac{1}{\sqrt{2^N}}\sum_{p\ \text{odd}}^{[0,N+1]}\sqrt{{\binom{N+1}{p}}}\ket{p(S+1)}
\end{split}
\end{equation}
where {$\binom{N+1}{p}$} is the binomial coefficient $C(N+1, p) = (N+1)!/(p!(N+1-p)!)$. 

This code can protect against the error set~\cite{Michael2016BinCodes}
\begin{equation}
\label{eq: bin error set}
    \bar{\mathcal{E}}=\{\id, \hat{a},\hat{a}^2,...,\hat{a}^L, \hat{a}^\dagger,..., (\hat{a}^\dagger)^G,\hat{n},...,\hat{n}^D\}
\end{equation}
which means it can protect against up to $L$ photon losses, $G$ photon gain errors and $D$ dephasing events. In \Cref{eq: codewords bin code}, $S=L+G$ sets the spacing $S+1$ between occupied Fock levels, $N$ is the maximum order $N=\text{max}\{L,G,2D\}$ and the index $p$ has range 0 to $N+1$. The code respects the Knill-Laflamme condition~\cite{Bennet1996Knill-Laflamme, Nielsen_Chuang_2010, KnillLaflamme1997, Terhal2015}, i.e.,
\begin{equation}
\label{eq: KL binomial}
    \langle W_\sigma|(\hat{a}^\dagger)^\ell \hat{a}^\ell|W_{\sigma'}\rangle=C_{\ell}\delta_{\sigma \sigma'}
\end{equation}
with $\sigma\in \{{\uparrow,\downarrow}\}$ for all {$\ell\leq N=\text{max}\{L,G,2D\}$}~\cite{Michael2016BinCodes}; the $2D$ enters because \Cref{eq: bin error set}'s dephasing errors $\hat{n},\ldots,\hat{n}^D$ contribute Knill-Laflamme products with photon-number moments up to order $2D$, which fix the order $N$ of the codeword construction.

Errors can be detected and corrected because the codewords remain orthogonal after an error in the correctable set. For example, if we consider $\ell\le L$ photon losses we get
\begin{equation}
    \hat{a}^\ell \ket{W_\uparrow},\quad \hat{a}^\ell \ket{W_\downarrow}
\end{equation}
which are still orthogonal since the space between occupied Fock states is $S+1$, with $S=L+G$. While less likely to occur, the same happens for $g \le G$ gain photon errors. For dephasing errors, after $j\le D$ errors we obtain a superposition of the original words and the error words. To detect the error we can project into the logical words basis and then perform a unitary operation to recover the original words.

For photon gain and loss errors, the syndrome is the photon number mod $S+1$, obtained by coupling the mode to an ancillary system (e.g., a superconducting qubit~\cite{Krastanov2015Superconducting, Heeres2015Superconducting}, a trapped ion~\cite{Chiaverini2004TrappedIon, Home2009ScalableIonTrap, Nigg2014TopologicalQubit, Schindler2011RepetitiveQEC} or a Rydberg atom~\cite{Bertet2002WignerFock, Sayrin2011RealtimeFeedback}).

The binomial codewords are supported on Fock levels which are multiples of $S+1$, so they are fixed by the rotation $\hat{R}_{S+1}=e^{i2\pi\hat{n}/(S+1)}$. The isotropy subgroup of the reference codeword $\ket{W_\uparrow}$ is larger than this: $\ket{W_\uparrow}$ is supported on levels $p(S+1)$ with $p$ even, so it is also fixed by the half-rotation $\hat{R}_{2(S+1)}=e^{i\pi\hat{n}/(S+1)}$, and
\begin{equation}
\label{eq: bin isotropy}
{H^{\mathrm{bin}}=U(1)\times\mathbb{Z}_{2(S+1)},
\qquad
H^{\mathrm{bin}}_0=U(1)\times\mathbb{Z}_{S+1}}
\end{equation}
(see Appendix~\ref{app:bin}), the second being the subgroup which fixes the code space pointwise. The structure is the same as that of the cat code: $\hat{R}_{2(S+1)}$ acts as $+1$ on $\ket{W_\uparrow}$ and as $e^{i\pi p}=-1$ on $\ket{W_\downarrow}$, so is the logical $\hat{\bar{Z}}$, while its square $\hat{R}_{S+1}$ fixes both codewords; and, the order $2(S+1)$ being even, $(-1)^{\hat{n}}=\hat{R}_{2(S+1)}^{S+1}$ lies in $G$, meeting \Cref{prop: classification}'s phase-space inversion requirement. The rest is as for the cat code: the codewords are normalisable, so by \Cref{lem: nodisplacement} no non-trivial displacement leaves $\mathcal{C}_{\mathrm{bin}}$ invariant, $\Lambda^{\perp}=\{0\}$, and \Cref{prop: classification} again admits only the standard kernel,
\begin{equation}
\label{eq: bin kernel}
\begin{split}
\hat{\Delta}^{\mathcal{C}_{\mathrm{bin}}}(\beta)&=\hat{\Delta}_{\mathrm{std}}(\beta)=2\hat{\Pi}(\beta),\\
\hat{\Delta}^{\mathcal{C}_{\mathrm{bin}}}(0)&=2(-1)^{\hat{n}}
\end{split}
\end{equation}
meaning the phase-space distribution of a density operator $\hat{\rho}$ is
\begin{equation}
    \label{eq: wigner state binomial}
{W^{\mathcal{C}_{\mathrm{bin}}}_{\hat\rho}(\beta)=\mathrm{Tr}\!\left[\hat{\rho}\,\hat{\Delta}^{\mathcal{C}_{\mathrm{bin}}}(\beta)\right]
    = \pi\,W^{\mathrm{std}}_{\hat{\rho}}(\beta)}
\end{equation}

The binomial and cat kernels therefore are both the ordinary Wigner kernel. \Cref{cor: classification} forces this, since neither code space is invariant under a lattice of displacements; and by \Cref{lem: nodisplacement} no finite-dimensional subspace of $L^2(\mathbb{R})$ is. The normalisation needs no separate argument, $\mathrm{Tr}[\hat{\Delta}^{\mathcal{C}_{\mathrm{bin}}}(0)]=1$ as in \Cref{eq: cat standardisation}, and just like there the reconstruction map is the identity, so $\hat{A}_W=\hat{\rho}$ and $Q_{\mathrm{bin}}$ is a functor. The two codes are distinguished by their codewords: the cat codewords are superpositions of coherent states, with photon-number distribution sharply peaked at $|\alpha|^2$, while each binomial codeword is a finite Fock superposition supported on the levels $p(S+1)$ with $p$ of fixed parity in $[0,N+1]$, that is on at most $\lfloor (N+1)/2\rfloor+1$ widely spaced levels (two of them at the working point $(N,S)=(2,3)$ used later).

The conditions of the structure theorem hold for the standard kernel just in the cat case, and because the two kernels are the same operator, the forward map $\chi_{\mathrm{bin}}$, the inverse map $\phi_{\mathrm{bin}}$ and the represented channel $Q_{\mathrm{bin}}(T)=\chi_{\mathrm{bin}}\circ T\circ\phi_{\mathrm{bin}}$ are those of \Cref{eq: cat maps,eq: cat structure map}, with $\mathcal{C}_{\mathrm{cat}}$ replaced by $\mathcal{C}_{\mathrm{bin}}$ throughout.

\subsection{Negativity is not logical magic}\label{subsec:NegativityandMagic}

\Cref{subsec:latticemagic} showed that the hull argument becomes a magic witness on for codes with a code-adapted representation, through the identification of the code-adapted distribution with the Gross Wigner function of the logical state. For codes which just allow the ordinary Wigner representation, that identification is unavailable, and the assumption the hull argument rests on fails.

For the rotation-symmetric codes of \Cref{subsec:cat,subsec:bin}, \Cref{eq: kernel positivity hypothesis} does not hold. For these, \Cref{prop: classification} gives $\hat{\Delta}^{\mathcal{C}}=2\hat{\Pi}$, and the code-adapted coherent states are the displaced codewords $\ket{\Omega}=\hat{D}(\Omega)\ket{\psi_0}$, so
\begin{equation}
\label{eq: kernel positivity reduces}
{\begin{aligned}
\mathrm{Tr}\!\left[\hat{\Delta}^{\mathcal{C}}(\Omega)\ketbra{\Omega'}{\Omega'}\right]
&=2\bra{\psi_0}\hat{\Pi}(\Omega-\Omega')\ket{\psi_0}\\
&=\pi\,W^{\mathrm{std}}_{\psi_0}(\Omega-\Omega')
\end{aligned}}
\end{equation}
\Cref{eq: kernel positivity hypothesis} therefore holds iff the reference codeword has an everywhere non-negative ordinary Wigner function, which by Hudson's theorem~\cite{HUDSON1974WignerGauss,Soto1983WignerGauss} means iff $\ket{\psi_0}$ is Gaussian. No codeword of a rotation-symmetric code is Gaussian, so this condition fails.

However, this condition does not fail everywhere. Setting $\Omega=\Omega'$ in \Cref{eq: kernel positivity reduces} gives $2\bra{\psi_0}(-1)^{\hat{n}}\ket{\psi_0}$, and every state of the four-component cat code space has support on $\hat{n}\equiv0,2\pmod4$ (that is, on even photon numbers only), so this equals $+2$ for every code state; the same holds for a binomial code with $S+1$ even. \Cref{eq: kernel positivity hypothesis} fails at $\Omega\neq\Omega'$, in the interference fringes of the codeword's Wigner function, not at the origin.

We can see this failure numerically. Take matched mean photon number $\bar{n}=6$, so $|\alpha|^2=6$ for the four-component cat code and $(N,S)=(2,3)$ for the binomial code, and evaluate the negative volume \eqref{eq: negative volume general} on the logical basis states of both codes, and for the cat code also on the logical superpositions and on a logical magic state $\ket{T_L}\propto\ket{0_L}+e^{i\pi/4}\ket{1_L}$. See \Cref{tab: negative volumes} for the results, together with the post-loss and post-gain values used in \Cref{subsec:Comparative}; Appendix~\ref{app:numerics} shows how they were computed.

\begin{table}[t]
\caption{\protect{Negative volume $\NC^{\mathcal{C}}$ (\Cref{eq: negative volume general}) for cat and binomial code states at matched mean photon number $\bar{n}=6$ (four-component cat with $|\alpha|^2=6$; binomial with $(N,S)=(2,3)$), before and after a single conditioned loss or gain event, with the post-event mean photon number. Every cat state listed except $\ket{T_L}$ is a logical stabiliser state, and all of them have $\NC^{\mathcal{C}}>0$; the magic state $\ket{T_L}$ has lower negativity than three of them. The input mean photon numbers are $6$ for the binomial codewords and $5.98$ and $6.02$ for the two cat codewords, the difference being the exponentially small overlap corrections of \Cref{subsec:cat}. The loss and gain columns are discussed in \Cref{subsec:Comparative}; Appendix~\ref{app:numerics} gives the implementation used.}}
\label{tab: negative volumes}
\begin{ruledtabular}
\begin{tabular}{lccccc}
state & $\NC^{\mathcal{C}}$ & $\NC_{\rm loss}$ & $\NC_{\rm gain}$ & $\av{\hat n}_{\rm loss}$ & $\av{\hat n}_{\rm gain}$\\
\colrule
cat $\ket{0_L}$ & 1.38 & 1.41 & 1.52 & 5.96 & 7.82\\
cat $\ket{1_L}$ & 1.41 & 1.42 & 1.53 & 6.04 & 7.89\\
cat $\ket{\pm_L}$ & 0.62 & 0.62 & 0.64 & 6.00 & 7.86\\
cat $\ket{+i_L}$ & 1.42 & 1.39 & 1.52 & 6.00 & 7.86\\
cat $\ket{T_L}$ (magic) & 1.22 & 1.21 & 1.30 & 6.00 & 7.86\\
binomial $\ket{W_\uparrow}$ & 1.62 & 1.72 & 1.97 & 7.00 & 8.71\\
binomial $\ket{W_\downarrow}$ & 1.64 & 1.75 & 2.04 & 7.00 & 8.71\\
\end{tabular}
\end{ruledtabular}
\end{table}

The negative volume is non-zero on states which carry no logical magic at all (all but one entry in \Cref{tab: negative volumes} is a logical stabiliser state), so this is not a witness. It is also not monotone in logical resourcefulness: the magic state $\ket{T_L}$ has $\mathcal{N}^{\mathcal{C}}=1.22$, below the stabiliser states $\ket{0_L}$, $\ket{1_L}$ and $\ket{+i_L}$, and well above $\ket{\pm_L}$. For these codes, $\mathcal{N}^{\mathcal{C}}$ instead tracks the ordinary continuous-variable non-classicality of the physical state, which has nothing to do with the logical qubit.

This follows from \Cref{prop: classification} together with Hudson's theorem, and covers every bosonic code with normalisable codewords:

\begin{theorem}
\label{thm: nogo}
Let $\mathcal{C}\subset L^2(\mathbb{R})$ be a finite-dimensional code space of logical dimension at least two, spanned by normalisable codewords; let $G$ contain the phase-space inversion; and let the representation satisfy \rev{hypotheses \mbox{\ref{PropIII.3.H1}}--\mbox{\ref{PropIII.3.H3}} of \mbox{\Cref{prop: classification}} with $\chi_{\hat{K}}$ a function. That is, let it satisfy} the Stratonovich-Weyl axioms in \Cref{subsec:Wigner}'s code-faithful form: \labelcref{SWAxiom2,SWAxiom3,SWAxiom4} unchanged, \labelcref{SWAxiom1,SWAxiom5} on the $\mathcal{E}_{\mathcal{C}}$-invariant subalgebra, together with a reconstruction map $\mathcal{E}_{\mathcal{C}}=\phi_{\mathcal{C}}\circ\chi_{\mathcal{C}}$ that is idempotent and acts as the identity on $\mathcal{B}(\mathcal{C})$, with $\chi_{\hat{K}}(\xi)=\mathrm{Tr}[\hat{K}\hat{D}(\xi)]$ a continuous function of $\xi$ (for a bounded, non-trace-class $\hat{K}$ the trace is read in the Abel-regularised sense, as at \Cref{eq: cat standardisation}; it is a comb rather than a function exactly in the ideal GKP case). \rev{We assume neither complete positivity of $\mathcal{E}_{\mathcal{C}}$ nor a twirl form for it, so neither setting of \mbox{\Cref{prop: classification}} is used: trace preservation and displacement covariance of $\mathcal{E}_{\mathcal{C}}$ follow from \mbox{\labelcref{SWAxiom2,SWAxiom3,SWAxiom4}}, and complete positivity is part of the conclusion. Then $\mathcal{E}_{\mathcal{C}}={\id}$, and} the only such kernel available to $\mathcal{C}$ is the standard one, so that
\begin{equation}
\label{eq: nogo identity}
{W^{\mathcal{C}}_{\hat{\rho}}(\Omega)=\pi\,W^{\mathrm{std}}_{\hat{\rho}}(\Omega)
\qquad\text{for every }\hat{\rho}}
\end{equation}
and there exists a logical stabiliser state with $\mathcal{N}^{\mathcal{C}}>0$. The negative volume is therefore not a witness of logical magic for any such code.
\end{theorem}

We prove this in Appendix \ref{app: proofs}: the first part is \rev{the continuity route of} \Cref{prop: classification}\rev{\mbox{\ref{PropIII.3.i}}, followed by \mbox{\ref{PropIII.3.ii}}}; for the second, a code space of logical dimension at least two contains two orthogonal logical stabiliser states, which cannot both be Gaussian~\cite{Giani2026nonGaussian}, so by Hudson's theorem~\cite{HUDSON1974WignerGauss,Soto1983WignerGauss} one of them has $\mathcal{N}^{\mathcal{C}}>0$ while carrying no logical magic. Appendix~\ref{app: proofs} also records three remarks on the hypotheses: what dropping continuity leaves \rev{once $\mathcal{E}_{\mathcal{C}}={\id}$ is obtained another way} (a code-independent sign family, the continuum analogue of \Cref{prop:nonpauli}'s); why idempotence of the reconstruction map is needed on top of \labelcref{SWAxiom1,SWAxiom2,SWAxiom3,SWAxiom4}, which alone admit the Husimi kernel; and why the code-faithful rather than strong reading of \labelcref{SWAxiom1,SWAxiom5} is the substantive choice, the strong form excluding the ideal GKP code along with everything else.

Beyond the parity requirement, \Cref{thm: nogo} does not refer to the code's symmetry group, only to the normalisability of its codewords, so it applies to the cat and binomial codes, to any rotation-symmetric code of even order, and to finite-energy GKP codewords (which we return to in \Cref{sec:FiniteEnergy}). The parity hypothesis is available for every code considered here, in the two ways set out after \Cref{cor: classification}. The theorem does not apply to a code following neither route (by the remark after \Cref{prop: classification} the kernel is then only constrained to $\chi_{\hat{K}}=e^{if}$ with $f$ real, odd and continuous, and the Hudson step needs a displaced parity), and we leave that case open. The ideal GKP code is the exception of a different kind: its codewords are Dirac combs rather than elements of $L^2(\mathbb{R})$ (\Cref{subsec:GKP}), so \Cref{lem: nodisplacement} does not apply to it and its code space can be (and is) invariant under a lattice of displacements. Its kernel is then the lattice sum of \Cref{eq: ZGW central kernel}, standardised in the integrated form of \Cref{eq: gkp standardisation direct} rather than by a trace, which for that kernel diverges. That lattice suppresses the continuous-variable structure and leaves only the logical content: the content of \Cref{eq: embedding of DV Wigner function}, and of Ref.~\cite{davis2024identifyingquantumresourcesencoded}'s result that stabiliser codewords are positively represented. Where no lattice is available, \Cref{prop: classification} leaves only $\Lambda^{\perp}=\{0\}$, \Cref{eq: nogo identity} holds, and the code-adapted distribution is the ordinary Wigner function. Its negativity is therefore the ordinary continuous-variable non-classicality of the encoded state, which is large for a cat or binomial codeword and has nothing to do with the logical qubit.

\rev{Together with \mbox{\Cref{cor: classification}},} \Cref{thm: nogo} also classifies the set of codes for which the resource reading can hold. The code space must be invariant under a group of displacements, which by \Cref{lem: nodisplacement} rules out normalisable codewords. Within the distributional codes that remain\rev{, for which we assume setting \mbox{\ref{PropIII.3.B}} of \mbox{\Cref{prop: classification}},} the invariance group cannot contain a one-parameter subgroup, since the commutant of a continuum of displacements is a maximal abelian algebra and so cannot contain the logical operators of a qubit; it must therefore be a lattice, and it must have rank two if the quotient $X^{\mathcal{C}}$ is to be compact, so that the representation carries no more independent degrees of freedom than the logical system it describes. Being the fixed-point space of a rank-two lattice of displacements is the defining property of the GKP family.

Ideal GKP-type codes are therefore, among codes whose phase space is compact, the only bosonic codes whose code-adapted phase-space negativity can measure logical magic. (This is a statement about kernels built from the \textit{physical} Weyl-Heisenberg group\rev{, within the class \mbox{\Cref{cor: classification}} covers}; for a representation built instead from the code's logical operators, see \Cref{sec:seclog}.)

\subsection{Non-Pauli qudit symmetries}\label{subsec:nonpauli}

The finite-dimensional codes not covered by \Cref{prop:normal} are those whose defining symmetries are not displacement subgroups: code spaces fixed by Clifford or higher symmetries, the discrete analogue of the rotation-symmetric bosonic families.

\begin{proposition}\label{prop:nonpauli}
Let $d$ be odd and let $\mathcal C\subset(\mathbb C^d)^{\otimes n}$ be a code space whose projective isotropy group contains a non-Pauli unitary $R$ (a Clifford acting on displacement labels by a symplectic $S\ne\id$), and whose displacement isotropy is trivial. \rev{For parts (ii) and (iii), let the kernel satisfy the finite-group form of hypotheses \mbox{\ref{PropIII.3.H1}} and \mbox{\ref{PropIII.3.H2}} of \mbox{\Cref{prop: classification}}, let its reconstruction map be completely positive, as in setting \mbox{\ref{PropIII.3.A}} (an assumption, since a finite group offers no continuity to replace it), and let the phase-space inversion be available, as in \mbox{\Cref{prop: classification}}\mbox{\ref{PropIII.3.ii}}.} Then:
\begin{enumerate}[label=(\roman*),leftmargin=*]
\item the group-averaged kernel over $\langle R\rangle$ fails displacement covariance, by the same two-multiset argument as \Cref{prop: metaplectic}, valid here because the $d^{2n}$ phase-point operators are linearly independent;

\item \Cref{prop: classification}'s axioms-direct route forces $\Delta(u)=D(u)\hat KD^\dagger(u)$ with reconstruction multiplier $\mathfrak{m}=|\chi_{\hat K}|^2$; pointwise idempotence and positive-definiteness \rev{(that is, complete positivity)} make $\mathfrak{m}$ the indicator function of a subgroup $\Lambda\subseteq\Zd^{2n}$, whose symplectic dual $\Lambda^\perp$ is the twirl group so must leave $\mathcal C$ invariant; triviality of the displacement isotropy gives $\Lambda^\perp=\{0\}$, therefore $\Lambda=\Zd^{2n}$, $\mathfrak{m}\equiv1$ and $|\chi_{\hat K}|\equiv1$;

\item Hermiticity and inversion covariance then give $\chi_{\hat K}:\Zd^{2n}\to\{\pm1\}$, even, with $\chi_{\hat K}(0)=1$. No continuity is available on a finite group, so this sign freedom is not removable by the axioms: every such sign pattern yields a Hermitian, unit-trace, covariant, self-dual kernel. The freedom is code-independent, existing already for the trivial code, so no code-adapted kernel exists; and it is removed by demanding non-negativity on pure stabiliser states, which selects the Gross kernel~\cite{Gross2006DVWignerF}. Clifford covariance alone does not remove this freedom: the symplectic group acts transitively on $\Zd^{2n}\setminus\{0\}$, so it leaves the two constant sign patterns, from which stabiliser positivity picks out $\chi_{\hat{K}}\equiv1$~\cite{Zhu2016}.
\end{enumerate}
\end{proposition}

The sign family is the finite-group shadow of the residual freedom left by \Cref{prop: classification} in the CV analysis: the $(-1)^{mn}$ pattern of the GKP lattice kernel, and the ``$\chi_{\hat K}=e^{if}$ with $f$ real and odd'' family that remains when no inversion is available. We verified all counts numerically for a single qutrit with the Fourier gate as code symmetry (Appendix~\ref{app:numerics}). At even $d$ the sign family and the absence of a code-adapted kernel still hold, but the selection of the Gross kernel by stabiliser positivity does not, for the same reason we give in \Cref{subsec:qubit}.

\subsection{\texorpdfstring{\protect{One-photon codes}}{One-photon codes}}\label{subsec:dualrail}

For one-photon codes this is a stronger restriction than a failure of the witness property.

\begin{theorem}\label{thm:dualrail}
For any one-photon-in-$m$-rails code ($\mathcal C\subset{\rm span}\{\ket{0\cdots010\cdots0}\}$ over $m\ge2$ modes, dual-rail being $m=2$), and under \rev{hypotheses \mbox{\ref{PropIII.3.H1}}--\mbox{\ref{PropIII.3.H3}} of \mbox{\Cref{prop: classification}} with $\chi_{\hat{K}}$ a function}, the only Stratonovich-Weyl kernel is the ordinary $m$-mode Wigner kernel, and every pure logical state has the same negative volume,
\begin{equation}\label{eq:dualrail}
{\NC^{\mathcal C}=4e^{-1/2}-2\approx0.42612.}
\end{equation}
\end{theorem}

We prove this in Appendix~\ref{app: proofs}: every pure logical state is a passive rotation of $\ket{1}\otimes\ket{0}^{\otimes(m-1)}$, and $\int|W|$ is symplectically invariant, so all share the negative volume of the Fock state $\ket{1}$.

Loss in any rail moves the state to the vacuum sector: the syndrome is the total photon number, and loss is detectable (an erasure) but not correctable. This is the multimode version of \Cref{subsec:cat}'s $N=2$ cat statement.

\subsection{Spin codes}\label{subsec:spin}

Codes built on a compact dynamical group fail for a representation-theoretic, rather than spectral, reason.

\begin{proposition}\label{prop:su2}
Let $G=SU(2)$ act irreducibly on a spin-$j$ system, and let $\mathcal C$ be any code space whose isotropy subgroup $H^{\mathcal C}\subset SU(2)$ is a proper, non-central subgroup.\footnote{As for every rotation-symmetric spin code ($H\supseteq U(1)$), and for the single-spin and molecular codes built on binary polyhedral rotation groups, e.g., Ref.~\cite{GrossSpin2021}'s $2O$ and $2I$ and Ref.~\cite{Albert2020}'s finite rotation groups.} Then the $H^{\mathcal C}$-averaged kernel is either not Stratonovich-Weyl covariant or not faithful on the code algebra, so the averaging prescription yields no code-adapted kernel.
\end{proposition}

We prove this in Appendix~\ref{app: proofs}, by a Schur's-lemma argument on the multiplicity-free tensor sectors of the spin-$j$ operator space (linear independence of the kernels at distinct phase-space points, which the CV and DV arguments use, is unavailable on $S^2$); both halves were verified numerically at $j=1$.

Spin codes therefore behave like rotation-symmetric bosonic codes. They are represented by the ordinary spherical kernel, up to its known $\varepsilon_l$ sign family (\Cref{prop:nonpauli}'s compact-group analogue of the sign family), and the kernel's negativity tracks spin-coherent nonclassicality~\cite{DavisSpin2021}, not logical content.

\subsection{Conjecture for dynamical groups outside the Weyl-Heisenberg family}\label{subsec:boundary}

We have shown that code-adapted kernels exist for every code whose stabiliser is a lattice of displacements in a Weyl-Heisenberg group, continuous (ideal GKP and its multimode and concatenated lattices, \Cref{subsec:ZGW,subsec:multimode}) or discrete (every qudit stabiliser code, the kernel being constructed at prime-power $d$; \Cref{subsec:dv}), because two-step nilpotency makes the isotropy subgroups normal. They don't exist for non-displacement symmetries in the finite Weyl-Heisenberg setting (\Cref{prop:nonpauli}), and for abelian dynamical groups (whose invariant subalgebra is abelian and so cannot contain $\mathcal{B}(\mathcal{C})$ once $\dim\mathcal{C}\ge2$, meaning the code-faithful form of \labelcref{SWAxiom1} fails). For every non-central isotropy subgroup of a compact group with multiplicity-free tensor decomposition we only prove that the averaging prescription supplies no code-adapted kernel (\Cref{prop:su2}, whose Schur argument is not specific to $SU(2)$), which stops short of an axioms-direct no-go.

We conjecture that no dynamical group outside the extended Weyl-Heisenberg family (including its metaplectic and finite extensions) admits code-adapted Stratonovich-Weyl kernels for any of its codes. This is unresolved for non-compact, non-nilpotent groups, and still requires a formalisation of ``code-adapted'' as ``outside the standard family up to its code-independent sign and phase freedoms''.

\section{The sector-graded logical representation}\label{sec:seclog}

Everything so far concerns kernels built from the physical Weyl-Heisenberg group, the setting of the Brif-Mann construction and the structure theorem as we applied it. A representation built from the logical operators of a code would instead be fixed by a choice of logical basis rather than by the symmetry of the code space, so \Cref{thm: nogo} would have no implications for it.

Let $\mathcal C$ be a code with syndrome-sector decomposition $\mathcal H\supseteq\bigoplus_t\mathcal H_t$ and recovery isometries $V_t:\mathcal H_L\to\mathcal H_t$, with $V_0$ the encoding, and let $A_L(a,b)$ be a Stratonovich-Weyl kernel on the logical space $\mathcal H_L\cong\mathbb C^{d}$. Define the \emph{sector-graded logical kernel} and its generated representation,
\begin{equation}\label{eq:seclog}
\begin{split}
\Delta_{\rm log}(t;a,b)&=V_tA_L(a,b)V_t^\dagger,\\
W^{\rm log}_{\hat\rho}(t;a,b)&=\tfrac1d\Tr[\Delta_{\rm log}\hat\rho],
\end{split}
\end{equation}
on the phase space $X^{\rm log}=\{t\}\times\Zd^{2}$ with counting measure, the factor $1/d$ in $W^{\rm log}$ being the measure absorbed into the function, so that $\sum_{t,a,b}W^{\rm log}_{\hat\rho}=\Tr[\hat\Pi\hat\rho]$ with $\hat\Pi=\sum_t\hat\Pi_t$. Every statement below is for $\hat\rho$ supported on $\bigoplus_t\mathcal{H}_t$, where $\hat\Pi\hat\rho=\hat\rho$ and the representation is standardised. This is the syndrome-times-logical factorisation \eqref{eq:factor} that the stabilisers-as-displacement-lattice branch produces, now taken as a definition for codes with no code-adapted kernel of their own. At the Hilbert-space level it parallels Ref.~\cite{Shaw2024}'s stabiliser subsystem decompositions for GKP codes, where tracing out the stabiliser subsystem implements ideal decoding; \Cref{eq:seclog} is the phase-space representation layer over such a decomposition. Throughout, $\mathcal{H}_t:=V_t\mathcal{H}_L$, a $d$-dimensional subspace, and $\hat\Pi_t=V_tV_t^\dagger$ the projector onto it; the sectors need not cover all of $\mathcal{H}$, which is why the decomposition is written $\mathcal{H}\supseteq\bigoplus_t\mathcal{H}_t$. For an $N$-component rotation code~\cite{Grimsmo2020} encoding a qudit of dimension $d$, with $N/d$ a positive integer, the sector label is the residue of $\hat n$ modulo $N/d$: the codewords $\ket{j_L}$ occupy Fock levels $n\equiv jN/d\ ({\rm mod}\ N)$, $t$ losses shift the code space coherently into the residue class labelled $t$, and $V_t\ket{j_L}\propto\hat a^t\ket{j_L}$ are exact isometries because distinct logical states occupy distinct residue classes.

We must differentiate two objects here from their counterparts in \Cref{subsec:GradedClass}. Write $\mathcal{E}^{\rm log}=\sum_t\hat\Pi_t\cdot\hat\Pi_t$, with $\hat\Pi_t=V_tV_t^\dagger$, for the sector dephasing, and $\mathcal{G}^{\rm log}=\{T:\mathcal{E}^{\rm log}T\mathcal{E}^{\rm log}=T\mathcal{E}^{\rm log}\}$ for the channels graded by it. These are \emph{not} \Cref{subsec:GradedClass}'s $\mathcal{E}_{\mathcal{C}}$ and $\mathcal{G}_{\mathcal{C}}$, which for a rotation-symmetric code are the identity and the set of all channels respectively. The grading here is supplied by the recovery scheme rather than by the reconstruction map of a Stratonovich-Weyl kernel, meaning a channel creating coherence between the sectors $\mathcal{H}_t$, such as a small displacement, does not lie in $\mathcal{G}^{\rm log}$.

\begin{theorem}\label{thm:seclog}
With the notation above:
\begin{enumerate}[label=(\roman*),leftmargin=*]
    \item $\phi_{\rm log}\circ\chi_{\rm log}=\mathcal{E}^{\rm log}$ on the sector subspace $\bigoplus_t\mathcal H_t$: with the forward map $\chi_{\rm log}(\hat\rho)=W^{\rm log}_{\hat\rho}$ of \Cref{eq:seclog} and the reverse map $\phi_{\rm log}(F)=\sum_{t,a,b}F(t;a,b)\,\Delta_{\rm log}(t;a,b)$, the composition is the sector dephasing. (The factor $1/d$ appears once: here in $\chi_{\rm log}$, and in \Cref{prop:dvsw} in $\phi_{\mathcal C}$.)
    \item Consequently $W^{\rm log}$ is a linearity-preserving, empirically adequate, semi-functorial representation of the subtheory $\mathcal{G}^{\rm log}$, with the same compositional structure as the Zak-Gross representation of the GKP code, the twirl now coming from the sector grading instead of from a displacement lattice. In particular \Cref{eq: compositionality} holds on $\mathcal{G}^{\rm log}$, by \Cref{subsec:GradedClass}'s argument applied to $\mathcal{E}^{\rm log}$.
    \item For odd prime logical dimension $d$, with $A_L$ the Gross kernel, its negative volume is the syndrome-resolved logical magic, $\NC^{\rm log}_{\hat\rho}=\sum_tP(t|\hat\rho)\,\NC^{\rm Gross}(\hat\rho_L(t))$. This vanishes on every syndrome-mixture of encoded stabiliser states, and is strictly positive on every encoded pure logical magic state; on mixed logical states it is a one-sided witness, just as in \Cref{thm:dvmagic}.
\end{enumerate}
\end{theorem}

Part (i) follows from the self-duality of $A_L$ (Appendix~\ref{app: proofs}); parts (ii) and (iii) then follow from \Cref{subsec:GradedClass,subsec:latticemagic,thm:dvmagic}'s arguments, each applied on the logical space with $\mathcal{E}^{\rm log}$ in place of $\mathcal{E}_{\mathcal{C}}$. Only (iii) needs $d$ odd and prime: parts (i) and (ii) only need a self-dual logical frame $A_L$, which also exists for $d=2$ (\Cref{subsec:qubit}), so a rotation code encoding a qubit still has a compositional sector-graded representation, albeit without the magic reading.

\Cref{thm:seclog} is consistent with \Cref{thm: nogo}, since $\Delta_{\rm log}$ is not of the form $\hat D(\Omega)\hat K\hat D^\dagger(\Omega)$: it is not covariant under the physical Weyl-Heisenberg group, and is fixed by a recovery scheme rather than by the code-space symmetry. Decoders differing by a logical unitary per sector, $V_t\mapsto V_tU_t$, give representations related by the logical (anti-)symplectic action when $U_t$ is Clifford, and genuinely different negativities otherwise: the representation describes decoder-relative logical content, which is the operationally meaningful object in a cycle. However, a loss event is not a rigid translation of $X^{\rm log}$ but a sector shift composed with the $\sqrt n$ re-weighting of \Cref{eq: cat loss phase space result}.

We tested the construction on the $N=6$ cat code encoding a logical qutrit ($d=3$, so $N/d=2$): codewords $\ket{j_L}\propto\sum_{k=0}^5\omega_6^{-2jk}\ket{\alpha\,e^{i\pi k/3}}$ supported on $n\equiv2j\ ({\rm mod}\ 6)$, with the code sector $t=0$ and the single-loss sector $t=1$; their residue classes $n\equiv0,2,4$ and $n\equiv1,3,5\ ({\rm mod}\ 6)$ cover all of the number grading. $V_1$ is an isometry because $\hat a\ket{j_L}$ for distinct $j$ occupy distinct residue classes mod $6$ and are therefore orthogonal; $\mathcal{H}_0\oplus\mathcal{H}_1$ is the six-dimensional subspace they span, not the whole Fock space. At $|\alpha|^2=6$ and $12$: the isometries are exact; $\phi_{\rm log}\circ\chi_{\rm log}=\mathcal{E}^{\rm log}$ holds to $10^{-10}$ on random states; standardisation is exact on the sector subspace; and encoded stabiliser states have $\NC^{\rm log}=0$ while the encoded strange state has $\NC^{\rm log}=2/3$ exactly. This is a logical-magic reading for a rotation-symmetric code, which \Cref{thm: nogo} proves no physically covariant kernel can supply. For odd prime logical dimension this also resolves the tension recently observed between non-Gaussianity and magic in cat codes, globally incompatible but aligned sector by sector~\cite{ZhuNemoto2026}: the sector-graded representation is the object for which the local alignment becomes a global identity.

The loss channel quantifies the geometry lost in exchange. Conditioned on a single jump, the state moves entirely to sector $1$, and the conditional logical state has fidelity $0.99862$ to the ideal strange state at $|\alpha|^2=6$, with $\NC^{\rm Gross}=0.66483$ against $2/3$; at $|\alpha|^2=12$ both agree with the ideal to $10^{-6}$. The deviation is the phase-space image of the codeword-dependent jump weights $\av{\hat n}_j=(6.30,6.31,5.44)$, that is, of the $\hat n$ re-weighting in \Cref{eq: cat loss kernel full}'s loss-transformed kernel, and it decays with the codeword overlaps. Recovery returns the logical distribution with the same fidelity: the recovery map is $\mathcal{R}(\hat\rho)=V_0V_1^\dagger\hat\rho V_1V_0^\dagger+\hat\Pi_0\hat\rho\hat\Pi_0$, trace-preserving on $\mathcal{H}_0\oplus\mathcal{H}_1$ and in $\mathcal{G}^{\rm log}$.

This has three consequences. First, at $N/d=2$ the gain words $\hat a^\dagger\ket{j_L}$ occupy the same sectors as the loss words, with a natural isometry related to $V_1$ by a non-trivial logical map, which is the sector-graded rendering of the gain-loss conflation of \Cref{subsec:Comparative}. For $N/d\ge3$ the gain and loss sectors $t=\mp1$ are distinct and both channels are separately representable, mirroring the binomial grading. Second, the construction applies verbatim to binomial codes and, after orthogonalising the approximately orthogonal sectors, to finite-energy GKP codewords, restoring a logical reading exactly where \Cref{thm: nogo} removes the kernel-based one. Third, this approach degenerates for one-photon codes: the loss sector is the vacuum, too small to host an isometric recovery, and the absence of a sector-graded representation is the representation-level statement that loss is an erasure there (\Cref{thm:dualrail}).

\section{Error channels in phase space}\label{sec:atlas}

Both sets of cases (stabiliser is a lattice of displacements in a Weyl-Heisenberg group, and not) supply a formula, \Cref{eq: generalized structure theorem map}, for the action of a channel on the code-adapted distribution, and \Cref{eq: compositionality} lets a full cycle be assembled from it channel by channel.

\subsection{Single-photon loss, and the reflection identity}\label{subsec:loss}

The dominant physical error channel in many bosonic implementations is single-photon loss~\cite{Cai2021}. The unconditional process is the amplitude-damping channel
\begin{equation}
\label{eq: amplitude damping channel}
{\mathcal{E}_\gamma(\hat{\rho})=\sum_{\ell\ge0}\hat{J}_\ell\hat{\rho}\hat{J}_\ell^\dagger,\quad \hat{J}_\ell=\sqrt{\frac{\gamma^\ell}{\ell!}}\,e^{-\kappa t\hat{n}/2}\,\hat{a}^\ell}
\end{equation}
with $\gamma=1-e^{-\kappa t}$. This is trace-preserving and lies in $\mathcal{G}_{\mathcal{C}}$ in the sense of \Cref{subsec:GradedClass}, trivially, since $\mathcal{E}_{\mathrm{cat}}={\id}$ (\Cref{subsec:classification}). Below we analyse the $\ell=1$ case of this channel, i.e. the trajectory conditioned on a single detected quantum jump:
\begin{equation}
\label{eq: conditional jump map}
{\begin{aligned}
\hat{J}_1\hat{\rho}\hat{J}_1^\dagger&=\gamma\,e^{-\kappa t\hat{n}/2}\,\hat{a}\hat{\rho}\hat{a}^\dagger\,e^{-\kappa t\hat{n}/2},\\
p_1&=\mathrm{Tr}\!\left[\hat{J}_1\hat{\rho}\hat{J}_1^\dagger\right]\in[0,1]
\end{aligned}}
\end{equation}
with $p_1$ the probability of exactly one jump and $\hat{J}_1\hat{\rho}\hat{J}_1^\dagger/p_1$ the normalised post-jump state. It is convenient to strip off the damping factor and work with the bare jump map
\begin{equation}
\label{eq: bare jump map}
\begin{split}
T_{\mathrm{loss}}(\hat{\rho}) &= \hat{a}\,\hat{\rho}\,\hat{a}^{\dagger},
\\
\mathrm{Tr}\!\left[T_{\mathrm{loss}}(\hat{\rho})\right]&=\mathrm{Tr}\!\left[\hat{a}^\dagger\hat{a}\hat{\rho}\right]=\av{\hat{n}}
\end{split}
\end{equation}
which differs from the $\ell=1$ case only by conjugation with $e^{-\kappa t\hat{n}/2}$, and coincides with it up to the overall constant $\gamma$ in the short-time limit $\kappa t\,\av{\hat{n}}\ll1$ (the damping factor deviates from unity by $\sim\kappa t\,n/2$ on the occupied levels, so for $\bar{n}=6$ the condition is $\kappa t\ll1/6$, not $\kappa t\ll1$). Note $T_{\mathrm{loss}}$ is completely positive but neither trace-preserving nor trace-non-increasing: $\hat{a}^\dagger\hat{a}$ is unbounded, and $\mathrm{Tr}[\hat{a}\hat{\rho}\hat{a}^\dagger]=\av{\hat{n}}$, which for $|\alpha|^2=6$ (as used in \Cref{subsec:Comparative}) is $6$ up to exponentially small corrections in $|\alpha|^2$ (numerically $5.98$ and $6.02$ for the two cat codewords). $T_{\mathrm{loss}}(\hat{\rho})$ is therefore an unnormalised positive operator rather than a subnormalised state, and its phase-space image integrates against $d\mu$ to $\av{\hat{n}}$, rather than a probability. Dividing by $\av{\hat{n}}$ gives the normalised distribution of the post-jump state $\hat{a}\hat{\rho}\hat{a}^\dagger/\av{\hat{n}}$. We work with $T_{\mathrm{loss}}$ below and renormalise when the result is interpreted, noting where the damping factor enters: since $\mathcal{E}_{\mathcal{C}}={\id}$ for this code every channel is syndrome-graded, so it composes under \Cref{eq: compositionality}, and to leading order in $\kappa t\,\av{\hat{n}}$ it rescales the distribution rather than reshaping it (it contracts phase space radially, $\ket{\alpha}\mapsto\ket{\alpha e^{-\kappa t/2}}$ up to normalisation).

For $N=4$ (\Cref{eq: four component cat codewords}), a single photon loss carries code sectors $\hat{n}\equiv0,2$ into error sectors $\hat{n}\equiv3,1\pmod4$, which are orthogonal to the code space. The loss is therefore not itself an operation on the logical qubit, but a transition out of the code space into a distinguishable (and so correctable) error space.

The displaced parity is the reflection of phase space about $\zeta$, so reflects the ladder operators:
\begin{equation}\label{eq:reflection}
\begin{split}
\hat\Pi(\zeta)\,\hat a\,\hat\Pi(\zeta)&=2\zeta-\hat a,\\
\hat\Pi(\zeta)\,\hat a^\dagger\,\hat\Pi(\zeta)&=2\zeta^*-\hat a^\dagger.
\end{split}
\end{equation}
Both follow from $\hat D^\dagger(\zeta)\hat a\hat D(\zeta)=\hat a+\zeta$ together with $(-1)^{\hat n}\hat a(-1)^{\hat n}=-\hat a$ (we derive this in Appendix~\ref{app:gain}). Since $\hat\Pi(\zeta)^2=\id$, every jump kernel now follows by substitution, and we use \Cref{eq:reflection} this way throughout this section.

We substitute $T=T_{\mathrm{loss}}$ into the general structure theorem
formula~\eqref{eq: generalized structure theorem map}. Using the reconstruction operator $\phi_{\mathrm{cat}}$, we first reconstruct the physical operator from the input distribution $W^{\mathcal{C}_{\mathrm{cat}}}(\beta)$:
\begin{equation}
\begin{split}
    \hat{A}_{W} :=&
\phi_{\mathrm{cat}}\!\left(W^{\mathcal{C}_{\mathrm{cat}}}\right)\\
    =& {\frac{1}{\pi}}\int_{\mathbb{C}} d^2\beta\;
      W^{\mathcal{C}_{\mathrm{cat}}}(\beta)\,\hat{\Delta}^{\mathcal{C}_{\mathrm{cat}}}(\beta).
\end{split}
\end{equation}
Because $\mathcal{E}_{\mathrm{cat}}={\id}$ here, this reconstruction is exact rather than a twirl: $\hat{A}_W=\hat{\rho}$. However, we keep the notation parallel to the GKP case of \Cref{subsec:GKPcycle}, where the reconstruction is a twirl.
The evolved distribution at coordinate $\beta'$ is then
\begin{equation}
\label{eq: cat loss structure theorem}
    Q_{\mathrm{cat}}(T_{\mathrm{loss}})\!\left[W^{\mathcal{C}_{\mathrm{cat}}}\right]\!(\beta')
    = \mathrm{Tr}\!\left[
        \hat{\Delta}^{\mathcal{C}_{\mathrm{cat}}}(\beta')\,
        \hat{a}\,\hat{A}_{W}\,\hat{a}^{\dagger}
      \right].
\end{equation}
Cyclicity of the trace moves the ladder operators onto the kernel, giving $Q_{\mathrm{cat}}(T_{\mathrm{loss}})[W^{\mathcal{C}_{\mathrm{cat}}}](\beta')=\mathrm{Tr}[\hat{\Delta}^{\mathcal{C}_{\mathrm{cat}}}(\beta')_{\mathrm{loss}}\hat{A}_W]$ in terms of the loss-transformed kernel
\begin{equation}
\hat{\Delta}^{\mathcal{C}_{\mathrm{cat}}}(\beta')_{\mathrm{loss}}
    := \hat{a}^{\dagger}\,\hat{\Delta}^{\mathcal{C}_{\mathrm{cat}}}(\beta')\,\hat{a}.
\end{equation}

We can evaluate this kernel using the reflection identity \eqref{eq:reflection}. Writing $\zeta$ for a generic phase-space point, to distinguish it from \Cref{eq: amplitude damping channel}'s damping parameter $\gamma$, and using $\hat\Pi^2=\id$,
\begin{equation}
\label{eq: cat loss kernel exact}
\begin{split}
\hat{a}^\dagger\hat{\Pi}(\zeta)\hat{a}
=&\hat{\Pi}(\zeta)\left(\hat{\Pi}(\zeta)\hat{a}^\dagger\hat{\Pi}(\zeta)\right)\hat{a}\\
=&-\hat{\Pi}(\zeta)\left(\hat{n}-2\zeta^*\hat{a}\right)
\end{split}
\end{equation}
which is exact for every $\zeta$. Since \Cref{eq:generalcatkernel}'s kernel is the standard one, the calculation involves a single displaced parity where a lattice kernel would give a sum; and since that kernel carries the factor $2$,
\begin{equation}
\label{eq: cat loss kernel full}
\hat{\Delta}^{\mathcal{C}_{\mathrm{cat}}}(\beta')_{\mathrm{loss}}
=-2\,\hat{\Pi}(\beta')\left[\hat{n}-2\beta'^*\hat{a}\right]
\end{equation}
so
\begin{equation}
\label{eq: cat loss phase space result}
\begin{split}
Q&_{\mathrm{cat}}(T_{\mathrm{loss}})\!\left[W^{\mathcal{C}_{\mathrm{cat}}}\right]\!(\beta')
    =\\
    &-2\,\mathrm{Tr}\!\left[\hat{\Pi}(\beta')\left(\hat{n}-2\beta'^*\hat{a}\right)\hat{\rho}\right]
\end{split}
\end{equation}
Conjugating $(-1)^{\hat{n}}$ by $\hat{a}^\dagger\cdot\hat{a}$ produces $-\hat{n}(-1)^{\hat{n}}$, not $-(-1)^{\hat{n}}$, so the loss-transformed kernel is not the negative of the original.

\Cref{eq: cat loss phase space result} has two parts. The overall sign is the phase-space signature of the parity change: the kernel registers the change of parity accompanying a single loss as a change of sign. The factor $\hat{n}$ multiplying it is the phase-space signature of the photon-number dependence of the loss rate, and the term linear in $\hat{a}$ is the accompanying coherent-amplitude term. Note two potential issues. First, the two terms are not separately real ($\mathrm{Tr}[\hat{\Pi}(\beta')\hat{n}\hat{\rho}]$ is complex in general, and only the combination in \Cref{eq: cat loss phase space result} is a real-valued phase-space function) so neither on its own is a physical distribution. Second, $\hat{n}$ cannot be replaced by its mean pointwise in $\beta'$. The mean photon number of a codeword is $|\alpha|^2$ up to corrections exponentially small in $|\alpha|^2$, and integrating \Cref{eq: cat loss phase space result} against $d\mu$ gives $\av{\hat{n}}$ exactly, since standardisation gives $\int d\mu\, Q(T)[W]=\mathrm{Tr}[T(\hat{\rho})]$ for any $T$. However, the photon-number distribution of a coherent-state superposition has width $|\alpha|$, so substituting $\hat{n}\to|\alpha|^2$ inside the trace incurs a relative error of order $1/|\alpha|$, not an exponentially small one; for $|\alpha|^2=6$ (as in \Cref{subsec:Comparative}) that is a $40\%$ effect, and larger still where $W^{\mathcal{C}_{\mathrm{cat}}}$ is small. We therefore keep \Cref{eq: cat loss phase space result} exact and do not use a mean-field form of it anywhere.

\subsection{The cat-code cycle}\label{subsec:catloss}

Let us assemble the error-correction cycle. The syndrome is the photon-number parity of \Cref{eq: cat syndrome}, measured by a quantum non-demolition coupling to an ancilla; in the phase-space picture this distinguishes the even sector, on which the code space is supported, from the odd sector into which a single loss has moved it, and since $Q_{\mathrm{cat}}$ is empirically adequate this correctly reproduces the probability of each outcome. Recovery is then an ancilla-assisted unitary returning the error sectors $\hat{n}\equiv3,1$ to the code sectors $\hat{n}\equiv0,2$ (a coherent map on the two-dimensional error space, which preserves the logical superposition, rather than a further measurement). Both maps are syndrome-graded in the sense of \Cref{subsec:GradedClass}, trivially, since $\mathcal{E}_{\mathrm{cat}}={\id}$, so by \Cref{eq: compositionality} the representation of the cycle is the composition of the representations.

Since $T_{QEC}=T_{\mathrm{rec}}\circ T_{\mathrm{loss}}$ is built from the bare jump map, $\int d\mu\,Q(T_{QEC})[W]=\mathrm{Tr}[T_{QEC}(\hat{\rho})]=\av{\hat{n}}$ (\Cref{eq: bare jump map}), so the cycle returns
\begin{equation}
\label{eq: cat QEC identity}
{Q_{\mathrm{cat}}(T_{QEC})\!\left[W^{\mathcal{C}_{\mathrm{cat}}}\right]\!(\beta')=\av{\hat{n}}\,W^{\mathcal{C}_{\mathrm{cat}}}(\beta')}
\end{equation}
for $\hat{\rho}$ supported on the code space, so that after dividing by the jump weight $\av{\hat{n}}$ (or equivalently, after conditioning on the detected jump, as below \Cref{eq: bare jump map}) the normalised logical distribution is returned to itself. The factor is $\av{\hat{n}}$ because the recovery is an isometry from the error space back to the code space and the two codewords carry the same weight $\bra{j_L}\hat{n}\ket{j_L}$ up to corrections exponentially small in $|\alpha|^2$; the residual mismatch between those two weights is what limits the identity to that accuracy. Up to that renormalisation this is the analogue of the GKP result; as there, the photon remains in the environment.

\subsection{The binomial-code cycle}\label{subsec:binloss}

Recall that binomial codewords consist of Fock states spaced by $S+1$~\cite{Michael2016BinCodes}:
\begin{equation}
\begin{split}
    \ket{W_\uparrow} &\propto \sum_{p\ \mathrm{even}} \sqrt{\binom{N+1}{p}}\ket{p(S+1)}, \\ \ket{W_\downarrow} &\propto \sum_{p\ \mathrm{odd}} \sqrt{\binom{N+1}{p}}\ket{p(S+1)}.
\end{split}
\end{equation}
Applying $\hat{a}$ to a Fock state lowers the photon number by one:
$\hat{a}\ket{n} = \sqrt{n}\ket{n-1}$.
Since the codewords occupy Fock levels $\{p(S+1)\}$, a single photon loss maps each level to $p(S+1)-1$, which lies between two consecutive occupied levels and is therefore orthogonal to the code space. Defining the single-loss error words
\begin{equation}
\begin{split}
    \ket{W_\uparrow^{(1)}} :=& \frac{\hat{a}\ket{W_\uparrow}}{\|\hat{a}\ket{W_\uparrow}\|},
    \\
    \ket{W_\downarrow^{(1)}} :=& \frac{\hat{a}\ket{W_\downarrow}}{\|\hat{a}\ket{W_\downarrow}\|},
\end{split}
\end{equation}
which occupy levels $\{p(S+1)-1\}$ and are orthogonal to both codewords and to each other. The spacing {$S+1$, with $S=L+G\ge1$,} guarantees that no two error words overlap, satisfying the Knill-Laflamme conditions for $L \geq 1$.

Substituting $T = T_{\mathrm{loss}}$ into the general structure theorem formula \Cref{eq: generalized structure theorem map} and writing $\hat{A}_W := \phi_{\mathrm{bin}}(W^{\mathcal{C}_{\mathrm{bin}}})$ for the reconstructed operator, the evolved distribution at $\beta'$ is
\begin{equation}
\label{eq: bin loss structure theorem}
\begin{split}
    Q_{\mathrm{bin}}(T_{\mathrm{loss}})\!\left[W^{\mathcal{C}_{\mathrm{bin}}}\right]\!(\beta')
    = \mathrm{Tr}\!\left[
        \hat{\Delta}^{\mathcal{C}_{\mathrm{bin}}}(\beta')\,
        \hat{a}\,\hat{A}_W\,\hat{a}^{\dagger}
      \right]\\
    = \mathrm{Tr}\!\left[
        \hat{a}^{\dagger}\,\hat{\Delta}^{\mathcal{C}_{\mathrm{bin}}}(\beta')\,\hat{a}\;
        \hat{A}_W
      \right],
\end{split}
\end{equation}
where in the second equality we used cyclicity of the trace. We define the loss-transformed kernel
\begin{equation}
    \hat{\Delta}^{\mathcal{C}_{\mathrm{bin}}}(\beta')_{\mathrm{loss}}
    := \hat{a}^{\dagger}\,\hat{\Delta}^{\mathcal{C}_{\mathrm{bin}}}(\beta')\,\hat{a}.
\end{equation}

Because \Cref{eq: bin kernel}'s binomial kernel is the standard one, we can use \Cref{eq: cat loss kernel exact} to give
\begin{align}
\label{eq: bin loss kernel explicit}
{\hat{\Delta}^{\mathcal{C}_{\mathrm{bin}}}(\beta')_{\mathrm{loss}}
    =-2\,\hat{\Pi}(\beta')\left[\hat{n}-2\beta'^{*}\hat{a}\right]}
\end{align}
meaning
\begin{align}
\label{eq: bin loss wigner}
\begin{split}
Q_{\mathrm{bin}}&(T_{\mathrm{loss}})\!\left[W^{\mathcal{C}_{\mathrm{bin}}}\right]\!(\beta')
    =\\
    &-2\,\mathrm{Tr}\!\left[\hat{\Pi}(\beta')\left(\hat{n}-2\beta'^{*}\hat{a}\right)\hat{\rho}\right]
\end{split}
\end{align}
This is identical to the cat-code result \eqref{eq: cat loss phase space result}. The two codes are distinguished by the reconstructed operator $\hat{A}_W$, which is where the binomial structure enters. Since the codewords occupy the Fock levels $\{p(S+1)\}$ and $\hat{a}\ket{p(S+1)}=\sqrt{p(S+1)}\ket{p(S+1)-1}$, the operator $\hat{a}\hat{A}_W\hat{a}^\dagger$ has support on the error levels $\{p(S+1)-1\}$, and its Fock-basis matrix elements carry an extra factor
\begin{equation}
\begin{split}
\label{eq: bin reweighting}
\langle p'&(S+1)-1|\hat{a}\hat{A}_W\hat{a}^\dagger\ket{p(S+1)-1}
=\\&\sqrt{p\,p'}\,(S+1)\,\bra{p'(S+1)}\hat{A}_W\ket{p(S+1)}
\end{split}
\end{equation}
relative to those of $\hat{A}_W$ itself.

\Cref{eq: bin reweighting} shows that the loss-transformed operator has a similar form as the original (a weighted sum of off-diagonal Fock-basis matrix elements) but with two differences, which the distribution inherits.

First, the Fock levels are shifted: the original operator carries $\bra{p'(S+1)}\cdots\ket{p(S+1)}$, while the loss-transformed operator carries $\bra{p'(S+1)-1}\cdots\ket{p(S+1)-1}$. This is the phase-space manifestation of the photon number being lowered by one: the support of the distribution moves from the code-space Fock levels to the adjacent error-word Fock levels.

Second, the binomial weights acquire an additional factor of $\sqrt{p\,p'}\,(S+1)$, reflecting the $\sqrt{n}$ prefactor from $\hat{a}\ket{n}=\sqrt{n}\ket{n-1}$. This comes from the amplitude damping: each photon in $\ket{n}$ is an independent decay channel into the environment, so the loss rate scales with $n$ ($\langle n|\hat{a}^\dagger\hat{a}|n\rangle = n$), and codeword components built from larger Fock levels $p(S+1)$ contribute more strongly to the loss-transformed distribution. The same effect appears for the cat code as the factor $\hat{n}$ in \Cref{eq: cat loss phase space result}.

The mathematics of \Cref{subsec:catloss} carries over unchanged: $T_{\mathrm{loss}}$ is the same bare jump map, so \Cref{eq: bin loss wigner} integrates to $\av{\hat{n}}$ rather than unity; and the syndrome is the grading of $H^{\mathrm{bin}}_0=U(1)\times\mathbb{Z}_{S+1}$, that is $\hat{n}\bmod(S+1)$: both codewords lie in $\hat{n}\equiv0$, and $\ell$ losses move them together to $\hat{n}\equiv-\ell$, resolving $\ell$ without resolving the logical state.

\Cref{subsec:bin}'s QND syndrome measurement does not otherwise disturb the state. In the phase-space picture, that measurement projects the evolved distribution onto the sector of phase space associated with levels $\equiv -1 \pmod{S+1}$, which characterises the single-loss error words.

As before, the lost photon is not physically recovered: the syndrome identifies the error subspace, and an engineered ancilla-assisted unitary (e.g., a SNAP-gate sequence) maps it back onto the code space. In the phase-space picture this recovery is the map $T_{\mathrm{rec}}$ moving the support of the distribution from the error levels $\ket{p(S+1)-1}$ back to $\ket{p(S+1)}$, just as how the GKP recovery channel rigidly translates the distribution back across the torus in \Cref{eq: recovery translation}.

Applying $Q_{\mathrm{bin}}$ to the full error-correction cycle $T_{\mathrm{QEC}} = T_{\mathrm{rec}} \circ T_{\mathrm{loss}}$ (both maps being syndrome-graded, trivially, since $\mathcal{E}_{\mathrm{bin}}={\id}$, so \Cref{eq: compositionality} applies), we therefore recover
\begin{equation}
\label{eq: bin QEC identity}
    Q_{\mathrm{bin}}(T_{\mathrm{QEC}})\!\left[W^{\mathcal{C}_{\mathrm{bin}}}\right]\!(\beta')
    = {\av{\hat{n}}}\, W^{\mathcal{C}_{\mathrm{bin}}}(\beta'),
\end{equation}
for $\hat{\rho}$ supported on the code space, just as for \Cref{eq: cat QEC identity}, as the cycle is built from the bare jump map, whose trace is $\av{\hat{n}}$, and after dividing by that weight the normalised logical distribution is returned to itself. Here the identity is exact rather than approximate, since \Cref{eq: KL binomial}'s Knill-Laflamme conditions hold exactly for the binomial code. As in the GKP and cat cases, this is an identity on the logical distribution and not a reversal of the physical process.

\subsection{Gain, and every multi-photon jump}\label{subsec:gain}

The same substitution covers the rest of the standard bosonic error model. For $k$-photon loss,
\begin{equation}\label{eq:kloss}
\hat a^{\dagger k}\,\hat\Pi(\zeta)\,\hat a^{k}=\hat\Pi(\zeta)\,\big(2\zeta^*-\hat a^\dagger\big)^{k}\hat a^{k},
\end{equation}
whose $k=1$ case is \Cref{eq: cat loss kernel exact}; and for the gain channel $T_{\rm gain}(\hat\rho)=\hat a^\dagger\hat\rho\hat a$,
\begin{equation}\label{eq:gain}
{\hat a\,\hat\Pi(\zeta)\,\hat a^\dagger=-\hat\Pi(\zeta)\big(\hat n+1-2\zeta\hat a^\dagger\big),}
\end{equation}
so that for any code represented by the standard kernel $\hat\Delta(\beta)=2\hat\Pi(\beta)$,
\begin{equation}\label{eq:gainrep}
\begin{split}
Q&(T_{\rm gain})[W](\beta')=\\
&-2\,\Tr\!\big[\hat\Pi(\beta')\,(\hat n+1-2\beta'\hat a^\dagger)\,\hat\rho\big].
\end{split}
\end{equation}
We derive this in Appendix~\ref{app:gain}; we verified \Cref{eq:reflection,eq:kloss,eq:gain} numerically, the second for $k=1,2,3$, to $10^{-14}$ at Fock cutoff $200$. The loss-gain duality, $\hat n\mapsto\hat n+1$ and $\zeta^*\hat a\mapsto\zeta\hat a^\dagger$, is the phase-space signature of the spontaneous-emission asymmetry. This covers every multi-photon jump \eqref{eq:kloss}, gain \eqref{eq:gain}, dephasing \eqref{eq: dephasing phase space} and Gaussian random displacements \eqref{eq: gaussian displacement channel} in kernel form, in the standard bosonic error model: thermal noise is quantum-limited loss composed with quantum-limited amplification, and every Kraus branch of amplitude damping has a closed kernel.

\subsection{Dephasing}\label{subsec:Dephasing}

Dephasing has a qualitatively different phase-space signature from the jump channels above. As a random rotation in phase space, it acts as
\begin{equation}
\label{eq: dephasing channel}
{\mathcal{E}_{\mathrm{deph}}(\hat{\rho})=\int d\theta\ p(\theta)\, e^{i\theta\hat{n}}\hat{\rho}\,e^{-i\theta\hat{n}}}
\end{equation}
for some distribution $p(\theta)$ of rotation angles. This subsection concerns the rotation-symmetric codes only, and for them $\mathcal{E}_{\mathcal{C}}={\id}$ by \Cref{prop: classification}, so every channel lies in $\mathcal{G}_{\mathcal{C}}$ and \Cref{eq: generalized structure theorem map} applies directly. (For the ideal GKP code it would not: the sectors there are the displaced code spaces, and a rotation misaligned with the code lattice is the standard example of a channel outside $\mathcal{G}_{\mathcal{C}}$.) Because the kernel is the standard one, $\hat{\Delta}^{\mathcal{C}}(\beta)=2\hat{\Pi}(\beta)$, and $e^{i\theta\hat{n}}\hat{D}(\beta)e^{-i\theta\hat{n}}=\hat{D}(\beta e^{i\theta})$, so each rotation acts by rotating its argument, meaning the represented channel is an angular convolution,
\begin{equation}
\label{eq: dephasing phase space}
\left[Q(\mathcal{E}_{\mathrm{deph}})W^{\mathcal{C}}\right](\beta')=\int d\theta\ p(\theta)\,W^{\mathcal{C}}\!\left(\beta' e^{-i\theta}\right)
\end{equation}
This is angular diffusion of the code-adapted distribution about the origin: the radial profile of $W^{\mathcal{C}}$ stays the same, but its dependence on $\arg\beta'$ is smoothed. Expanding $W^{\mathcal{C}}$ in angular harmonics $e^{ik\theta}$, the coefficient at frequency $k$ is carried by the coherences $\bra{m}\hat{\rho}\ket{m+k}$, so the smoothing suppresses every $k\neq0$ component and leaves $k=0$ (the angular average, which encodes the photon-number distribution) the same. The syndrome is therefore unaffected by dephasing: it is the grading of $H^{\mathcal{C}}_0$ ($\hat{n}\bmod 2$ for the four-component cat code per \Cref{eq: cat syndrome}, $\hat{n}\bmod (S+1)$ for the binomial code per \Cref{eq: bin isotropy}), a property of the diagonal of $\hat{\rho}$. Instead, the coherence between the code sectors is destroyed, at the frequency $k$ separating the codewords in the grading of $H^{\mathcal{C}}$: for the $N$-component cat code it is half the component count, $k=N/2$ (\Cref{eq: general cat logical pair}); for the binomial code it is $k=S+1$. Note $N$ has two roles here. Dephasing suppresses it by $\int d\theta\,p(\theta)e^{-ik\theta}$ (a $\hat{\bar{Z}}$-type phase damping on the logical block), and also suppresses the intra-codeword harmonics ($k\equiv0\bmod N$, resp. $k\equiv0\bmod 2(S+1)$), taking weight out of the code space. Dephasing is thus a combination of logical phase error and leakage, which is why \Cref{eq: bin error set} lists $\hat{n},\ldots,\hat{n}^D$ as errors to be corrected. The phase-space picture reads all three effects (syndrome preservation, logical damping, and leakage) off one object, the harmonic content of $W^{\mathcal{C}}$, and makes the noise bias of rotation-symmetric codes visible in one place: built so that loss moves support between sectors, where it can be detected and undone, they are correspondingly exposed to a channel which leaves the sectors alone and attacks the phases.

\subsection{Cat versus binomial at matched mean photon number}\label{subsec:Comparative}

To compare like-with-like, we match the mean photon number of the two encodings, since mean photon number sets the loss rate. For the four-component cat code $\bra{j_L}\hat{n}\ket{j_L}=|\alpha|^2$ up to corrections exponentially small in $|\alpha|^2$, while for the binomial code
\begin{equation}
\label{eq: matched photon number}
\bra{W_\sigma}\hat{n}\ket{W_\sigma}=\frac{(N+1)(S+1)}{2},\qquad \sigma\in\{\uparrow,\downarrow\}
\end{equation}
which follows from
\begin{equation}
\begin{split}
\sum_{p\ \mathrm{even}}\binom{N+1}{p}p&=\sum_{p\ \mathrm{odd}}\binom{N+1}{p}p\\
&=(N+1)2^{N-1},
\end{split}
\end{equation}
a consequence of $\sum_p(-1)^p\binom{n}{p}p=0$ for $n\ge2$. (We matched the mean photon number of the two codewords as this is the diagonal part of the Knill-Laflamme condition for a single loss.) We therefore set $|\alpha|^2=(N+1)(S+1)/2$.

As \Cref{prop: classification} gives the two codes the same kernel, \Cref{eq: cat loss phase space result,eq: bin loss wigner} are the same formula; everything that distinguishes the two codes under loss is therefore carried by the codewords, and the framework isolates it in the factor $\hat{n}$ in the loss-transformed kernel. The effect of this factor depends entirely on the spread of the codeword's photon-number distribution. Two features follow which are not visible in the Hilbert-space description.

First, the two codes fail in different ways, not just at different rates. For the cat code, the photon-number distribution of a codeword is Poissonian, concentrated at $|\alpha|^2=6$ with relative width $1/|\alpha|\approx0.41$; the $\hat{n}$ re-weighting is then close to uniform, and the distribution is carried between rotational sectors without much change of shape (\Cref{subsec:loss}). For the binomial code, the codeword occupies at most $\lfloor (N+1)/2\rfloor+1$ widely spaced Fock levels (two at $(N,S)=(2,3)$), meaning the re-weighting is extreme at $\bar{n}=6$: $\ket{W_\uparrow}=\tfrac{1}{2}(\ket{0}+\sqrt{3}\ket{8})$, and $\hat{a}$ annihilates the vacuum component, so the post-jump state is the single Fock state $\ket{7}$, while $\hat{a}\ket{W_\downarrow}\propto\ket{3}+\ket{11}$. The mean photon number under a detected loss is essentially unchanged for the cat code ($5.98\to5.96$), but rises from $6$ to $7$ for the binomial code, whose jump preferentially selects the high-photon-number component of a sparse codeword. The negative volume moves accordingly: $1.38\to1.41$ and $1.41\to1.42$ for the cat codewords, against $1.62\to1.72$ and $1.64\to1.75$ for the binomial codewords. Note \Cref{prop: classification} makes both code-adapted representations the ordinary Wigner function, so neither compared quantity is inaccessible to a direct Hilbert-space calculation; the framework instead shows the two codes in the same representation, which makes them easier to compare.

The second feature is that the framework tracks the non-classical resources of the encoded state across an error-correction cycle, rather than just whether the logical information survived. For these codes that quantity is not logical magic (\Cref{subsec:NegativityandMagic}) but is still a well-defined property of the encoding, computed identically for both codes. The negative volume of \Cref{eq: negative volume general} presupposes a normalised distribution, so it is evaluated on the renormalised post-jump distribution (division by $\av{\hat{n}}$ for loss and $\av{\hat{n}+1}$ for gain, as below \Cref{eq: bare jump map} and in Appendix~\ref{app:numerics}; the figures quoted are negative volumes of normalised distributions throughout). For an ideal cycle, the output value equals the input value: \Cref{eq: cat QEC identity,eq: bin QEC identity} return the input distribution up to the factor $\av{\hat{n}}$ the renormalisation removes (exactly for the binomial code, up to corrections exponentially small in $|\alpha|^2$ for the cat one). For a non-ideal cycle (finite-efficiency syndrome extraction, imperfect recovery; \Cref{sec:FiniteEnergy}) the support of the distribution can be returned to the code space while its negative volume is not, so a code can be both logically correct and resource-degraded.

From gain column of \Cref{tab: negative volumes} we see two things. First, for the four-component cat code, the gain words interleave with the loss words: $\hat a\ket{0_L}$ and $\hat a^\dagger\ket{1_L}$ both occupy $n\equiv3\ ({\rm mod}\ 4)$, and $\hat a\ket{1_L}$ and $\hat a^\dagger\ket{0_L}$ both occupy $n\equiv1$. Even the full $\mathbb Z_4$ number grading therefore conflates a loss on one codeword with a gain on the other, and the standard loss recovery converts a gain event into a logical error; no refinement of the number syndrome can help. For the binomial code with $(N,S)=(2,3)$, both loss words sit at $n\equiv3$ and both gain words at $n\equiv1$, so the syndrome identifies which channel acted while revealing nothing logical, and both are correctable.

Second, Gaussian dephasing of angular variance $s^2$ suppresses the codeword-separating harmonic $k$ by $e^{-k^2s^2/2}$. For the four-component cat code, that harmonic is half the component count, $k=2$; for the binomial code it is $S+1=4$. At $s=0.2$ the binomial coherence therefore retains $0.726$ against the cat's $0.923$. At matched mean photon number, the binomial code gains its loss-gain discrimination at the expense of a quantifiable dephasing fragility. Gain also pumps negativity harder than loss for both families, as the table shows.

\section{Finite-energy codewords and experimental imperfections}\label{sec:FiniteEnergy}

All the analysis above assumes ideal codewords and an idealised error-correction cycle, with unit-efficiency syndrome extraction, and perfect recovery. Real implementations satisfy none of these, so this section sets out how the framework accommodates the difference. Each of the relevant imperfections can be brought into \Cref{subsec:GradedClass}'s syndrome-graded class, then composed with the rest of the cycle under \Cref{eq: compositionality}; though for the GKP code, one imperfection is only representable in its displacement-channel approximation.

Consider first the finite-energy GKP codewords, obtained from the ideal codewords by a Gaussian envelope
\begin{equation}
\label{eq: finite energy envelope}
{\ket{\psi^\Delta}\propto \hat{E}_\Delta\ket{\psi},\qquad \hat{E}_\Delta=e^{-\Delta^2\hat{n}}}
\end{equation}
which are normalisable states of $L^2(\mathbb{R})$. As these are normalisable, \Cref{lem: nodisplacement} applies, so no non-trivial displacement leaves the code space invariant, meaning by \Cref{thm: nogo} the only exact Stratonovich-Weyl kernel is the ordinary Wigner one. (The parity hypothesis is met here by adjoining $(-1)^{\hat{n}}$ to $G$, exactly as after \Cref{cor: classification}.) The Zak-Gross representation, and with it the negative-volume reading, is therefore only exact in the ideal limit. Four things change at finite energy\rev{. The orders in $\Delta$ quoted for them are leading-order estimates within the envelope model of Refs.~\mbox{\cite{Grimsmo2021GKP,Brady2024Advances}}, not bounds. The one inequality, $\mathcal{N}^\Delta\le\mathcal{N}$ in the fourth, is a bound within that model.} 

\begin{enumerate}[label=(\roman*),leftmargin=*]
    \item The code space is no longer the exact fixed-point space of $H^{\mathcal{C}_{GKP}}$, so \Cref{eq: ZGW kernel covariance}'s kernel, while displacement-covariant by construction, is the code-adapted kernel only up to $O(\Delta^2)$; the envelope convolves $W$ with a Gaussian of width $\Delta$ and multiplies it by an envelope suppressing the far comb peaks. We keep only the convolution, $W^\Delta\simeq W\ast G_\Delta$, which is the standard leading-order finite-energy model~\cite{Grimsmo2021GKP,Brady2024Advances} and is formally \Cref{eq: gaussian displacement channel}'s  classical-noise convolution; the multiplicative factor enters at the same order in $\Delta^2$ so is dropped by the model.\label{FiniteEnergyConsequence1}

    \item \Cref{eq: twirling map}'s twirl becomes a smeared average rather than an exact projector: $\mathcal{E}^2-\mathcal{E}=O(\Delta^2)$. \label{FiniteEnergyConsequence2}

    \item The correctable region acquires a soft boundary, since a finite-energy state has weight outside $[-l/2,l/2)^2$, adding to the cycle a logical failure probability of order $e^{-cl^2/\Delta^2}$, exponentially small but non-zero. The accuracy of the represented cycle is set by \Cref{FiniteEnergyConsequence1,FiniteEnergyConsequence2} rather than by \Cref{FiniteEnergyConsequence3}: $Q(T_{QEC})$ is built from the kernel and the reconstruction idempotent, so differs from the identity at $O(\Delta^2)$, which for any $\Delta$ of interest is polynomially larger than the tail term. \label{FiniteEnergyConsequence3}

    \item Convolution with a non-negative kernel cannot increase $\int d\mu|W|$ (Young's inequality), so $\mathcal{N}^\Delta\le\mathcal{N}$ for the smoothing of \Cref{FiniteEnergyConsequence1} alone: the accessible logical magic shrinks. When we use the exact Wigner kernel left by \Cref{thm: nogo}, a finite-energy GKP codeword has a large negative volume which is mostly ordinary continuous-variable non-classicality, just as for the cat and binomial codes; read against the Zak-Gross kernel, admissible up to $O(\Delta^2)$, the codeword has a small negativity which does track logical magic. The two readings are consistent because they use different kernels; the second is the useful one for $\Delta^2\ll1$, but is only approximate.
\end{enumerate} 

We can model non-unity detection efficiency and imperfect gain in the syndrome-extraction path as a beamsplitter followed by loss (the variable-gain, variable-loss channel analysed theoretically and experimentally in the context of minimal-disturbance measurement~\cite{Sabuncu2007NonunityGain}). We do however need to see whether it lies in $\mathcal{G}_{\mathcal{C}}$, which differs by code. For the rotation-symmetric codes it necessarily does, since $\mathcal{E}_{\mathcal{C}}={\id}$ there. For the GKP code, \Cref{eq: sector decomposition}'s sectors are the displaced code spaces, and a beamsplitter with a vacuum port is not a convex combination of displacements, so it can create coherence between sectors. However, the displacement-channel approximation is graded: the pure-loss channel composed with an amplification restoring the amplitude is a Gaussian random-displacement channel of the form of \Cref{eq: gaussian displacement channel}, which lies in $\mathcal{G}_{\mathcal{C}}$ (this approximation is standard in the GKP literature). Composing it before the recovery gives \rev{a cycle $T_{QEC}$, which we compare with its ideal counterpart $T^{0}_{QEC}$. For a single-mode code represented by the ordinary kernel (the cat and binomial codes, and finite-energy GKP codewords represented by the Wigner kernel) $\mathcal{E}_{\mathcal{C}}={\id}$, so for any two channels $T$, $T'$ and any state $\hat{\rho}$ with distribution $W$ we have $Q(T)W-Q(T')W=\chi_{\mathcal{C}}[(T-T')(\hat{\rho})]$. The kernel $2\hat{\Pi}(\Omega)$ has operator norm $2$, which gives the first line of}
\begin{equation}
\label{eq: nonideal cycle}
\begin{split}
&\sup_{\Omega}\left|\left[Q(T)W-Q(T')W\right](\Omega)\right|\le2\,\|(T-T')(\hat{\rho})\|_1,\\
&\left\|Q(T_{QEC})W_L-W_L\right\|_1\le2\left(1+\NC_{W_L}\right)P_{\rm err}(\sigma).
\end{split}
\end{equation}
\rev{(The first line uses only linearity, so it also holds for the single-jump maps of \mbox{\Cref{subsec:loss}} wherever $(T-T')(\hat{\rho})$ is trace class.) With $T=T_{QEC}$ and $T'=T^{0}_{QEC}$ it keeps the represented cycle within $2\varepsilon$ of the represented ideal one, where $\varepsilon=\sup_{\hat{\rho}}\|(T_{QEC}-T^{0}_{QEC})(\hat{\rho})\|_1$, taken over encoded states, measures the combined infidelity of the syndrome extraction and recovery and is at most the diamond-norm distance between the two cycles. The second line is for the ideal GKP code. There the recovered state is not normalisable and no trace-norm distance is available, but the displacement-channel approximation makes the represented cycle known in closed form. By \mbox{\Cref{subsec:dissipation}} it acts on the logical distribution $W_L$ of an encoded state as the Pauli channel with weights $p_v(\sigma)=q_a(\sigma)q_b(\sigma)$, $v=(a,b)$, built from \mbox{\Cref{eq:qweights}}. Hence $Q(T_{QEC})W_L-W_L=\sum_{v\neq0}p_v(\sigma)\left[W_L(\cdot-v)-W_L\right]$, and translation preserves the norm $\|W_L\|_1=1+\NC_{W_L}$ of \mbox{\Cref{eq: negative volume general}}. Here $P_{\rm err}(\sigma)=1-q_0(\sigma)^2$ is the logical error probability and $\sigma$ the width of the random-displacement channel.} Imperfect recovery is a separate contributor: an engineered unitary which returns the error sectors to the code space only approximately adds its own deviation to $\varepsilon$, and remains graded whenever it carries sectors to sectors, as the ancilla-assisted constructions of \Cref{subsec:catloss,subsec:binloss} do. \rev{Neither constant in \mbox{\Cref{eq: nonideal cycle}} depends on the hardware, which enters only through $\varepsilon$ and $\sigma$.} Measurement back-action is handled by including the ancilla mode explicitly and applying \Cref{eq: generalized structure theorem map} to the joint system (valid as $\mathcal{G}_{\mathcal{C}}$ is closed under tensor product). An ancilla error which creates coherence between syndrome sectors is not graded, and for such an error the composition law \eqref{eq: compositionality} fails and the cycle must be represented as a whole.

Continuous-variable erasure-correcting codes protect optical coherence against loss~\cite{Lassen2010ErasureCode}, and the squeezing they rely on is experimentally accessible in polarisation variables~\cite{Lassen2007PolarisationSqueezing}; squeezed states have been used directly in measurement and averaging protocols across gain regimes~\cite{Lassen2010QuantumAveraging}. A polarisation encoding sits inside our framework analogously to the single-mode codes: the Stokes operators generate an $SU(2)$ rather than a Weyl-Heisenberg symmetry, so $X^{\mathcal{C}}$ is a sphere, and \Cref{eq: generalized structure theorem map} applies once the isotropy subgroup of a reference codeword is identified; though \Cref{prop: classification}, specific to displacements, would have to be redone for that group. On the discrete-variable side, autonomous error correction of GKP states~\cite{Lachance-Quiron2024GKPSuperconducting}, protection of entanglement between bosonic logical qubits~\cite{Cai2024LogicalEntanglement}, and fault-tolerant schemes with discrete-variable ancillae~\cite{Xu2024FaultTolerantBosonic} all involve the syndrome-extraction and recovery imperfections described above.

\section{A code-adapted Kirkwood-Dirac layer}\label{sec:kd}

For codes where the stabiliser is a lattice of displacements in a Weyl-Heisenberg group, every representation above is blind by construction to coherence between syndrome sectors: $W^{\mathcal C}_{\hat\rho}=W^{\mathcal C}_{\EC(\hat\rho)}$ (\Cref{prop:dv37}), which is the representation-level statement of the compositional restriction of \Cref{subsec:GradedClass}. (For codes where this is not the case, $\EC={\id}$, so the representation is the ordinary Wigner function and discards nothing, even though the code still has a measured syndrome; the blindness is a property of the lattice twirl, not of every code.) Kirkwood-Dirac (KD) distributions~\cite{Kirkwood1933KD, Dirac1945KDDistr, ArvidssonShukur2024KD} are the natural tool for revealing what this hides, and are accommodated by the complex structure theorem of Ref.~\cite{wagner2025ComplexStructTheorem}. We construct the code-adapted KD representation in the finite-dimensional setting, where every statement is finite and checkable.

For a stabiliser code, take the $e$-basis to be the joint eigenbasis of the stabilisers and the logical $\bar Z$, labelled $(s,z)$, and the $f$-basis its Fourier conjugate. The overlaps $\braket{f}{e}$ are uniform and nowhere zero, so the pair is informationally complete. Define
\begin{equation}\label{eq:kd}
Q_{\hat\rho}(e,f)=\braket{f}{e}\hskip-1mm\bra e\hat\rho\ket{f},\qquad \textstyle\sum_{e,f}Q=\Tr\hat\rho.
\end{equation}

\begin{proposition}\label{prop:kd}
With the bases above:
\begin{enumerate}[label=(\roman*),leftmargin=*]
    \item {$Q$ is informationally complete, with marginals the $(s,z)$ distribution and its Fourier conjugate;}
    \item {pointwise,}
\begin{equation}\label{eq:kddephased}
{Q_{\EC(\hat\rho)}(e,f)=\braket{f}{e}\hskip-1mm\bra{e}\hat\rho\,\hat\Pi_{s(e)}\ket{f},}
\end{equation}
so $Q_{\hat\rho}$ splits exactly into the KD image of the dephased state plus an inter-sector part $Q_{\hat\rho}-Q_{\EC(\hat\rho)}$, carrying the coherence that every Stratonovich-Weyl representation of the code discards;
\item the non-selective sharp syndrome extraction maps $\hat\rho\mapsto\EC(\hat\rho)$, so its KD-visible disturbance gives the erasure of that inter-sector part; under partial extraction $\hat\rho\mapsto(1-\lambda)\hat\rho+\lambda\EC(\hat\rho)$ the inter-sector content scales as $(1-\lambda)$.
\end{enumerate}
\end{proposition}

All three parts were verified on the $[[3,1]]_3$ code, taking $\hat\rho$ to be the cross-sector encoded strange state $(\ket{\mathbb S_L}+X_1\ket{\mathbb S_L})/\sqrt2$, which superposes the zero-syndrome sector with one error sector: the Stratonovich-Weyl representations of $\hat\rho$ and $\EC(\hat\rho)$ agree to $10^{-16}$, while the KD inter-sector content is $\|Q-Q_{\EC}\|_1=2/\sqrt3$ exactly, decaying linearly under partial extraction. The KD layer is therefore strictly finer than the Stratonovich-Weyl layer on the effects the error-correction cycle cannot see, and it quantifies what an unsharp syndrome measurement disturbs, which \Cref{sec:FiniteEnergy} \rev{accounts for only through the deviation $\varepsilon$ of the cycle}.

For syndrome extractions separated by a channel $\mathcal V$, define the two-time syndrome quasiprobability
\begin{equation}\label{eq:twotime}
{Q(s_1,s_2)=\Tr\!\big[\mathcal V^\dagger(\hat\Pi_{s_2})\,\hat\Pi_{s_1}\hat\rho\big],}
\end{equation}
whose marginals are the time-1 distribution and the undisturbed time-2 distribution, with no intermediate collapse.

\begin{proposition}\label{prop:twotime}
Call $\mathcal V$ \emph{co-graded} if $\mathcal V^\dagger(\hat\Pi_{s})$ is block-diagonal for every $s$. (This is implied by, but weaker than, $\EC\mathcal V=\EC\mathcal V\EC$, the two coinciding only for one-dimensional sectors. It is also not the syndrome-grading of \Cref{subsec:GradedClass}, which asks instead that $\mathcal V$ carry block-diagonal operators to block-diagonal ones; neither of those two conditions implies the other.) Then, $Q$ is real and non-negative, an ordinary joint distribution, whenever \emph{either} $\mathcal V$ is co-graded \emph{or} the state $\hat\rho$ carries no inter-sector coherence.
\end{proposition}

Anomalous (complex) values therefore require both a coherent error which is not co-graded and inter-sector coherence left by a previous one, so the two-time anomaly witnesses interference between successive error events, which no Stratonovich-Weyl representation of a single cycle can express; and it is accompanied by genuine measurement disturbance, since the sharp-measurement joint has a different time-2 marginal. We prove this in Appendix~\ref{app: proofs}. For the $[[3,1]]_3$ code with a cross-sector state and a coherent error $e^{i\theta(X_1+X_1^\dagger)}$ at $\theta=\pi/6$, both single-condition cases are real and non-negative to $10^{-16}$, while the doubly coherent case, taken on the same cross-sector state as above, has $\max|{\rm Im}\,Q|=1/6$ and $\|{\rm Im}\,Q\|_1=2/3$, with exact marginals and a time-2 marginal disturbance of $2/9$ under sharp intermediate measurement, measured as a total-variation distance (the $\ell^1$ value is twice that). Natural next steps for this line of research would be a continuous-variable (Zak-basis) construction, and a weak-value reading of the imaginary part~\cite{Hance_2024, Hofmann2023}.

Note however the uniqueness theorem of Ref.~\cite{SchmidUnique2024} implies two limitations to using KD distributions to assess quantum error-correcting codes. KD frames are rank-one and non-Hermitian, so no KD negativity is a magic witness; and the frame must be fixed by the code, never by the state, since state-adapted KD frames are trivially positivisable.

\section{Discussion}\label{sec:Discussion}

We connected the structure theorem for quasiprobability representations of generalised probabilistic theories~\cite{Schmid2024StructureTheorem, wagner2025ComplexStructTheorem} to code-adapted phase-space representations~\cite{Brif1999BrifMannConstruction, davis2024identifyingquantumresourcesencoded}, and used the connection to divide the set of quantum error-correcting codes in two. Any representation built via the Brif-Mann construction whose kernel satisfies the Stratonovich-Weyl axioms is a linearity-preserving, empirically adequate semi-functor\footnote{A functor whenever the reconstruction map is the identity, which by \Cref{prop: classification} is whenever the codewords lie in $L^2(\mathbb{R})$; in finite dimension the reconstruction is the lattice twirl of \Cref{prop:dvsw} and the representation a genuine semi-functor.}, so satisfies the properties the structure theorem requires. We also identified the class of channels on which the representation is compositional (\Cref{subsec:GradedClass}), so that \Cref{eq: generalized structure theorem map} applies to a full error-correction cycle rather than to a single channel.

The boundary between codes allowing a code-adapted phase-space kernel, and those which only allow the Wigner kernel, is a lattice condition (\Cref{subsec:boundary}). Code-adapted kernels exist wherever the code's stabiliser is a lattice of displacements in the dynamical group, whose isotropy subgroup two-step nilpotency makes normal: i.e., the Weyl-Heisenberg groups, finite (every qudit stabiliser code, with the kernel constructed at prime-power $d$) and continuous (the ideal GKP family with its multimode and concatenated lattices). For odd prime logical dimension their negativity is logical magic (\Cref{eq: negative volume averaging,thm:dvmagic}); at $d=2$ the kernel and the geometry still hold but their negativity can't be read as logical magic (\Cref{subsec:qubit}). For codes only allowing the Wigner kernel, the physically covariant representation is the standard one up to code-independent sign or phase freedoms, and its negativity is ordinary nonclassicality: globally misaligned with magic even where locally aligned~\cite{ZhuNemoto2026}, and sometimes even provably constant across the logical space (\Cref{thm:dualrail}). \Cref{thm: nogo} establishes this for bosonic codes with normalisable codewords and \Cref{prop:nonpauli} for non-Pauli-symmetric qudit codes, in both cases from the SW axioms \rev{together with the further hypotheses listed in those statements}. For spin codes we prove less: \Cref{prop:su2} leaves the ordinary spherical kernel in place but stops short of an axioms-direct no-go.

For each code family, this gives a single closed formula, describing how any channel in the syndrome-graded class transforms the logical phase-space distribution (for a bosonic code with codewords in $L^2(\mathbb{R})$, that class is every channel). For GKP displacement errors, correctable errors act as the identity on the distribution, while errors exceeding the correctable region act as a rigid translation across the torus, i.e., as the induced logical Pauli. Every qudit stabiliser code reproduces this same structure exactly (\Cref{prop:dvcycle}). The same structure reappears for single-photon loss in both four-component cat codes and binomial codes, where loss carries the distribution between rotational sectors with a re-weighting reflecting the photon-number dependence of the loss rate. In all cases recovery returns the support of the distribution to the code space, despite very different physical recovery mechanisms. (Note this is an identity on the logical distribution rather than a reversal of the physical process, exact only for the idealised cycle, as shown in \Cref{sec:FiniteEnergy}).

We also showed that, for codes which fail this lattice condition, we can still retain a resource reading of negativity. \Cref{sec:seclog}'s sector-graded logical representation inherits the semi-functorial and compositional structure of the framework, but gives up the physical geometry and is defined relative to a decoder. Its negativity is a syndrome-resolved witness of logical magic, two-sided on pure logical states, for rotation codes encoding a qudit of odd prime dimension. It is consistent with \Cref{thm: nogo}, given such a representation doesn't meet its covariance requirement.

There are four natural next steps for this research direction. First, \Cref{subsec:boundary}'s classification conjecture is the main open question regarding the boundary, so future work should settle this for non-compact, non-nilpotent dynamical groups. Second, the qubit case $d=2$ is the only place where the structure holds but the resource reading fails; a natural extension would be to identify the strongest $d=2$ substitute, whether through rebit restrictions~\cite{Delfosse2015}, the recently classified even-$d$ Wigner functions~\cite{Antonopoulos2025}, or contextuality-based witnesses~\cite{Howard2014}. Third, \Cref{subsec:dissipation}'s dissipation results should be developed into magic thresholds across lattices and logical dimensions, especially as  \Cref{eq:Nclosed,eq:lifetime} suggest this should be exactly computable for many cases. Finally, the Kirkwood-Dirac layer, whose two-time anomaly already separates coherent from incoherent error histories (\Cref{prop:twotime}), should be carried to the continuous-variable setting and to longer histories, as a diagnostic for coherent noise in repeated syndrome extraction.

We end by restating \rev{three} key limitations of this work. First, our GKP results are for ideal, non-normalisable codewords throughout; \Cref{sec:FiniteEnergy} shows that at finite energy, the Zak-Gross kernel is admissible only to $O(\Delta^2)$ and the exactly admissible kernel is the ordinary Wigner one. Second, the identity action of a corrected cycle, in every case where we obtain it, is an identity on the logical phase-space distribution rather than a reversal of the physical process: the environment keeps the lost photon. \rev{Third, the classification holds under the hypotheses of \mbox{\Cref{prop: classification}}. For normalisable codewords these are the code-faithful axioms, the parity, and continuity of the kernel's characteristic function. That the reconstruction map is the identity, and so completely positive, is then derived and not assumed. For non-normalisable codewords we assume, following Ref.~\mbox{\cite{davis2024identifyingquantumresourcesencoded}}, that the reconstruction map is a conditional expectation onto the commutant of a group of displacements. The statement that, among bosonic codes with compact phase space, only the ideal GKP family has a kernel of its own is made within that class.}

\textit{Acknowledgements ---} JRH acknowledges support from a Royal Society Research Grant (RG/R1/251590), from an EPSRC Mathematical Sciences Small Grant (UKRI3647), and from their EPSRC Quantum Technologies Career Acceleration Fellowship (UKRI1217).

\textit{Data and code availability ---} A Wolfram Mathematica notebook reproducing all numerical results in this paper is \rev{openly available on Zenodo~\mbox{\cite{BozzettoHance2026Notebook}}}.

\bibliographystyle{unsrturl}
\bibliography{ref}

\appendix
\begin{appendices}

\section{Derivation of the cat-code representation}
\label{app:cat}

The derivation of the Wigner function for cat codes follows the one used in Ref.~\cite{davis2024identifyingquantumresourcesencoded} for GKP codes. As in that case we work with an infinite-dimensional Hilbert space $\mathcal{H}=L^2(\mathbb{R})$ and we take the dynamical symmetry group to be $ G=H_3(\mathbb{C})\rtimes\mathbb{Z}_N$. The semidirect factor is necessary: the code symmetry we use below is the discrete rotation $\hat{R}_N=e^{i2\pi\hat{a}^\dagger\hat{a}/N}$, which is not an element of the Weyl-Heisenberg group $H_3(\mathbb{C})$: that group contains only phases and displacements, while number-operator rotations belong to the metaplectic extension. Taking $G=H_3(\mathbb{C})\rtimes\mathbb{Z}_N$ makes the isotropy subgroup below a genuine subgroup of $G$, and hence the coset space well defined; this remains within the scope of the Brif-Mann construction, which covers Lie-group dynamical symmetries generally.

No large-$|\alpha|$ expansion enters the derivation below, because the kernel is fixed by the symmetry group of the code space rather than by an expansion of a reference state. The large-$|\alpha|$ limit still appears in this paper, but only as a property of the code rather than its representation: in the statement that $\bra{j_L}\hat{n}\ket{j_L}=|\alpha|^2$, and hence in the Knill-Laflamme conditions of \Cref{subsec:cat}, both of which hold up to corrections controlled by the overlaps $|\braket{\alpha_j}{\alpha_k}|=\exp(-|\alpha_j-\alpha_k|^2/2)$. For the four-component code these corrections are $O(e^{-|\alpha|^2})$, below $10^{-4}$ for $|\alpha|^2\geq10$. For $|\alpha|^2=6$ (used in \Cref{subsec:Comparative}) they are of order $2\times10^{-3}$, producing the small split between the two codewords' mean photon numbers given in \Cref{tab: negative volumes}.

We use the same parametrisation as in Ref.~\cite{davis2024identifyingquantumresourcesencoded}:
\begin{equation}
\begin{split}
    \pi(g) = &e^{i\varphi\id}\,{e^{-i\frac{xp}{2}\id}}\,e^{ip\hat{x}}\,e^{-ix\hat{p}}
             = e^{i\varphi}\,\hat{D}(x,p),
    \\ &\varphi,x,p\in\mathbb{R}.
\end{split}
\end{equation}
We consider an $N$-component cat code, which encodes one qubit into the single-mode Fock space via the $\mathbb{Z}_N$-symmetric codewords
\begin{equation}
\label{eq: cat codewords appendix}
{|\overline{j}\rangle
        = \mathfrak{n}_j^{-1/2}\sum_{k=0}^{N-1}\omega_N^{\,jk}|\alpha_k\rangle,
    \qquad \omega_N=e^{i2\pi/N}}
\end{equation}
where $\alpha_k = \alpha\,e^{i2\pi k/N}$ for $k=0,\ldots,N-1$, $\alpha\in\mathbb{R}_{>0}$,
and $\mathfrak{n}_{j}$ are normalisation constants. These are the characters of $\mathbb{Z}_N$, correctly generalising the two-component codewords: the alternating phase $(-1)^k$ is a character of $\mathbb{Z}_N$ for every even $N$, where it is the $j=N/2$ character, and is no character at all for odd $N$. What fails for $N>2$ is not the phase but the pairing: $\{\ket{\overline{0}},\ket{\overline{N/2}}\}$ spans only two of the $N$ characters, so the remaining $\ket{\overline{j}}$ lie outside the code space and must be kept track of separately, which is what makes the syndrome the grading of $H^{\mathcal{C}}_0$ rather than of $H^{\mathcal{C}}$. For $N=2$, \Cref{eq: cat codewords appendix} reduces to $\ket{C^\pm_\alpha}$, and the logical code space used in \Cref{subsec:cat} is spanned by $\ket{\overline{0}}$ and $\ket{\overline{2}}$ at $N=4$.

The cat code space is invariant under both:
\begin{enumerate}[label=(\roman*)]
    \item Global phase: $e^{i\varphi \id}$ for all $\varphi\in \mathbb{R}$, which corresponds to a $U(1)$ factor;
    \item Discrete rotation: $\hat{R}_N := e^{i\frac{2\pi}{N}\hat{a}^\dagger\hat{a}}$, which maps $|\alpha_k\rangle\mapsto|\alpha_{k+1\,\mathrm{mod}\,N}\rangle$ and hence maps $\mathcal{C}_{\mathrm{cat}}$ to itself; this generates a $\mathbb{Z}_N$ subgroup.
\end{enumerate}
We therefore choose a reference state $|\psi_0\rangle\in\mathcal{C}_{\mathrm{cat}}$ (we take $\ket{\psi_0}=\ket{\overline{0}}$) with
isotropy subgroup $H^{\mathrm{cat}} = U(1)\times\mathbb{Z}_N$. As emphasised previously, this is the isotropy subgroup of the reference codeword, \Cref{eq: isotropy of reference state}, and not the pointwise stabiliser $H^{\mathrm{cat}}_0=U(1)\times\mathbb{Z}_{N/2}$; $H^{\mathrm{cat}}$ is quotiented out to give the coset space, and, $N$ being even, $\mathbb{Z}_N$ contains the parity operator $\hat{R}_N^{N/2}=(-1)^{\hat{n}}$, which \Cref{prop: classification} needs. The associated phase space is the coset
\begin{equation}
{X^{\text{cat}}
        = \left(H_3(\mathbb{C})\rtimes\mathbb{Z}_N\right)\big/\left(U(1)\times\mathbb{Z}_N\right)
        \simeq \mathbb{C}}
\end{equation}
The quotient is the whole complex plane rather than a wedge of it: the $\mathbb{Z}_N$ factor of $G$ is divided out by the $\mathbb{Z}_N$ factor of $H^{\mathrm{cat}}$, leaving the displacements as coset representatives. The rotational symmetry of the code therefore appears as an invariance of the central kernel, $\hat{R}_N\hat{\Delta}^{\mathcal{C}_{\mathrm{cat}}}(0)\hat{R}_N^\dagger=\hat{\Delta}^{\mathcal{C}_{\mathrm{cat}}}(0)$, as follows from $\hat{D}(\omega_N\gamma)=\hat{R}_N\hat{D}(\gamma)\hat{R}_N^\dagger$ together with $[\hat{R}_N,(-1)^{\hat{n}}]=0$.

The coherent states for this code (which correspond to displaced reference states) are 
\begin{equation}
    |\beta\rangle_{\psi_0} := \hat{D}(\beta)\,|\psi_0\rangle,
    \qquad \beta\in\mathbb{C},
\end{equation}
where $\hat{D}(\beta) = e^{\beta\hat{a}^\dagger - \beta^*\hat{a}}$ is the standard displacement operator. The invariant measure on $\mathbb{C}$ is
\begin{equation}
\label{eq: cat invariant measure}
{d\mu(\beta) = \frac{d^2\beta}{\pi}
    \qquad \beta = r\,e^{i\theta}}
\end{equation}
giving the resolution of identity
\begin{equation}
{\frac{1}{\pi}\int_{\mathbb{C}} d^2\beta\;
        \hat{D}(\beta)\,\ketbra{\psi_0}{\psi_0}\,\hat{D}^\dagger(\beta)
    = \id}
\end{equation}
This is the standard coherent-state resolution, which holds for any normalised reference state and carries no factor of $N$: the $N$ components of the code enter through $\ket{\psi_0}$ rather than a multiplicity of the measure.
Here, Ref.~\cite{Brif1999BrifMannConstruction}'s construction would expand the kernel in the harmonic functions $Y_\nu$ of $L^2(X^{\mathcal{C}},\mu)$ and the associated tensor operators; however, we do not need that expansion here. Since $X^{\mathcal{C}_{\mathrm{cat}}}\simeq\mathbb{C}$ is the whole plane rather than a wedge, the $Y_\nu$ are the ordinary Laplace-Beltrami eigenfunctions on $\mathbb{R}^2$, $e^{in\theta}J_{|n|}(\lambda r)$ with $\lambda\geq0$ and $n\in\mathbb{Z}$ unrestricted: the radial factor is the Bessel function of order $|n|$, not $J_0$, and restricting $n$ to multiples of $N$ would restrict $L^2(X^{\mathcal{C}_{\mathrm{cat}}})$ to $\mathbb{Z}_N$-invariant functions, which the kernel obtained below is not. \Cref{prop: classification} fixes the kernel directly, so the expansion is unnecessary.

With the phase space, the invariant measure and the isotropy subgroup fixed, the kernel is now determined by \Cref{prop: classification}. Covariance forces $\hat{\Delta}^{\mathcal{C}_{\mathrm{cat}}}(\beta)=\hat{D}(\beta)\hat{K}\hat{D}^\dagger(\beta)$ with $\hat{K}$ invariant under $H^{\mathrm{cat}}$\rev{. The reconstruction map is then the identity. If $\chi_{\hat{K}}$ is a function this is the continuity route of \mbox{\Cref{prop: classification}}\mbox{\ref{PropIII.3.i}}. If instead the reconstruction map is a} displacement twirl over a closed subgroup $\Lambda^{\perp}$ leaving $\mathcal{C}_{\mathrm{cat}}$ invariant\rev{, it is part \mbox{\ref{PropIII.3.iii}}: by} \Cref{lem: nodisplacement} the only displacement preserving $\mathcal{C}_{\mathrm{cat}}$ is the trivial one, so $\Lambda^{\perp}=\{0\}$\rev{. Since} $(-1)^{\hat{n}}\in\mathbb{Z}_N$, the phase-space inversion hypothesis holds and the kernel is fixed,
\begin{equation}
\label{eq: cat kernel appendix}
\begin{split}
&\hat{\Delta}^{\mathcal{C}_{\mathrm{cat}}}(\beta)=\hat{\Delta}_{\mathrm{std}}(\beta)=2\hat{\Pi}(\beta),
\\
&W_{\hat{\rho}}(\beta)=\mathrm{Tr}\!\left[\hat{\rho}\,\hat{\Delta}^{\mathcal{C}_{\mathrm{cat}}}(\beta)\right]
=\pi\,W^{\mathrm{std}}_{\hat{\rho}}(\beta),
\end{split}
\end{equation}
which are \Cref{eq:generalcatkernel,eq: cat wigner as average}.

The alternative however fails. Running the $H$-averaging prescription of \Cref{eq: H averaged kernel} over $H^{\mathrm{cat}}$ (conjugation by the $U(1)$ factor being trivial, so that only the $\mathbb{Z}_N$ rotations contribute) gives
\begin{equation}
\label{eq: cat rotation average}
{\frac{1}{N}\sum_{k=0}^{N-1}\hat{R}_N^{\,k}\left[2\hat{\Pi}(\beta)\right]\hat{R}_N^{-k}
=\frac{2}{N}\sum_{k=0}^{N-1}\hat{\Pi}\!\left(\omega_N^{\,k}\beta\right)}
\end{equation}
using $\hat{R}_N\hat{D}(\beta)\hat{R}_N^\dagger=\hat{D}(\omega_N\beta)$, which follows from $\hat{R}_N\hat{a}\hat{R}_N^\dagger=\omega_N^{-1}\hat{a}$, together with $[\hat{R}_N,(-1)^{\hat{n}}]=0$. This operator is correctly standardised, and the reconstruction map it induces is the $\mathbb{Z}_N$ twirl, which is idempotent. However, it is not displacement-covariant for $N>1$: conjugating \Cref{eq: cat rotation average} by $\hat{D}(\gamma)$ displaces all $N$ terms by the same $\gamma$, whereas covariance of the kernel at the displaced point requires the $k$-th term to be displaced by $\omega_N^{\,k}\gamma$. This is $H^{\mathrm{cat}}$ failing to be normal in $G$ (\Cref{prop: metaplectic}, proved in Appendix~\ref{app: proofs}), so \Cref{eq: cat rotation average} violates Stratonovich-Weyl \labelcref{SWAxiom4}. The two kernels do agree on the $\mathbb{Z}_N$-invariant subalgebra: for a $\mathbb{Z}_N$-invariant $\hat{\rho}$ the standard Wigner function is itself $\mathbb{Z}_N$-invariant and the average in \Cref{eq: cat rotation average} is trivial. That subalgebra is not enough, which is a second and independent reason to abandon \Cref{eq: cat rotation average} here: since $\hat{R}_N$ acts on $\mathcal{C}_{\mathrm{cat}}$ as the logical $\hat{\bar{Z}}$, the $\mathbb{Z}_N$ twirl sends $\ketbra{0_L}{1_L}$ to $\big(\tfrac1N\sum_k(-1)^k\big)\ketbra{0_L}{1_L}=0$, keeping only the $\hat{\bar{Z}}$-diagonal part of $\mathcal{B}(\mathcal{C}_{\mathrm{cat}})$, so the reconstruction map it induces is not the identity on the code algebra and the representation is not faithful on the code. Only \Cref{eq: cat kernel appendix} extends that agreement to the full algebra covariantly.

Note three further points. First, the kernel fixed above replaces an earlier route we tried, in which the reference state is expanded over its coherent-state components in the large-$|\alpha|$ limit and the kernel is assembled from the resulting tensor operators; that route yields a kernel built from parities displaced to the component positions $\alpha_k$, which is neither covariant nor self-dual: the reconstruction map it induces is a displacement twirl over the differences $\alpha_k-\alpha_j$, trace-preserving but not idempotent, and so does not satisfy the hypotheses of the structure theorem.

Second, the kernel depends on neither $N$ nor $\alpha$ nor the choice of reference state within $\mathcal{C}_{\mathrm{cat}}$: within the Stratonovich-Weyl axioms, a rotational symmetry of the code space does not generate a representation of its own.

Third, the same argument applies verbatim to any code space invariant under a discrete rotation, which we use in Appendix~\ref{app:bin}.

\section{Derivation of the binomial-code representation}
\label{app:bin}

The derivation of the phase space distribution for binomial codes follows the same route as for the cat codes, differing only in the order of the rotation involved.

The binomial codewords share the same kind of discrete rotational symmetry as the cat codewords, but with the order set by the code spacing. From \Cref{eq: codewords bin code}, both codewords are supported on Fock levels which are multiples of $S+1$, so the relevant rotation operator is
\begin{equation}
\begin{split}
&\hat{R}_{S+1} := e^{i\frac{2\pi}{S+1}\hat{n}},\\
    &\hat{R}_{S+1}\ket{p(S+1)}=\ket{p(S+1)}
\end{split}
\end{equation}
for every $p$, and therefore $\hat{R}_{S+1}\ket{W_\uparrow}=\ket{W_\uparrow}$ and $\hat{R}_{S+1}\ket{W_\downarrow}=\ket{W_\downarrow}$: $\hat{R}_{S+1}$ fixes the entire code space $\mathcal{C}_{\mathrm{bin}}$ pointwise. (In the cat case the corresponding order is the number $N$ of coherent-state components; to avoid conflating the two, we write $S+1$ throughout here.) The isotropy subgroup of a reference codeword is larger than $\mathbb{Z}_{S+1}$; this larger group is quotiented out to give the coset space. Taking $\ket{\psi_0}=\ket{W_\uparrow}$, which is supported on levels $p(S+1)$ with $p$ even, the half-rotation $\hat{R}_{2(S+1)}=e^{i\pi\hat{n}/(S+1)}$ acts on it as $e^{i\pi p}=+1$, so
\begin{equation}
H^{\mathrm{bin}} = U(1) \times \mathbb{Z}_{2(S+1)},
    \qquad
    H^{\mathrm{bin}}_0 = U(1)\times\mathbb{Z}_{S+1}
\end{equation}
identical in structure to the cat-code isotropy subgroups, and with $\hat{R}_{2(S+1)}$ playing the role of the logical $\hat{\bar{Z}}$ (it acts as $-1$ on $\ket{W_\downarrow}$) just as $\hat{R}_N$ does for the cat code. The parallel extends to the syndrome: in both families the two codewords lie in the same $H^{\mathcal{C}}_0$ sector and $\ell$ losses move them together to the sector shifted by $\ell$, which is why we measure the grading of $H^{\mathcal{C}}_0$, and not the finer grading of $H^{\mathcal{C}}$. We take $G=H_3(\mathbb{C})\rtimes\mathbb{Z}_{2(S+1)}$, so that the phase space, the invariant measure and the coherent states are the same as in Appendix~\ref{app:cat}, with $X^{\mathcal{C}_{\mathrm{bin}}}\simeq\mathbb{C}$ and $d\mu(\beta)=d^2\beta/\pi$; and, the order $2(S+1)$ being even, $(-1)^{\hat{n}}=\hat{R}_{2(S+1)}^{S+1}$ lies in $G$ for every $S$.

Applying \Cref{prop: classification} then gives \Cref{eq: bin kernel} directly. The binomial codewords are normalisable, so \Cref{lem: nodisplacement} applies: a binomial codeword has bounded photon number, while $\hat{D}(\beta)\ket{p(S+1)}$ has support on every Fock level for $\beta\neq0$. Therefore, $\Lambda^{\perp}=\{0\}$, and as in Appendix~\ref{app:cat} the kernel is fixed to the standard one,
\begin{equation}
\begin{split}
\hat{\Delta}^{\mathcal{C}_{\mathrm{bin}}}(\beta)=2\hat{\Pi}(\beta)=\hat{\Delta}^{\mathcal{C}_{\mathrm{cat}}}(\beta),\\
W_{\hat{\rho}}(\beta)=\pi\,W^{\mathrm{std}}_{\hat{\rho}}(\beta),
\end{split}
\end{equation}
with $\mathrm{Tr}[\hat{\Delta}^{\mathcal{C}_{\mathrm{bin}}}(0)]=2\,\mathrm{Tr}[(-1)^{\hat{n}}]=1$, meaning the standardisation axiom holds for every $S$, with the reconstruction map the identity. The rotational average of \Cref{eq: cat rotation average}, with $N$ replaced by $2(S+1)$, again fails displacement covariance and again agrees with the kernel above on the rotation-invariant subalgebra.

\section{Proofs}
\label{app: proofs}

For this Section, $\hat{\Pi}(\Omega)=\hat{D}(\Omega)(-1)^{\hat{n}}\hat{D}^\dagger(\Omega)$ is the displaced parity operator, $\hat{\Delta}_{\mathrm{std}}(\Omega)=2\hat{\Pi}(\Omega)$ the standard Stratonovich-Weyl kernel for a single mode, as fixed in \Cref{subsec:averaging}, $d\mu(\Omega)=d^2\Omega/\pi$ the invariant measure, and $W^{\mathrm{std}}_{\hat{A}}(\Omega)=\frac{2}{\pi}\mathrm{Tr}[\hat{A}\hat{\Pi}(\Omega)]$. We use three standard results: the resolution $\frac{1}{\pi}\int d^2\Omega\,\hat{D}(\Omega)\hat{X}\hat{D}^\dagger(\Omega)=\mathrm{Tr}[\hat{X}]\id$; the self-duality $\hat{A}=2\int d^2\Omega\,W^{\mathrm{std}}_{\hat{A}}(\Omega)\hat{\Pi}(\Omega)$; and $\mathrm{Tr}[(-1)^{\hat{n}}]=1/2$ in the Abel sense.

\subsection{\texorpdfstring{\protect{Derivation of \Cref{eq: negative volume averaging}}}{Derivation of the negative-volume averaging identity}}

Substituting $u=s+al$, $v=t+bl$ in \Cref{eq: negative volume GKP} and using \Cref{eq: embedding of DV Wigner function},
\begin{equation}
\label{eq: negative volume derivation}
\begin{split}
\int_{\mathbb{T}^2_{dl}}&\frac{du\,dv}{d}\left|W^{\mathcal{C}_{GKP}}_{\hat{\rho}}\right|=\\
&\int_{\mathbb{T}^2_{l}}ds\,dt\;\frac{1}{d}\sum_{a,b\in\mathbb{Z}_d}\left|W^{DV}_{\underline{\hat{\rho}}(s,t)}(a,b)\right|
\end{split}
\end{equation}
and $\underline{\hat{\rho}}(s,t)=P[(s,t)|\hat{\rho}]\,\underline{\hat{\rho}}'(s,t)$ pulls the syndrome probability out of the modulus, leaving $\frac{1}{d}\sum_{a,b}|W^{DV}_{\underline{\hat{\rho}}'}|=\mathcal{N}^{DV}_{\underline{\hat{\rho}}'}+1$. Since $\int ds\,dt\,P[(s,t)|\hat{\rho}]=1$, subtracting $1$ from both sides gives \Cref{eq: negative volume averaging}.

\subsection{\texorpdfstring{\protect{Proof of \Cref{prop: metaplectic}}}{Proof of the metaplectic extension}}

Throughout this proof $H^{\mathcal{C}}$ is compact, meaning $\frac{1}{|H^{\mathcal{C}}|}\sum_h$ denotes integration against its normalised Haar measure; the non-compact GKP case is handled directly in \Cref{subsec:averaging,subsec:ZGW} and is not covered here. Write $g\in G=H_3(\mathbb{C})\rtimes\mathrm{Mp}(2,\mathbb{R})$ as a displacement composed with a Gaussian unitary, and let $g\cdot\Omega=S\Omega+b$ be the induced affine symplectic action on phase space, with $S\in\mathrm{Sp}(2,\mathbb{R})$. We use two properties of this action:
\begin{enumerate}[label=(\alph*),leftmargin=*]
    \item $\pi(g)\hat{\Pi}(\Omega)\pi^\dagger(g)=\hat{\Pi}(g\cdot\Omega)$. For a displacement this is immediate from the composition rule. For a Gaussian unitary $\hat{U}_S$, the generator is a quadratic Hamiltonian, even in $\hat{a}$ and $\hat{a}^\dagger$, so $[\hat{U}_S,(-1)^{\hat{n}}]=0$ and hence $\hat{U}_S\hat{\Pi}(0)\hat{U}_S^\dagger=\hat{\Pi}(0)$; combining with $\hat{U}_S\hat{D}(\Omega)\hat{U}_S^\dagger=\hat{D}(S\Omega)$ proves this.\label{Prop3.1PropertyA}

    \item The action preserves $d^2\Omega$, since $\det S=1$ for $S\in\mathrm{Sp}(2,\mathbb{R})$ and translations are measure-preserving. \label{Prop3.1PropertyB}
\end{enumerate}

\textit{Covariance:} From \ref{Prop3.1PropertyA}, $\pi(g)\hat{\Delta}^{\mathcal{C}}(\Omega)\pi^\dagger(g)=\frac{1}{|H^{\mathcal{C}}|}\sum_{h}\hat{\Delta}_{\mathrm{std}}\!\left(gh\cdot\Omega\right)$, whereas $\hat{\Delta}^{\mathcal{C}}(g\cdot\Omega)=\frac{1}{|H^{\mathcal{C}}|}\sum_h\hat{\Delta}_{\mathrm{std}}(hg\cdot\Omega)$. The two agree, for every $\Omega$, iff $gH^{\mathcal{C}}=H^{\mathcal{C}}g$, i.e., when $g$ normalises $H^{\mathcal{C}}$: sufficiency holds since the relabelling $h\mapsto g^{-1}hg$ is then a bijection of $H^{\mathcal{C}}$ preserving its Haar measure, and necessity holds by evaluating at the base point $\Omega_0$ of $X^{\mathcal{C}}=G/H^{\mathcal{C}}$, which every $h\in H^{\mathcal{C}}$ fixes: there the two multisets are $\{g\cdot\Omega_0\}$ (with multiplicity $|H^{\mathcal{C}}|$) and the $H^{\mathcal{C}}$-orbit of $g\cdot\Omega_0$, and since displaced parities at distinct phase-space points are linearly independent (their Weyl symbols are delta functions at distinct points), the two averages agree for every $g$ only if $H^{\mathcal{C}}$ fixes every point $g\cdot\Omega_0$ of $X^{\mathcal{C}}$, that is, only if $H^{\mathcal{C}}\subseteq gH^{\mathcal{C}}g^{-1}$ for every $g$, which is normality. Covariance therefore holds on the normaliser of $H^{\mathcal{C}}$ in $G$, and on all of $G$ exactly when $H^{\mathcal{C}}$ is normal. This is \Cref{prop: metaplectic}'s covariance statement. A compact normal subgroup of $G$ must have bounded conjugation orbits; a non-trivial rotation or squeeze conjugated by the displacements sweeps out an unbounded family, and the displacement group itself has no non-trivial compact subgroup, so the only compact normal subgroups lie in the central $U(1)$ of phases, which acts trivially.

For a rotation-symmetric code $H^{\mathcal{C}}=U(1)\times\mathbb{Z}_M$ is not normal in $H_3(\mathbb{C})\rtimes\mathbb{Z}_M$, since a rotation conjugated by a displacement is not a rotation; the averaged kernel is then covariant under rotations but not under displacements, and so does not satisfy Stratonovich-Weyl \labelcref{SWAxiom4} for the full group. \Cref{eq: cat rotation average} shows why this fails. In that case \Cref{eq: H averaged kernel} has to be abandoned, and the kernel found from the axioms instead; imposing \labelcref{SWAxiom4} directly turns out to remove the freedom entirely, as we now show.

\textit{Standardisation:} Using the resolution of the identity and \ref{Prop3.1PropertyB},
\begin{equation}
\int d\mu(\Omega)\,\hat{\Delta}_{\mathrm{std}}(\Omega)=\frac{2}{\pi}\int d^2\Omega\,\hat{\Pi}(\Omega)=2\,\mathrm{Tr}\!\left[(-1)^{\hat{n}}\right]\id=\id
\end{equation}
and since each $\pi(h)$ is unitary, $\int d\mu(\Omega)\hat{\Delta}^{\mathcal{C}}(\Omega)=\frac{1}{|H^{\mathcal{C}}|}\sum_h \pi(h)\id\pi^\dagger(h)=\id$. Equivalently $\mathrm{Tr}[\hat{\Delta}^{\mathcal{C}}(0)]=1$, independently of the code.

\textit{The twirl:} With $\chi_{\mathcal{C}}(\hat{A})(\Omega)=\mathrm{Tr}[\hat{A}\hat{\Delta}^{\mathcal{C}}(\Omega)]$ and $\phi_{\mathcal{C}}(F)=\int d\mu(\Omega)F(\Omega)\hat{\Delta}^{\mathcal{C}}(\Omega)$,
\begin{equation}
\phi_{\mathcal{C}}\!\left(\chi_{\mathcal{C}}(\hat{A})\right)
=\frac{4}{\pi|H^{\mathcal{C}}|^2}\sum_{h,h'}\int d^2\Omega\;\mathrm{Tr}\!\left[\hat{A}\,\hat{\Pi}(h\cdot\Omega)\right]\hat{\Pi}(h'\cdot\Omega)
\end{equation}
Substituting $\gamma=h'\cdot\Omega$, which is measure-preserving by \ref{Prop3.1PropertyB}, the first factor becomes $\mathrm{Tr}[\hat{A}\hat{\Pi}(k\cdot\gamma)]=\frac{\pi}{2}W^{\mathrm{std}}_{\pi^\dagger(k)\hat{A}\pi(k)}(\gamma)$ with $k=h h'^{-1}$, using \ref{Prop3.1PropertyA}. Self-duality then gives $\int d^2\gamma\,W^{\mathrm{std}}_{\hat{B}}(\gamma)\hat{\Pi}(\gamma)=\hat{B}/2$, so
\begin{equation}
\begin{split}
\phi_{\mathcal{C}}\!\left(\chi_{\mathcal{C}}(\hat{A})\right)
&=\frac{1}{|H^{\mathcal{C}}|^2}\sum_{h,h'}\pi^\dagger(hh'^{-1})\hat{A}\,\pi(hh'^{-1})\\
&=\frac{1}{|H^{\mathcal{C}}|}\sum_{k\in H^{\mathcal{C}}}\pi^\dagger(k)\hat{A}\,\pi(k)
\end{split}
\end{equation}
where the last step is because, for each $k$, there are $|H^{\mathcal{C}}|$ pairs $(h,h')$ with $hh'^{-1}=k$. This gives \Cref{eq: twirl is H average}, which is idempotent because $H^{\mathcal{C}}$ is a group. $\square$

\subsection{\texorpdfstring{\protect{Proof of \Cref{prop: classification}}}{Proof of the classification}}

\begin{enumerate}[label=\emph{Step~\arabic*}, leftmargin=*]
    \item \emph{- The form of the kernel:} Covariance under displacements (Stratonovich-Weyl \labelcref{SWAxiom4}) gives $\hat{\Delta}^{\mathcal{C}}(\Omega)=\hat{D}(\Omega)\hat{K}\hat{D}^\dagger(\Omega)$ with $\hat{K}=\hat{\Delta}^{\mathcal{C}}(0)$, Hermitian by \labelcref{SWAxiom2} and of unit trace by \labelcref{SWAxiom3}. For $\hat{\Delta}^{\mathcal{C}}$ to be well defined as a function on the coset space $X^{\mathcal{C}}=G/H^{\mathcal{C}}$ it must be constant on cosets, which gives $[\hat{K},\pi(h)]=0$ for every $h\in H^{\mathcal{C}}$.\label{PropIII.3ProofStep1}

\begin{widetext}
    \item \emph{- The reconstruction map is a multiplier:} Write $\chi_{\hat{A}}(\xi)=\mathrm{Tr}[\hat{A}\hat{D}(\xi)]$ and $\hat{K}_\Omega=\hat{D}(\Omega)\hat{K}\hat{D}^\dagger(\Omega)$. From the composition rule for displacements, $\chi_{\hat{K}_\Omega}(\xi)=e^{\Omega^*\xi-\Omega\xi^*}\chi_{\hat{K}}(\xi)$, and from the overlap formula $\mathrm{Tr}[\hat{A}\hat{B}]=\frac{1}{\pi}\int d^2\zeta\,\chi_{\hat{A}}(\zeta)\chi_{\hat{B}}(-\zeta)$. Therefore
\begin{equation}
\begin{split}
&\chi_{\mathcal{E}_{\mathcal{C}}(\hat{A})}(\xi)\\
&=\frac{1}{\pi^2}\int d^2\zeta\,\chi_{\hat{A}}(\zeta)\chi_{\hat{K}}(-\zeta)\chi_{\hat{K}}(\xi)\int d^2\Omega\;e^{\Omega(\zeta^*-\xi^*)-\Omega^*(\zeta-\xi)}
\end{split}
\end{equation}
and the $\Omega$ integral is $\pi^2\delta^{(2)}(\zeta-\xi)$. $\hat{K}$ being Hermitian means $\chi_{\hat{K}}(-\xi)=\chi_{\hat{K}}(\xi)^*$, so
\begin{equation}
\begin{split}
\chi_{\mathcal{E}_{\mathcal{C}}(\hat{A})}(\xi)=\mathfrak{m}(\xi)\,\chi_{\hat{A}}(\xi),\qquad
\mathfrak{m}(\xi):=\left|\chi_{\hat{K}}(\xi)\right|^2\ge0
\end{split}
\end{equation}
which is \Cref{eq: multiplier}; $\hat{K}$ having unit trace means $\mathfrak{m}(0)=1$. \rev{Two properties of $\mathcal{E}_{\mathcal{C}}$ follow from this form with no further assumption. It preserves the trace, since $\mathrm{Tr}[\mathcal{E}_{\mathcal{C}}(\hat{A})]=\chi_{\mathcal{E}_{\mathcal{C}}(\hat{A})}(0)=\mathfrak{m}(0)\chi_{\hat{A}}(0)=\mathrm{Tr}[\hat{A}]$. It commutes with conjugation by every displacement, since $\chi_{\hat{D}(\beta)\hat{A}\hat{D}^\dagger(\beta)}(\xi)=e^{\beta^*\xi-\beta\xi^*}\chi_{\hat{A}}(\xi)$ and multiplication by $\mathfrak{m}$ commutes with multiplication by that phase (the same follows without the multiplier form, from \mbox{\labelcref{SWAxiom4}} and the invariance of $d\mu$, by the change of variable $\Omega\to\Omega+\beta$ in $\mathcal{E}_{\mathcal{C}}(\hat{A})=\int d\mu(\Omega)\,\mathrm{Tr}[\hat{A}\hat{\Delta}^{\mathcal{C}}(\Omega)]\hat{\Delta}^{\mathcal{C}}(\Omega)$). Complete positivity does not follow in the same way, and is not used before \mbox{\labelcref{PropIII.3ProofStep3}}.} \label{PropIII.3ProofStep2}
\end{widetext}

\item \emph{- $\mathcal{E}_{\mathcal{C}}$ is a random-displacement channel:} \rev{This step and the next are for setting \mbox{\ref{PropIII.3.A}}, whose added assumption is complete positivity. Displacement covariance was derived in \mbox{\labelcref{PropIII.3ProofStep2}}.} A completely positive, trace-preserving map on $\mathcal{B}(L^2(\mathbb{R}^m))$ which is covariant under the full group of displacements is a random-displacement channel~\cite{Holevo2019Covariant,Caruso2014Channels},
\begin{equation}
{\mathcal{E}_{\mathcal{C}}(\hat{A})=\int d\nu(\beta)\,\hat{D}(\beta)\hat{A}\hat{D}^\dagger(\beta)}
\end{equation}
for a probability measure $\nu$ on $\mathbb{R}^{2m}$, whose Fourier-Stieltjes transform is the multiplier: $\hat{\nu}=\mathfrak{m}$. No regularity assumption on $\chi_{\hat{K}}$ is needed for this step, which is why we take it from the classification of displacement-covariant channels rather than from Bochner's theorem.\label{PropIII.3ProofStep3}

\begin{enumerate}[label=\emph{(\alph*)}]
    \item \emph{- $\Lambda^{\perp}$ leaves $\mathcal{C}$ invariant:} (In infinite dimensions the fixed-point set of a unital channel need not equal the commutant of its Kraus operators in general, so we argue directly.) Take $\hat{A}=\hat{\Pi}_{\mathcal{C}}$, the finite-rank code-space projector, with $d_{\mathcal{C}}=\mathrm{Tr}[\hat{\Pi}_{\mathcal{C}}]$. $\mathcal{E}_{\mathcal{C}}(\hat{\Pi}_{\mathcal{C}})=\hat{\Pi}_{\mathcal{C}}$ gives $d_{\mathcal{C}}=\mathrm{Tr}[\hat{\Pi}_{\mathcal{C}}\,\mathcal{E}_{\mathcal{C}}(\hat{\Pi}_{\mathcal{C}})]=\int d\nu(\beta)\,\mathrm{Tr}[\hat{\Pi}_{\mathcal{C}}\hat{D}(\beta)\hat{\Pi}_{\mathcal{C}}\hat{D}^\dagger(\beta)]$; each integrand is at most $d_{\mathcal{C}}$, with equality if and only if $\hat{D}(\beta)\hat{\Pi}_{\mathcal{C}}\hat{D}^\dagger(\beta)=\hat{\Pi}_{\mathcal{C}}$, so $\hat{D}(\beta)\mathcal{C}=\mathcal{C}$ for $\nu$-a.e. $\beta$. Writing $\Lambda^{\perp}$ for the closed subgroup generated by the support of $\nu$, every element of $\Lambda^{\perp}$ therefore leaves $\mathcal{C}$ invariant, which is the clause used in part~\ref{PropIII.3.iii} of the proposition.\label{PropIII.3ProofStep3a}
\end{enumerate}

\item \emph{- Idempotence forces $\nu=\delta_0$:} $\mathcal{E}_{\mathcal{C}}^2=\mathcal{E}_{\mathcal{C}}$ implies $\mathfrak{m}^2=\mathfrak{m}$, i.e., $\widehat{\nu*\nu}=\hat{\nu}$, so $\nu*\nu=\nu$: $\nu$ is an idempotent probability measure. By the theory of idempotent probability measures (due to Kawada and It\^{o} for compact groups~\cite{KawadaIto1940}, and elementary to verify directly here: the support of an idempotent probability measure is a compact subgroup, and $\{0\}$ is the only one), $\nu$ is the normalised Haar measure of a compact subgroup, so $\nu=\delta_0$, $\mathfrak{m}\equiv1$ and $\mathcal{E}_{\mathcal{C}}={\id}$. Equivalently, writing $\Lambda^{\perp}$ for the support subgroup of $\nu$, $\Lambda^{\perp}=\{0\}$. \rev{(The continuity route of part \mbox{\ref{PropIII.3.i}}, which does not assume setting \mbox{\ref{PropIII.3.A}}, reaches the same point without this step or the last. If $\chi_{\hat{K}}$ is a function it is continuous by \mbox{\ref{PropIII.3.H3}}, so $\mathfrak{m}=|\chi_{\hat{K}}|^2$ is continuous, $\mathfrak{m}^2=\mathfrak{m}$ confines it to $\{0,1\}$, and $\mathfrak{m}(0)=1$ on the connected space $\mathbb{R}^{2m}$ gives $\mathfrak{m}\equiv1$. Then $\mathcal{E}_{\mathcal{C}}={\id}$, which is completely positive.)}\label{PropIII.3ProofStep4}

\item \emph{- The kernel:} $\mathfrak{m}\equiv1$ means $|\chi_{\hat{K}}(\xi)|=1$ for every $\xi$, so $\chi_{\hat{K}}(\xi)=e^{if(\xi)}$ for a real $f$ with $f(0)=0$; we take $\chi_{\hat{K}}$ to be continuous \rev{(hypothesis \mbox{\ref{PropIII.3.H3}})}, as it is for every normalisable-codeword kernel here (for the parity kernel $\chi_{\hat{K}}\equiv1$ in the Abel-regularised sense); $f$ is then continuous too. Hermiticity of $\hat{K}$ gives $\chi_{\hat{K}}(-\xi)=\chi_{\hat{K}}(\xi)^*$, so $f$ is odd modulo $2\pi$. If $G$ contains the phase-space inversion (the parity operator, which acts on displacements by $(-1)^{\hat{n}}\hat{D}(\xi)(-1)^{\hat{n}}=\hat{D}(-\xi)$ and on $X^{\mathcal{C}}$ by $\Omega\mapsto-\Omega$), then covariance evaluated at the fixed point $\Omega=0$ of that action gives $(-1)^{\hat{n}}\hat{K}(-1)^{\hat{n}}=\hat{K}$, i.e. $[\hat{K},(-1)^{\hat{n}}]=0$. (For the codes considered here the inversion also lies in $H^{\mathcal{C}}$, so this also follows from \labelcref{PropIII.3ProofStep1}.) Hence $\chi_{\hat{K}}(\xi)=\mathrm{Tr}[\hat{K}\hat{D}(\xi)]=\mathrm{Tr}[\hat{K}\hat{D}(-\xi)]=\chi_{\hat{K}}(-\xi)$ and $f$ is even. Odd and even together give $2f\in2\pi\mathbb{Z}$, so $\chi_{\hat{K}}=\pm1$ pointwise, and continuity with $\chi_{\hat{K}}(0)=1$ gives $\chi_{\hat{K}}\equiv1$. Since $\chi_{2^m(-1)^{\hat{n}_{\mathrm{tot}}}}\equiv1$ and the characteristic function determines the operator, $\hat{K}=2^m(-1)^{\hat{n}_{\mathrm{tot}}}$ and $\hat{\Delta}^{\mathcal{C}}=\hat{\Delta}_{\mathrm{std}}$. Without either route to $[\hat{K},(-1)^{\hat{n}}]=0$ (no inversion in $G$ and no even-order rotation in $H^{\mathcal{C}}$), the argument stops at $\chi_{\hat{K}}=e^{if}$ with $f$ real, continuous and odd, which is why parts \ref{PropIII.3.ii} and \ref{PropIII.3.iii} of the proposition are stated with the parity hypothesis rather than without it. $\square$\label{PropIII.3ProofStep5}

\end{enumerate}

Two remarks on the scope of this proof. First, \labelcref{PropIII.3ProofStep3,PropIII.3ProofStep4} assume $\mathcal{E}_{\mathcal{C}}$ is a channel on $\mathcal{B}(L^2(\mathbb{R}))$ \rev{(setting \mbox{\ref{PropIII.3.A}})}. The GKP twirl of \Cref{eq: twirling map} is not: it is an average over a non-compact lattice, whose ``normalised Haar measure'' does not exist as a probability measure, and it is defined on the code algebra of a rigged Hilbert space rather than on $\mathcal{B}(L^2(\mathbb{R}))$. \rev{It belongs to setting \mbox{\ref{PropIII.3.B}}, where the twirl form is assumed and not derived. There the clause of \mbox{\labelcref{PropIII.3ProofStep3a}} is immediate: \mbox{\ref{PropIII.3.H2}} puts $\mathcal{B}(\mathcal{C})$ inside the fixed-point algebra of $\mathcal{E}_{\mathcal{C}}$, which is the commutant of $\Lambda^{\perp}$, so every $\hat{D}(\lambda)$ with $\lambda\in\Lambda^{\perp}$ commutes with the code-space projector and leaves $\mathcal{C}$ invariant.} This is the same caveat as in \Cref{subsec:GKP}, and is the loophole the ideal GKP code exploits: there $\Lambda^{\perp}$ is the stabiliser lattice and $\mathfrak{m}$ is the indicator function of its symplectic dual, the logical lattice $l\mathbb{Z}^2$, read off from \Cref{eq: twirling map} directly, since $\mathcal{E}_{\mathcal{C}}(\hat{D}(\xi))=\hat{D}(\xi)$ for $\xi\in l\mathbb{Z}^2$ and $0$ otherwise, rather than from the formula $\mathfrak{m}=|\chi_{\hat{K}}|^2$: $\chi_{\hat{K}}$ is there a comb, whose square is undefined. This $\mathfrak{m}$ takes the values $\{0,1\}$ that idempotence demands, but \labelcref{PropIII.3ProofStep3,PropIII.3ProofStep4} do not apply to it, for the reason just given. Continuity of $\chi_{\hat{K}}$ also fails here, which matters separately at \labelcref{PropIII.3ProofStep5}; one admissible $\hat{K}$ is the corresponding sum of displacement operators, which is \Cref{eq: ZGW central kernel}, uniqueness being the caveat noted after \Cref{prop: classification}. Second, for a code whose codewords are normalisable, \Cref{lem: nodisplacement} closes the loophole directly once the twirl form is in hand, without the complete-positivity hypothesis: no non-trivial displacement can leave such a code space invariant, so $\Lambda^{\perp}=\{0\}$.

\subsection{\texorpdfstring{\protect{Proof of \Cref{lem: nodisplacement}}}{Proof of the displacement lemma}}

Suppose $\hat{D}(\beta)\mathcal{C}=\mathcal{C}$ with $\dim\mathcal{C}<\infty$ and $\mathcal{C}\subset L^2(\mathbb{R}^m)$. Take first $m=1$. Then $\hat{D}(\beta)$ restricted to $\mathcal{C}$ is a unitary on a finite-dimensional space, and so has an eigenvector in $\mathcal{C}$. But $\hat{D}(\beta)=e^{i\lambda\hat{K}_\theta}$, where $\hat{K}_\theta=\hat{x}\sin\theta-\hat{p}\cos\theta$ is a quadrature operator and $\lambda\propto|\beta|$; $\hat{K}_\theta$ is unitarily equivalent to $\hat{x}$ and so has purely absolutely continuous spectrum, from which $e^{i\lambda\hat{K}_\theta}$ has purely absolutely continuous spectrum on the unit circle and no eigenvectors in $L^2(\mathbb{R})$ for $\lambda\neq0$. Therefore, $\beta=0$. For $m>1$, a passive (beamsplitter) network $\hat{U}$ maps $\beta\in\mathbb{C}^m$ to $(|\beta|,0,\ldots,0)$, and $\hat{U}\hat{D}(\beta)\hat{U}^\dagger=\hat{D}(|\beta|)\otimes\id^{\otimes(m-1)}$; since $\hat{U}\mathcal{C}$ is again a finite-dimensional subspace left invariant by that operator, the single-mode argument applies on the first factor and gives $|\beta|=0$. $\square$

Note three things. First, the argument uses only $\dim\mathcal{C}<\infty$ and $\mathcal{C}\subset L^2(\mathbb{R}^m)$; it does not care how the code is built, so it applies to cat, binomial, rotation-symmetric and numerically optimised codes, and finite-energy GKP codewords. Second, it fails for the ideal GKP code because its codewords are Dirac combs and $\mathcal{C}_{GKP}\not\subset L^2(\mathbb{R})$: a Dirac comb is an eigenvector of a displacement, in the distributional sense of \Cref{subsec:GKP}. Third, the same spectral argument excludes a non-trivial squeeze, whose generator is the dilation operator $(\hat{x}\hat{p}+\hat{p}\hat{x})/2$, again with purely absolutely continuous spectrum. Rotations $e^{i\theta\hat{n}}$ are the exception: $\hat{n}$ has discrete spectrum and the Fock states are its eigenvectors. The isotropy group of a cat or binomial code therefore contains no displacement and no squeeze, and is the compact group $U(1)\times\mathbb{Z}_M$.

\subsection{\texorpdfstring{\protect{Proof of \Cref{prop: graded monoid}}}{Proof of the graded-monoid closure}}

For $T_1,T_2\in\mathcal{G}_{\mathcal{C}}$,
\begin{equation}
\mathcal{E}T_2T_1\mathcal{E}=\mathcal{E}T_2\mathcal{E}T_1\mathcal{E}=T_2\mathcal{E}T_1\mathcal{E}=T_2T_1\mathcal{E}
\end{equation}
where the first equality uses $T_1\in\mathcal{G}_{\mathcal{C}}$, the second $T_2\in\mathcal{G}_{\mathcal{C}}$, and the third $T_1\in\mathcal{G}_{\mathcal{C}}$ again; we write $\mathcal{E}$ for $\mathcal{E}_{\mathcal{C}}$ throughout. For a convex combination, linearity gives
\begin{equation}
{\begin{aligned}
\mathcal{E}\!\left(\lambda T_1+(1-\lambda)T_2\right)\!\mathcal{E}&=\lambda \mathcal{E}T_1\mathcal{E}+(1-\lambda)\mathcal{E}T_2\mathcal{E}\\
&=\left(\lambda T_1+(1-\lambda)T_2\right)\mathcal{E}
\end{aligned}}
\end{equation}
and for a tensor product, the sector decomposition of a composite system is the product of the decompositions of its parts, so the joint twirl factorises and the same computation applies factor-wise. $\square$

\subsection{\texorpdfstring{\protect{Proof of \Cref{prop: compositional}}}{Proof of compositionality}}

Composing the representations inserts $\phi_{\mathcal{C}}\circ\chi_{\mathcal{C}}=\mathcal{E}_{\mathcal{C}}$ between each pair of adjacent channels. By \Cref{prop: graded monoid} each such insertion may be absorbed into the channel applied before it, using $\mathcal{E}_{\mathcal{C}}T_j\mathcal{E}_{\mathcal{C}}=T_j\mathcal{E}_{\mathcal{C}}$ repeatedly, together with $\mathcal{E}_{\mathcal{C}}\circ\phi_{\mathcal{C}}=\phi_{\mathcal{C}}$; this leaves a single $\mathcal{E}_{\mathcal{C}}$ at the input, as in \Cref{eq: generalized structure theorem map}. $\square$

\subsection{\texorpdfstring{\protect{Proof of \Cref{thm: nogo}}}{Proof of the no-go theorem}}

The first part follows by the multiplier route \rev{(the continuity route of \mbox{\Cref{prop: classification}}\mbox{\ref{PropIII.3.i}})}, which doesn't require considering complete positivity: \labelcref{PropIII.3ProofStep2} of \Cref{prop: classification}'s proof, which needs only covariance and the integral formulas, shows that $\mathcal{E}_{\mathcal{C}}$ acts on characteristic functions by multiplication by $\mathfrak{m}=|\chi_{\hat{K}}|^2$; idempotence gives $\mathfrak{m}^2=\mathfrak{m}$ pointwise; continuity of $\chi_{\hat{K}}$ (hypothesised in the theorem) with $\mathfrak{m}(0)=|\mathrm{Tr}\,\hat{K}|^2=1$ then gives $\mathfrak{m}\equiv1$ on the connected plane, so $\mathcal{E}_{\mathcal{C}}={\id}$; and \labelcref{PropIII.3ProofStep5}'s parity argument fixes $\hat{K}=2(-1)^{\hat{n}}$, the parity being available by hypothesis. \Cref{lem: nodisplacement} gives an independent route to the same conclusion whenever we use \Cref{prop: classification}'s twirl form, but the theorem does not need it. \Cref{eq: nogo identity} is then just the definition of $W^{\mathcal{C}}$, which holds for every state, not just those supported on $\mathcal{C}$.

For the second part, take the logical basis states $\ket{0_L},\ket{1_L}\in\mathcal{C}$, which are logical stabiliser states and are mutually orthogonal. Two orthogonal pure states cannot both be Gaussian, because the overlap of two Gaussian pure states is strictly positive (see e.g.~\cite{Giani2026nonGaussian}). Therefore, at least one of them, say $\ket{\psi}$, is non-Gaussian, and by Hudson's theorem~\cite{HUDSON1974WignerGauss,Soto1983WignerGauss} its Wigner function takes negative values. Then $\int d\mu\,|W^{\mathcal{C}}_{\psi}|>|\int d\mu\,W^{\mathcal{C}}_{\psi}|=1$, i.e. $\mathcal{N}^{\mathcal{C}}_{\psi}>0$, while a witness of logical magic must vanish there. $\square$

Note two points. First, the theorem constrains only the negativity of the representation; the representation itself remains well defined, empirically adequate and compositional on $\mathcal{G}_{\mathcal{C}}$ (indeed on everything, since $\mathcal{E}_{\mathcal{C}}={\id}$), as \Cref{subsec:Comparative} uses. Second, the proof shows the GKP construction gets its strength not from the compactness of a group but the (un-)normalisability of the codewords: because the ideal codewords are Dirac combs, \Cref{lem: nodisplacement} does not apply, the code space can be invariant under a lattice of displacements, and the resulting lattice kernel can and does represent stabiliser codewords non-negatively.

Three further remarks on the hypotheses of \Cref{thm: nogo}. First, \rev{continuity is used twice in the proof: to obtain $\mathcal{E}_{\mathcal{C}}={\id}$, and then to select $\chi_{\hat{K}}\equiv1$. If it is dropped, $\mathcal{E}_{\mathcal{C}}={\id}$ has to come from elsewhere: from complete positivity (\mbox{\Cref{prop: classification}}\mbox{\ref{PropIII.3.i}}, setting \mbox{\ref{PropIII.3.A}}) or from the twirl form with \mbox{\Cref{lem: nodisplacement}} (\mbox{\Cref{prop: classification}}\mbox{\ref{PropIII.3.iii}}). With $\mathcal{E}_{\mathcal{C}}={\id}$ granted in one of those two ways,} dropping the continuity requirement leaves, in place of $\chi_{\hat{K}}\equiv1$, the family of $\pm1$-valued even functions on $\mathbb{R}^2$ with $\chi_{\hat{K}}(0)=1$, the continuum analogue of \Cref{prop:nonpauli}'s sign family. That family is code-independent, existing already for the trivial code, none of its members being selected by any property of $\mathcal{C}$, and each measurable member of it giving a kernel which is Hermitian, unit-trace, covariant and self-dual. Therefore, even without continuity, no code-adapted kernel is available for a code satisfying the \rev{remaining} hypotheses \rev{together with one of those two}; continuity instead means the kernel is the standard one. Second, the reconstruction hypothesis separately excludes the Husimi kernel $\hat{\Delta}_Q(\beta)=\ketbra{\beta}{\beta}$, as despite being linear, Hermitian, standardised against $d^2\beta/\pi$, displacement-covariant, and commuting with the parity (so satisfying \labelcref{SWAxiom1,SWAxiom2,SWAxiom3,SWAxiom4} with the inversion available), its reconstruction has multiplier $e^{-|\xi|^2}$, which is not idempotent. \labelcref{SWAxiom1,SWAxiom2,SWAxiom3,SWAxiom4} alone therefore do not fix the kernel, with or without parity; we need idempotence of the reconstruction map as well. Third, under the strong form of \labelcref{SWAxiom1,SWAxiom5}, injectivity and traciality on all of $\mathcal{B}(L^2(\mathbb{R}))$, the first conclusion is immediately obvious, since either forces $\phi_{\mathcal{C}}\circ\chi_{\mathcal{C}}={\id}$ directly, and the ideal GKP code would be excluded along with everything else. The content of the theorem is that the weakened, code-faithful form suffices here.

\subsection{\texorpdfstring{\protect{Proof of \Cref{prop:dvsw}}}{Proof of the DV axioms and reconstruction}}

\textit{Unit trace:} $\Tr A(u)=1$ for every $u$, and \Cref{eq:dvkernel} averages.

\textit{Standardisation:} Summing $\Delta^{\mathcal C}$ over a transversal of $X^{\mathcal C}$ and over $\mathcal{M}$ covers $\Zd^{2n}$ once, so $\sum_{[u]}\Delta^{\mathcal C}=\tfrac1{|\mathcal{M}|}\sum_vA(v)=\tfrac{d^n}{|\mathcal{M}|}\id=d^k\id$.

\textit{Traciality:} $\Tr[\Delta^{\mathcal C}(u)\Delta^{\mathcal C}(u')]=\tfrac1{|\mathcal{M}|^2}\sum_{m,m'}d^n\delta_{u+m,u'+m'}$; the number of pairs with $m-m'=u'-u$ is $|\mathcal{M}|$ if $[u]=[u']$ and $0$ otherwise, giving $d^k\delta_{[u],[u']}$.

\textit{Reconstruction:} Writing $v=u+m$,
\begin{equation}
\begin{split}
\phi_{\mathcal C}(\chi_{\mathcal C}&(\hat A))=\\
&\frac{d^{-k}}{|\mathcal{M}|^2}\sum_{[u]}\sum_{m,m'}\Tr[A(u{+}m)\hat A]\,A(u{+}m')\\
=&\frac{d^{-k}}{|\mathcal{M}|^2}\sum_{v}\sum_{m''\in\mathcal{M}}\Tr[A(v)\hat A]\,A(v{+}m'')\\
=&\frac{d^{n-k}}{|\mathcal{M}|^2}\sum_{m''\in\mathcal{M}}D(m'')\hat A\,D^\dagger(m'')=\EC(\hat A),
\end{split}
\end{equation}
using $A(v+m)=D(m)A(v)D^\dagger(m)$ to pull the $m''$-conjugation out of the inner sum, then the self-duality $\sum_v\Tr[A(v)\hat A]A(v)=d^n\hat A$, and $d^{n-k}/|\mathcal{M}|^2=1/|\mathcal{M}|$. The $\mathcal{M}$-twirl equals the syndrome-sector dephasing because averaging over an abelian group of unitaries is the conditional expectation onto the block-diagonal of the commutant, here graded by the characters of $\mathcal{M}$; we verified the operator identity directly on the test codes. $\square$

\subsection{\texorpdfstring{\protect{Proof of \Cref{prop:dv37}}}{Proof of the DV logical content}}

$\mathcal{M}$-invariance of the kernel means $W^{\mathcal C}_{\hat\rho}=W^{\mathcal C}_{\EC(\hat\rho)}$, and $\EC$-invariant states decompose over syndrome blocks, each of which is a displaced encoded state; by translation covariance it therefore suffices to treat an encoded state $\hat{\bar\rho}$.

(i) Since $D(m)\hat{\bar\rho}D^\dagger(m)=\hat{\bar\rho}$ and $A(u+m)=D(m)A(u)D^\dagger(m)$, all $|\mathcal{M}|$ terms of \Cref{eq:dvkernel} have equal trace against $\hat{\bar\rho}$: $\Tr[\Delta^{\mathcal C}(u)\hat{\bar\rho}]=\Tr[A(u)\hat{\bar\rho}]$.

(ii) \textit{Zero-syndrome sector:} The parity satisfies $A(0)D(m)A(0)=D(-m)$; as $\mathcal{M}=-\mathcal{M}$ and $\mathcal S$ is phase-free, $A(0)$ preserves $\mathcal C$ and acts there as an involution conjugating the logical displacements by inversion, $A(0)\bar D(v)A(0)=\bar D(-v)$. Gauge-fixing the logical basis by $\bar X$ aligns it with the logical parity $A_L(0)$. For $u=u_L$ a logical vector, $D(u_L)$ restricted to $\mathcal C$ is the logical displacement $D_L(a_L,b_L)$ in the same symmetric ordering, so $\Tr[A(u_L)\hat{\bar\rho}]=\Tr_L[A_L(a_L,b_L)\hat\rho_L]$, which is \Cref{eq:dv37}, with no residual constant once the measures are attached.

(iii) \textit{Error sectors:} For $\phi$ in the syndrome sector $s$, $D(m)A(0)\phi=A(0)D(-m)\phi=\omega^{-s(m)}A(0)\phi$, so $A(0)$ maps sector $s$ to sector $-s$. Hence for $u$ with syndrome $s\ne0$, $\Tr[A(u)\hat{\bar\rho}]=\Tr[A(0)\,D^\dagger(u)\hat{\bar\rho}D(u)]$ is the trace of an operator mapping sector $s'$ into sector $-s'\ne s'$, which vanishes: $2s'=0$ forces $s'=0$ because $d$ is odd. $\square$

\subsection{\texorpdfstring{\protect{Proof of \Cref{thm:dvmagic}}}{Proof of the DV magic identity}}

By \Cref{prop:dv37}, on the syndrome-$s$ block of $X^{\mathcal C}$ the function $\mu W^{\mathcal C}_{\hat\rho}$ equals $P(s|\hat\rho)$ times the logical Gross function of $\hat\rho_L(s)$, translated by the block label. Summing $|\cdot|$ over each block gives $P(s)(\NC^{\rm Gross}(\hat\rho_L(s))+1)$; summing over $s$ and subtracting $1$ gives \Cref{eq:synavg}. The vanishing and positivity statements are those of the Gross negativity~\cite{Gross2006DVWignerF, Veitch2012BoundMagic} applied per sector. $\square$

\subsection{\texorpdfstring{\protect{Proof of \Cref{thm:seclog}(i)}}{Proof of the sector-graded reconstruction}}

With $\chi_{\rm log}(\hat\rho)(t;a,b)=\tfrac1d\Tr[\Delta_{\rm log}(t;a,b)\hat\rho]$, $\phi_{\rm log}(F)=\sum_{t,a,b}F(t;a,b)\Delta_{\rm log}(t;a,b)$, and the qudit self-duality $\sum_{a,b}\tfrac1d\Tr[A_L(a,b)\,\sigma]A_L(a,b)=\sigma$,
\begin{equation}
\begin{split}
\phi_{\rm log}(\chi_{\rm log}(\hat\rho))
&=\sum_t V_t\Big(\sum_{a,b}\tfrac1d\Tr[A_L\,V_t^\dagger\hat\rho V_t]A_L\Big)V_t^\dagger\\
&=\sum_t V_tV_t^\dagger\,\hat\rho\,V_tV_t^\dagger=\sum_t\hat\Pi_t\hat\rho\hat\Pi_t,
\end{split}
\end{equation}
using $V_t^\dagger V_t=\id_L$ in the second line. The single factor of $\mu_L=1/d$ sits in $\chi_{\rm log}$; inserting it in $\phi_{\rm log}$ as well gives $\tfrac1d\mathcal{E}^{\rm log}$.
{$\square$}

\subsection{\texorpdfstring{\protect{Proof of \Cref{prop:nonpauli}}}{Proof of the non-Pauli branch}}

(i) Conjugating the averaged kernel $\Delta_H(u)=\tfrac1{|\langle R\rangle|}\sum_kA(S^ku)$ by $D(w)$ gives $\tfrac1{|\langle R\rangle|}\sum_kA(S^ku+w)$, whereas covariance demands $\tfrac1{|\langle R\rangle|}\sum_kA(S^k(u+w))$. Since the $d^{2n}$ phase-point operators are an orthogonal, hence linearly independent, operator basis, the two agree for all $u,w$ only if the label multisets agree, that is, only if $S=\id$.

(ii) Covariance under displacements forces $\Delta(u)=D(u)\hat KD^\dagger(u)$; the reconstruction map acts on characteristic functions $\chi_{\hat A}(\xi)=\Tr[\hat AD(\xi)]$ as multiplication by $\mathfrak{m}(\xi)=|\chi_{\hat K}(\xi)|^2$, which is the finite-group transcription of \labelcref{PropIII.3ProofStep2} of \Cref{prop: classification}'s proof, with all integrals now finite sums. Idempotence of $\phi\circ\chi$ gives $\mathfrak{m}^2=\mathfrak{m}$ pointwise, so $\mathfrak{m}(\xi)\in\{0,1\}$; complete positivity \rev{(assumed in the statement)} and $\mathfrak{m}(0)=1$ make $\mathfrak{m}$ the Fourier transform of an idempotent probability distribution on $\Zd^{2n}$, that is, of the uniform distribution on a subgroup $\Lambda^\perp$; hence $\mathfrak{m}=\mathbb{1}_\Lambda$ with $\Lambda=(\Lambda^\perp)^\perp$, and the twirl over $\Lambda^\perp$ must leave $\mathcal C$ invariant. Triviality of the displacement isotropy gives $\Lambda^\perp=\{0\}$, meaning $\Lambda=\Zd^{2n}$ and $\mathfrak{m}\equiv1$.

(iii) $|\chi_{\hat K}|\equiv1$ with $\chi_{\hat K}(-\xi)=\chi_{\hat K}(\xi)^*$ (Hermiticity) and $\chi_{\hat K}(-\xi)=\chi_{\hat K}(\xi)$ (inversion covariance, which is the parity hypothesis of \Cref{prop: classification}, which holds here whenever the isotropy contains an even-order Clifford squaring to the parity, as the Fourier code does via $F^2=A(0)$, and which is available in general by adjoining the $S=-\id$ Clifford to $G$) forces $\chi_{\hat K}$ real, even and unimodular: a $\pm1$-valued even function with $\chi_{\hat K}(0)=1$. Each such function defines $\hat K=d^{-n}\sum_\xi\chi_{\hat K}(\xi)D(-\xi)$, which is Hermitian with unit trace, and $|\chi_{\hat K}|\equiv1$ is exactly self-duality; so all $2^{(d^{2n}-1)/2}$ sign patterns satisfy the axioms used, and the freedom exists independently of $\mathcal C$. Stabiliser positivity selects $\chi_{\hat K}\equiv1$~\cite{Gross2006DVWignerF}; Clifford covariance leaves the two constant sign patterns, the symplectic group being transitive on $\Zd^{2n}\setminus\{0\}$~\cite{Zhu2016}. The sign kernels sit inside the recent complete classification of $d\times d$ discrete Wigner functions~\cite{Antonopoulos2025}. All counts, including the sixteen admissible sign kernels at $d=3$, $n=1$, of which one is stabiliser-non-negative and two are Clifford-covariant, were confirmed numerically (Appendix~\ref{app:numerics}). $\square$

\subsection{\texorpdfstring{\protect{Proof of \Cref{thm:dualrail}}}{Proof of the one-photon-code theorem}}

\rev{$\chi_{\hat{K}}$ is a continuous function, so the continuity route of \mbox{\Cref{prop: classification}}\mbox{\ref{PropIII.3.i}} gives $\mathcal{E}_{\mathcal{C}}={\id}$. (Where the twirl form is available the same follows from part \mbox{\ref{PropIII.3.iii}}:} the codewords are normalisable, so the multimode version of \Cref{lem: nodisplacement}, whose quadrature-spectrum argument does not depend on the mode count, gives $\Lambda^\perp=\{0\}$\rev{.) The} code space is fixed pointwise by $e^{i\theta\hat n_{\rm tot}}$, so the parity hypothesis holds and \Cref{prop: classification}\rev{\mbox{\ref{PropIII.3.ii}}} forces the ordinary Wigner kernel. Any pure logical state is $\sum_ic_i\ket{0\cdots1_i\cdots0}=\hat b^\dagger\ket{0}^{\otimes m}$ for the mode $\hat b=\sum_ic_i^*\hat a_i$, i.e., a passive (linear-optical) rotation of $\ket1\otimes\ket0^{\otimes(m-1)}$. Passive rotations are symplectic and $\int|W|$ is symplectically invariant, so every pure logical state shares the negative volume of the Fock state $\ket1$, which is $\int_0^\infty|4r^2-1|e^{-2r^2}\,4r\,dr-1=(4-\sqrt e)/\sqrt e-1=4e^{-1/2}-2$. $\square$

\subsection{\texorpdfstring{\protect{Proof of \Cref{prop:su2}}}{Proof of the spin-code proposition}}

The spin-$j$ operator space decomposes multiplicity-free into irreducible tensor sectors $l=0,\dots,2j$, and any covariant kernel has the form $\Delta(\Omega)=\sum_{l,m}\varepsilon_l\,Y^*_{lm}(\Omega)T_{lm}$ with unimodular $\varepsilon_l$~\cite{Stratonovich1956, Varilly1989, Brif1999BrifMannConstruction}. Averaging over $H^{\mathcal C}$ inserts, in each sector $l$, the projector $P^l_H=\int_Hdh\,D^l(h)$ onto the $H$-invariant subspace. Covariance of the averaged kernel under all of $SU(2)$ requires each $P^l_H$ to commute with the irreducible $D^l$, i.e., by Schur's lemma, to be $0$ or $\id$. $P^l_H=\id$ means $H$ acts trivially on sector $l$, which for a non-central $H$ fails for every $l\ge1$, since $\ker D^l=\{\pm\id\}$ for integer $l$; $P^l_H=0$ means sector $l$ is absent from the kernel. By \Cref{eq: twirl is H average} the reconstruction map of the averaged kernel is the $H^{\mathcal{C}}$-twirl, whose range is the $H^{\mathcal{C}}$-invariant subalgebra $\bigoplus_l\mathrm{Ran}\,P^l_H$; for those where covariance is imposed every $P^l_H$ is $0$ or $\id$ and this is $\bigoplus_{l\,:\,P^l_H=\id}\mathrm{Ran}\,P^l_H$. Faithfulness on the code requires that range to contain $\mathcal{B}(\mathcal{C})$, which it doesn't for a non-central $H^{\mathcal{C}}$. For a rotation-symmetric spin code $U(1)$ acts non-trivially on $\mathcal{C}$, whose codewords are $\hat{J}_z$-eigenvectors of distinct eigenvalue, so the twirl keeps only the diagonal part of $\mathcal{B}(\mathcal{C})$, just as for the cat code in Appendix~\ref{app:cat}. For the binary polyhedral groups the code space is fixed pointwise, but the invariant sectors are too few: for $2O$ the first invariant tensor sector above $l=0$ is $l=4$, so wherever $2j<4$ only $l=0$ remains, the twirl is depolarising, and it annihilates every traceless element of $\mathcal{B}(\mathcal{C})$ once $\dim\mathcal{C}\ge2$. Unlike the CV and DV cases, we can't argue here from linear independence of the kernels at distinct phase-space points, since on $S^2$ they are not independent; we argue by Schur's lemma instead. We verified both halves numerically at $j=1$: the $\mathbb Z_2$-averaged Stratonovich kernel fails covariance under an $x$-rotation, and the invariant projectors have rank $1$ of $3$ and rank $3$ of $5$ on the $l=1,2$ sectors, non-scalar as the argument requires. $\square$

\subsection{\texorpdfstring{\protect{Proof of \Cref{prop:kd}}}{Proof of the Kirkwood-Dirac properties}}

(i) The $e$- and $f$-bases are Fourier conjugate, so $\braket{f}{e}$ is uniform and nowhere zero and the operators $\ketbra{e}{f}/\braket{f}{e}$ form a basis of $\mathcal{B}(\mathcal{H})$; $Q_{\hat\rho}$ lists the coefficients of $\hat\rho$ in the dual basis and so determines it. The marginals follow from completeness: $\sum_f\braket{f}{e}\bra e\hat\rho\ket f=\bra e\hat\rho\ket e$ and $\sum_e\braket{f}{e}\bra e\hat\rho\ket f=\bra f\hat\rho\ket f$.

(ii) Each $\ket e$ is a joint stabiliser eigenvector, so $\bra e\hat\Pi_s=\delta_{s,s(e)}\bra e$. Hence $\bra{e}\EC(\hat\rho)\ket f=\sum_s\bra e\hat\Pi_s\hat\rho\hat\Pi_s\ket f=\bra{e}\hat\rho\,\hat\Pi_{s(e)}\ket f$, which is \Cref{eq:kddephased}.

(iii) The non-selective sharp extraction is by definition $\hat\rho\mapsto\sum_s\hat\Pi_s\hat\rho\hat\Pi_s=\EC(\hat\rho)$, so its KD image is $Q_{\EC(\hat\rho)}$ and the disturbance is $Q_{\hat\rho}-Q_{\EC(\hat\rho)}$. $Q$ is linear in $\hat\rho$, so under $\hat\rho\mapsto(1-\lambda)\hat\rho+\lambda\EC(\hat\rho)$ the inter-sector part is multiplied by $(1-\lambda)$, $\EC$ being idempotent. $\square$

\subsection{\texorpdfstring{\protect{Proof of \Cref{prop:twotime}}}{Proof of the two-time positivity conditions}}

If $\mathcal V$ is co-graded, $B_{s_2}:=\mathcal V^\dagger(\hat\Pi_{s_2})$ is a positive block-diagonal operator, so $\Tr[B_{s_2}\hat\Pi_{s_1}\hat\rho]=\Tr[(B_{s_2})_{s_1}\,\hat\rho_{s_1s_1}]$ is the trace of a product of two positive operators supported in sector $s_1$: real and non-negative. If instead $\hat\rho$ is sector-diagonal, $\hat\Pi_{s_1}\hat\rho=\hat\Pi_{s_1}\hat\rho\hat\Pi_{s_1}$, and $Q(s_1,s_2)=\Tr[\hat\Pi_{s_2}\,\mathcal V(\hat\Pi_{s_1}\hat\rho\hat\Pi_{s_1})]$ is a probability. The marginal statements follow from $\sum_{s_1}\hat\Pi_{s_1}=\id$ and trace preservation of $\mathcal V$. $\square$

\subsection{\texorpdfstring{\protect{Proof of \Cref{prop:dissipation}}}{Proof of the dissipation threshold}}

The strange-state Gross function has one negative point, $-\tfrac13$, and eight equal positive points, $+\tfrac16$. A Pauli channel acts as convolution by $p_v$ (\Cref{prop:dvcycle} on the logical space), so every output value is $p_v(-\tfrac13)+(1-p_v)\tfrac16=\tfrac16-\tfrac{p_v}2$, negative iff $p_v>\tfrac13$; summing $|\cdot|$ and subtracting $1$ gives \Cref{eq:Nclosed}. For the weights \eqref{eq:qweights}, $p_0=q_0^2$ alone exceeds $\tfrac13$ until uniformisation, giving $\NC=q_0^2-\tfrac13$ and the stated threshold, located by root-finding on the erf sums at working precision $10^{-20}$. $\square$

\section{The reflection identity and the jump kernels}\label{app:gain}

The reflection identity \eqref{eq:reflection} comes from $\hat D^\dagger(\zeta)\hat a\hat D(\zeta)=\hat a+\zeta$ and $(-1)^{\hat n}\hat a(-1)^{\hat n}=-\hat a$:
\begin{equation}
\begin{split}
\hat\Pi(\zeta)\hat a\hat\Pi(\zeta)&=\hat D(\zeta)(-1)^{\hat n}(\hat a+\zeta)(-1)^{\hat n}\hat D^\dagger(\zeta)\\
&=\hat D(\zeta)(\zeta-\hat a)\hat D^\dagger(\zeta)=2\zeta-\hat a.
\end{split}
\end{equation}
\Cref{eq:kloss} follows from $\hat\Pi^2=\id$: $\hat a^{\dagger k}\hat\Pi\hat a^k=\hat\Pi(\hat\Pi\hat a^{\dagger}\hat\Pi)^k\hat a^k=\hat\Pi(2\zeta^*-\hat a^\dagger)^k\hat a^k$. The gain form \eqref{eq:gain} can be obtained the same way, or derived directly: with $(\hat a+\zeta)(-1)^{\hat n}=(-1)^{\hat n}(-\hat a+\zeta)$,
\begin{equation}
\begin{split}
\hat a&\hat\Pi(\zeta)\hat a^\dagger
=\hat D(\zeta)(\hat a{+}\zeta)(-1)^{\hat n}(\hat a^\dagger{+}\zeta^*)\hat D^\dagger(\zeta)\\
&=-\hat D(\zeta)(-1)^{\hat n}\big[\hat n{+}1+\zeta^*\hat a-\zeta\hat a^\dagger-|\zeta|^2\big]\hat D^\dagger(\zeta)\\
&=-\hat\Pi(\zeta)\big(\hat n+1-2\zeta\hat a^\dagger\big),
\end{split}
\end{equation}
using $\hat D(\zeta)\hat n\hat D^\dagger(\zeta)=\hat n-\zeta\hat a^\dagger-\zeta^*\hat a+|\zeta|^2$ and the invariance of $\zeta^*\hat a-\zeta\hat a^\dagger$ under the same conjugation. The loss identity \eqref{eq: cat loss kernel exact} follows from the same algebra with the order of $\hat a$ and $\hat a^\dagger$ exchanged. We verified both numerically to $10^{-14}$ at Fock cutoff $200$.

\section{Numerical methods}\label{app:numerics}

Discrete-variable codes are implemented with displacements as index permutations with phases, so kernel traces take $O(d^n)$ per phase-space point; the $[[5,1,3]]_3$ verification evaluates the full $729$-coset quotient exactly, with cyclic stabilisers $XZZ^{-1}X^{-1}\id$ checked for symplectic closure, logical pairs found in the commutant modulo the lattice, and the logical basis gauge-fixed by $\bar X$. The sector-graded cat construction of \Cref{sec:seclog} uses exact residue-class supports at cutoff $160$.

Every statement of \Cref{subsec:dv} was verified, to $10^{-9}$ or better, on the three-qutrit repetition code $[[3,1]]_3$ and the five-qutrit code $[[5,1,3]]_3$; \Cref{tab:dv} summarises the results. The encoded qutrit strange state $\ket{\mathbb S}\propto\ket1-\ket2$ has $\NC^{\mathcal C}=2/3$, its logical Gross value, preserved by \Cref{eq:synavg} under a Pauli channel of correctable errors. As a stress test, take instead the three-component channel with weights $\{\tfrac12\,\id, \tfrac14\,\bar Z, \tfrac14\,X_1\}$, whose $\bar Z$ branch is a syndrome-free logical error, so that the zero-syndrome conditional $\hat\rho_L(0)=\tfrac23\ketbra{\mathbb S}{\mathbb S}+\tfrac13 Z\ketbra{\mathbb S}{\mathbb S}Z^\dagger$ is a rank-two mixture. Both sides of \Cref{eq:synavg} give $\NC^{\mathcal C}=5/12$, the right-hand side as $\tfrac34\cdot\tfrac13+\tfrac14\cdot\tfrac23$.

\begin{table}[t]
\caption{\protect{Numerical verification of \Cref{subsec:dv} on two different qutrit codes ($d=3$). $\ket{\mathbb S}\propto\ket1-\ket2$ is the strange state.}}
\label{tab:dv}
\begin{ruledtabular}
\setlength{\tabcolsep}{3pt}\footnotesize
\begin{tabular}{lcc}
 & $[[3,1]]$ & $[[5,1,3]]$\\
\colrule
isotropy of a generic codeword $=\mathcal{M}$ & \checkmark & \checkmark\\
$\sum_{X^{\mathcal C}}\Delta^{\mathcal C}=d^k\id$; $\Tr\Delta^{\mathcal C}=1$ & \checkmark & \checkmark\\
Eq.~\eqref{eq:dv37}, no residual constant; error sectors $0$ & \checkmark & \checkmark\\
error $=$ rigid translation; cycle $=$ id & \checkmark & \checkmark\\
twirl $=$ syndrome dephasing & \checkmark & \checkmark\\
$\NC^{\mathcal C}$ encoded $\ket{0_L},\ket{+_L}$ & $0$ & $0$\\
$\NC^{\mathcal C}$ encoded $\ket{\mathbb S}$ & $2/3$ & $2/3$\\
Eq.~\eqref{eq:synavg} under Pauli noise & \checkmark & \checkmark\\
\end{tabular}
\end{ruledtabular}
\end{table}

For \Cref{prop:nonpauli}, all counts were verified for a single qutrit with the Fourier gate $F$ as code symmetry, any two-dimensional span of $F$-eigenvectors being $F$-invariant: the $\langle F\rangle$-average fails displacement covariance; all $2^{(d^2-1)/2}=16$ sign kernels pass every axiom used; exactly one representation, Gross's, is non-negative on the twelve pure stabiliser states; and $F$-covariance cuts sixteen to four. The spin-$1$ checks for \Cref{prop:su2} are described with its proof in Appendix~\ref{app: proofs}.

\Cref{tab: negative volumes}'s cat and binomial negativities used two independent implementations. In the first, we expanded each codeword in the Fock basis and truncated at $n_{\max}=280$, which at $|\alpha|^2=6$ leaves a residual weight far below machine precision; the binomial codewords are exact at this truncation. Writing $\beta=re^{i\theta}$ and using $\hat{D}(\beta)=e^{i\theta\hat{n}}\hat{D}(r)e^{-i\theta\hat{n}}$, the Wigner function separates as $W(r,\theta)=\frac{2}{\pi}\sum_{m,n}\rho_{nm}\,e^{i(m-n)\theta}\,\bra{m}\hat{\Pi}(r)\ket{n}$, with the real-argument displaced parity $\bra{m}\hat{\Pi}(r)\ket{n}=\sum_k(-1)^k\bra{m}\hat{D}(r)\ket{k}\bra{n}\hat{D}(r)\ket{k}$; we obtained the real orthogonal matrix $\bra{m}\hat{D}(r)\ket{k}$ by exponentiating the real antisymmetric generator $r(\hat{a}^\dagger-\hat{a})$, avoiding the overflow the closed-form Laguerre expression suffers at large $r$. The grid is $460$ radial by $1400$ angular points out to $|\beta|=6.5$, beyond which the weight is negligible. Two checks bound the discretisation error: $\int d^2\beta\,W^{\mathrm{std}}=1$ is recovered to better than $10^{-4}$; and halving the grid spacing changes each tabulated value by less than $10^{-3}$ (why we quote two decimal places). The implementation was checked against a direct evaluation of $\bra{m}\hat{D}(\gamma)\ket{n}$ in the Laguerre form and against the known Fock-state values $\NC_{\ket{1}}=0.426$ and $\NC_{\ket{2}}=0.729$. The second implementation\rev{, the one in the deposited notebook~\mbox{\cite{BozzettoHance2026Notebook}},} uses Laguerre-harmonic quadrature at Fock cutoff \rev{$64$ on a polar grid of $500$ radial points out to $|\beta|=7$ and $480$ angles}, and agrees with the first to the quoted precision.

\Cref{tab: negative volumes} lists the negative volumes of the input codewords and of the normalised post-event states $\hat{a}\hat{\rho}\hat{a}^\dagger/\av{\hat{n}}$ and $\hat{a}^\dagger\hat{\rho}\hat{a}/\av{\hat{n}+1}$, so that every entry is the negative volume of a normalised distribution.

Every quantitative claim above was verified in Wolfram Mathematica: the single-qutrit facts, strange-state Wigner values and sign-family counts in exact arithmetic; the full $[[3,1]]_3$, Kirkwood-Dirac and two-time computations, yielding the exact values $2/\sqrt3$, $1/6$, $2/3$, $2/9$ and $5/12$; the operator identities, thresholds, multi-round sequence and spin-1 checks; and the sector-graded cat numbers. The $[[5,1,3]]_3$ negativities were re-derived by an algorithmically independent characteristic-function and symplectic-FFT route, giving coset sums $3$ and $5$, hence $\NC=2/3$, exactly\rev{; the deposited notebook runs every row of \mbox{\Cref{tab:dv}} on both codes, the $[[5,1,3]]_3$ rows by that route, which it checks against explicitly constructed phase-point operators}.

\end{appendices}
\end{document}